\documentclass[11pt,twoside]{article}
\newcommand{\chapter}[1]{\section{#1}}

\usepackage{graphicx,color}

\usepackage{amsmath}
\usepackage{amsfonts}
\usepackage{amssymb}

\usepackage{a4wide}

\usepackage[thmmarks,amsmath,amsthm]{ntheorem}

\usepackage[mathscr]{eucal}

\usepackage{fancyhdr}

\theoremstyle{nonumberplain}

\theoremstyle{plain}
\newtheorem{thm}{Theorem}[section]

\newtheorem{lemma}[thm]{Lemma}
\newtheorem{cor}[thm]{Corollary}

\theoremstyle{remark}
\newtheorem{claim}{Claim}

\theoremsymbol{\ensuremath{_\square}}
\renewtheorem*{proof}{Proof\setcounter{claim}{0}}
\theoremsymbol{\ensuremath{_\dashv}}
\newtheorem*{clproof}{Proof}

\newcommand{\nc}[1]{\expandafter\newcommand\csname#1\endcsname}

\nc{gX}[1]{\nc{g#1}{\mathfrak #1}} 

\nc{aX}[1]{\nc{a#1}{\mathcal #1}} 

\gX G
\gX H
\gX C
\gX S
\gX A

\aX H

\newcommand{\gneut}{0}

\newcommand{\m}[1]{\mathcal{#1}}
\newcommand{\mfam}[1]{\mathbf{#1}} 

\newlength{\ppshort}
\newcommand{\idfunc}{\mathbf{id}}

\newcommand{\Nat}{ \mathbb{N}}
\newcommand{\Int}{ \mathbb{Z}}
\newcommand{\Real}{ \mathbb{R}}

\newcommand{\NP}{\textup{NP}}

\newcommand{\scalp}[1]{\ensuremath{\langle #1\rangle}}
\newcommand{\mspan}[1]{\ensuremath{\langle #1\rangle}}

\newcommand{\Tle}{ \ensuremath{\le} }  
\newcommand{\Tequiv}{ \ensuremath{\equiv} }  
\newcommand{\rank}[1]{\ensuremath{\mathrm{rank}\, #1}} 
\newcommand{\abs}[1]{{\ensuremath{\textup{abs}\left({#1}\right)}}} 

\newcommand{\tr}[1]{\ensuremath{\mathrm{tr}(#1)}} 

\newcommand{\ord}{\text{ord}} 

\newcommand{\sdef}[1]{\emph{#1}}
\newcommand{\sdefi}[2]{\emph{#1}}

\newcommand{\sdefisub}[3]{\emph{#1}}

\newcommand{\sdefis}[3]{\emph{#1}}
\newcommand{\idxsymb}[1]{#1}
\newcommand{\idxsymbm}[1]{#1}

\newcommand{\marc}[1]{}
\newcommand{\draftOK}[1]{}

\newcommand{\PP}{\textup{P}}

\newcommand{\cnt}{\textup{COUNT}}

\newcommand{\dlen}[1]{\textup{len}(#1)}

\renewcommand{\vec}[1]{\mathbf{#1}}

\newcommand{\iu}{\textup{i}} 
\newcommand{\itpi}{2\pi\textup{i}} 

\newcommand{\cj}[1]{\ensuremath{\overline{#1}}} 

\newcommand{\eval}{\textup{EVAL}}

\newcommand{\evalk}{\textup{EVAL}^{\textsf{pin}}}

\newcommand{\absent}{\ensuremath{*}}
\newcommand{\row}[1]{\ensuremath{_{#1,*}}}
\newcommand{\col}[1]{\ensuremath{_{*,#1}}}
\newcommand{\C}{\mathbb{C}}
\newcommand{\Q}{\mathbb{Q}}
\renewcommand{\Int}{\mathbb{Z}}

\usepackage{mathrsfs}

\newcommand{\cngc}[1]{\llbracket#1\rrbracket}

\usepackage{stmaryrd}

\newcommand{\twres}[1]{[#1]}

\newcommand{\vcfg}{\sigma} 

\newcommand{\indeg}{\text{deg}^-}
\newcommand{\outdeg}{\text{deg}^+}
\newcommand{\grade}{\partial}

\newcommand{\mmod}[1]{\;(\text{mod } #1)}

\newenvironment{condition}[1]
{\begin{list}{} {\setlength{\leftmargin}{40pt} 
                \setlength{\itemindent}{-6pt}}
\item[\textbf{#1}]} 
{\end{list}}

\newcommand{\cond}[1]{\textup{\textbf{(#1)}}}

\newcommand{\df}{\textup{def}}  

\newcommand{\rw}{\textup{row}}
\newcommand{\cl}{\textup{col}}

\newcommand{\U}{\mathbb{U}} 
\newcommand{\Ualg}{\mathbb{U}_\mathbb{A}} 

\newcommand{\vpin}{\phi} 

\newcommand{\Qu}{\mathbb{Q}}

\newcommand{\wset}{\mathcal{W}} 

\newcommand{\vdeg}{d} 

\newcommand{\diag}{\textup{diag}}

\newcommand{\bil}[1]{\langle #1 \rangle_{\denom}} 
\newcommand{\rcwf}{\rho} 

\newcommand{\caniso}{\mathscr X}

\newcommand{\denom}{\omega}

\newcommand{\Ralg}{\mathbb{R}_{\mathbb{A}}}
\newcommand{\Calg}{\mathbb{C}_{\mathbb{A}}}

\begin{document}
\title{The Complexity of Partition Functions on Hermitian matrices}
\author{ Marc Thurley\footnote{This work was supported in part 
by the Deutsche Forschungsgemeinschaft within
the research training group ’Methods for Discrete Structures’ (GRK 1408)
and by a fellowship within the Postdoc-Programme of the German Academic Exchange Service (DAAD).} \\
Computer Science Division \\
University of California \\
Berkeley CA 94720-1776, U.S.A. \\
thurley@eecs.berkeley.edu
}

\maketitle

\begin{abstract}
\noindent
Partition functions of certain classes of ``spin glass'' models in statistical 
physics show strong connections to combinatorial graph invariants. Also known
as \emph{homomorphism functions} they allow for the representation of many such
invariants, for example, the number of independent sets of a graph or the number
nowhere zero $k$-flows.

Contributing to recent developments on the complexity of partition functions 
\cite{golgrojerthu09,caichelu09} we study the complexity 
of partition functions with complex values. These functions are usually 
determined by a square matrix $A$ and it was shown in \cite{golgrojerthu09} that
for each real-valued symmetric matrix, the corresponding partition function is
either polynomial time computable or $\#\PP$-hard.

Extending this result, we give a complete description of the complexity of 
partition functions definable by Hermitian matrices. These can also be 
classified into polynomial time computable and $\#\PP$-hard ones. Although 
the criterion for polynomial time computability is not describable in a single 
line, we give a clear account of it in terms of structures associated with
Abelian groups.
\end{abstract}

\newpage 

\tableofcontents

\pagestyle{headings}

\section{Introduction}
We study the complexity of \emph{partition functions} 
(or \emph{homomorphism functions}) on Hermitian\footnote{That is, $A_{ij} = \cj{ A_{ji}}$ for all $i,j$, where $\cj a$ denotes complex conjugate.}
matrices --- a class of
partition functions which has been
studied recently in \cite{lovsch08}.
Let $A \in \C^{m \times m}$ be a Hermitian matrix
and $D \in \Real^{m \times m}$
a diagonal matrix with positive diagonal. Let $G = (V,E)$ be a directed graph with
possibly multiple edges.
The \emph{partition function} $Z_{A,D}$ on $G$ is defined by
\begin{equation}\label{eq:pf_intro_one}
Z_{A,D}(G) = \sum_{\vcfg: V \rightarrow [m]} \prod_{uv \in E} A_{\vcfg(u),\vcfg(v)} \prod_{v \in V} D_{\vcfg(v),\vcfg(v)}.
\end{equation}
It is important to note that the product over edges $uv \in E$ 
counts multiplicities as well.
If the matrix $A$ is symmetric, the direction of edges in $G$ does not affect 
the value of the partition function. We may then think of $G$ 
likewise as an undirected graph. In general, however, the partition function is
well-defined only on digraphs.
We refer to $D$ also as a \emph{diagonal matrix of vertex weights} and 
usually we omit it from the subscript of $Z_{A,D}$ if it is the identity 
matrix. Elements of $[m]$ are called \sdef{spins} and
mappings $\vcfg: V \rightarrow [m]$ which 
assign a spin to every vertex of $G$ are \sdef{configurations}.
Let $\Calg$ denote the set of algebraic numbers and let $\Ralg$
be the algebraic reals. 

In this paper we study the complexity of partition 
functions of the form $Z_{A,D}$ for a Hermitian matrix 
$A \in \Calg^{m \times m}$ and a diagonal matrix $D \in \Ralg^{m \times m}$
with positive diagonal. Our main result 
(cf. Theorem~\ref{thm:hermitian_main}, below) is a complete complexity 
classification of these functions.

\subsection{Examples.}
There are several well-known examples of graph-invariants 
representable by partition functions. 
Let $G = (V,E)$ be a digraph. The number of independent 
sets of $G$ is given by $Z_S(G)$ where $S$ is a $2\times 2$ matrix 
which satisfies $S_{11} = 0$ and it is one in all other positions.
If $A$ is the adjacency matrix of a $k$-clique without self-loops, then $Z_A(G)$
is the number of $k$-colorings of $G$.

It will
be convenient now and in the following to consider matrices likewise as weighted
(directed) graphs such that entry $A_{ij}$ is the weight of edge $ij$ and this 
value being zero denotes absence of an edge.
Viewed this way, $Z_A(G)$ is the number of homomorphisms from $G$ to 
(the digraph represented by) $A$.

There is an important connection between partition functions and the 
\emph{Tutte polynomial}. Let $G=(V,E)$ be a graph. 
For some subset $A\subseteq E$ of edges, let $c(A)$ denote the number of 
components of the graph $(V,A)$, then the \emph{Tutte polynomial} of $G$, with
parameters $q,v$ is defined by
\[
T(G;q,v) = \sum_{A \subseteq E} q^{c(A)}v^{|A|}.
\]
If the value $q$ is a positive integer then this is also known as the partition
function of the
\emph{$q$-state Potts model}. 
By expanding the above sum over components of $G$, it can also be shown that,
for $A = J_q + v\cdot I_q$, we have $T(G;q,v) = Z_{A}(G)$. Here,
$J_{q}$ denotes the all-ones $q\times q$ matrix 
and $I_q$ the $q\times q$ identity matrix.
In fact this is also an exact correspondence since (see \cite{frelovsch07}) 
$q \in \Nat$ is the only way for which $T(G;q,v) = Z_{A}(G)$ for a symmetric 
real-valued $A$.

\paragraph*{Flows and Hermitian Matrices.} Let us turn to an
example motivating partition functions on Hermitian matrices.
We start with a simple example (which can be found e.g. in \cite{harjon93}), 
given by the matrices
\begin{equation}\label{eq:def_eul}
U=\left(\begin{array}{r r}
  1&-1\\
  -1&1
\end{array}\right)
\quad \quad 
D_U =\left(\begin{array}{r r}
  1/2 & 0\\
  0 & 1/2
\end{array}\right).
\end{equation}
It is not hard to see that, for connected $G$, the value $Z_{U,D_U}(G)$ equals the characteristic function of Eulerian graphs. That is, it is $1$ if $G$ is Eulerian and $0$ otherwise.
We will see a proof of this in the more general setting of \emph{flows} which we will introduce now.

Usually the representability of \emph{nowhere zero flows} by partition functions
is derived via the Tutte polynomial which has strong connections to both.
Deviating from this, a direct derivation of this correspondence has been given in
\cite{frelovsch07}.

Let $\gG$ be a finite Abelian group and $S \subseteq \gG$ any subset. For some oriented graph $G = (V,E)$ a mapping $\phi: E \to S$ and a $v \in V$ let
\begin{equation}\label{eq:3105091351}
\partial_\phi(v) = \sum_{uv \in E} \phi(uv) - \sum_{vu \in E} \phi(vu)
\end{equation}
Such an assignment $\phi$ is called an \sdef{$S$-flow}, if $\partial_\phi(v) = \gneut$ holds for every vertex $v \in V$. 
For some oriented graph $G$ let $f_{S,\gG}(G)$ denote the number of $S$-flows (w.r.t. $\gG$) in $G$. Let $\gG^*$ denote the character group of $\gG$. 
The functions $f_{S,\gG}$ relate to partition functions in the following way.
Define two matrices with indices in $\gG^*$: a matrix $A$ and a diagonal matrix $D$ given by
\begin{equation}\label{eq:3105091655}
D_{\chi\chi} = \vert \gG \vert^{-1} \quad \text{ and } 
A_{\chi,\chi'} = \sum_{g \in S} \cj{\chi(g)}\cdot \chi'(g)
\end{equation}
It is not too hard to show that (see \cite{frelovsch07} for details)  
for every oriented graph $G$ we have $Z_{A,D}(G) = f_{S,\gG}(G)$. 
If we stipulate that $S$ be closed under inversion
then $A$ is real-valued and symmetric. The orientation of $G$ is then
inessential. Particularly, if $S= \gG \setminus \{0\}$ then $Z_{A,D}(G)$ is
the number of \sdef{nowhere-zero $\gG$-flows} of $G$. Since this value 
depends only on the order $k = \vert \gG \vert$ of $\gG$, this counts
\sdef{nowhere-zero $k$-flows} of $G$.

If $S$ is not closed under inversion, then generally
$A$ is not merely a real symmetric matrix: 
Let, for example $\gG = \Int_3$ and $S = \{2\}$. Define $\zeta = \exp(\itpi/ 3)$, and consider the character group $\gG^*$ given by $\chi_0 = (1,1,1)$, $\chi_1 = (1, \zeta,\zeta^2)$ and $\chi_2 = (1, \zeta^2,\zeta)$. The definition of $A$ is given by $A_{i,j} = \cj{\chi_{i}(2)}\cdot \chi_{j}(2)$ which yields
\begin{equation*}\label{eq:herm_counting_flows}
A = \left( \begin{array}{c c c}
                1 & \zeta^2 & \zeta \\
                \zeta^{-2} & 1 & \zeta^{-1} \\
                \zeta^{-1} & \zeta & 1
           \end{array}\right).
\end{equation*}

\subsection{Complexity}

A general study of the complexity of partition functions has developed only 
quite recently, although
some instances of these functions can be found among the first problems studied in the context of counting complexity. It follows, for example, from a result of Simon \cite{sim77} that computing the number of independent sets in a graph is $\#\PP$-hard. This has also been derived independently
by Provan and Ball \cite{probal83}. The $\#\PP$-completeness of counting $k$-colorings for all $k \ge 3$ is shown by Linial in \cite{lin86}. 

More general complexity results which have a connection to partition functions have been obtained by
Jerrum and Sinclair \cite{jersin93} who show in particular that the partition function of the $2$-state Potts model is $\#\PP$-hard to compute. Related to this, Jaeger, Vertigan and Welsh \cite{jaeverwel90} study the complexity of computing the Tutte polynomial. 

Results directly related to our work have strong motivational connections
to research in constraint satisfaction problems
(CSP) (see e.g. \cite{fedvar98,Bul06,BarKoz09}). As a matter of fact partition
functions and related objects are instances of weighted generalizations of the
\emph{counting} constraint satisfaction problem (\#CSP).

Of particular interest are \emph{dichotomy results} in the spirit of 
Schaefer's \cite{sch78} classical result on Boolean CSP:
Depending on the structure of the constraints allowed in a CSP instance,
each of these problems is either in $\PP$ or $\NP$-complete.
That such results are not self-evident was shown by Ladner 
\cite{lad75}: if $\NP$-hard problems are not polynomial time computable, then 
there is an infinite hierarchy of problems of increasing difficulty between these
extremes.
The situation is analogous with counting problems and it is assumed, 
for example, that counting graph isomorphisms is 
neither in $\PP$ not $\#\PP$-complete (cf. \cite{mat79}).

Feder and Vardi \cite{fedvar98} conjectured that such intermediate complexity is
not possible for CSPs. And in fact, the analogous conjecture 
for the $\#$CSP problem has been shown
to be true very recently by Bulatov \cite{bul08} (see also the new proof by 
Dyer and Richerby \cite{dyeric10}).
Note that the counting analog of Schaefer's result has been obtained much earlier
by Creignou and Hermann \cite{creher96} and several weighted extensions of this 
have been studied recently 
(see \cite{dyegoljer09,buldyegoljr08,cailuxia09,dyegoljalric09}).

The line of dichotomy results for partition functions starts with
Dyer and Greenhill \cite{dyegre00} who studied
partition functions $Z_A$ on symmetric $\{0,1\}$-matrices.
This is a counting analog of a result by Hell and Ne\v{s}et\v{r}il \cite{helnes90}. 
It turns out that
these functions are polynomial time computable only if every component of $A$ 
is either a complete bipartite graph, or a complete non-bipartite graph with a 
loop at every vertex. In all other cases the problem is $\#\PP$-hard. 

Bulatov and Grohe \cite{bulgro05} extended this to a dichotomy for all non-negative 
symmetric matrices. To state this result, we need to introduce the notion of 
\emph{blocks} of a matrix $A$. Considering $A$ as
a graph, blocks arise naturally as follows: each non-bipartite component 
corresponds to one
block and each bipartite one corresponds to two blocks.

\begin{thm}[Bulatov \& Grohe \cite{bulgro05}]\label{thm:BG}
Let $A \in \Ralg^{m \times m}$ be a non-negative symmetric matrix. The 
following holds.
\begin{description}\itemsep = 0pt
 \item[(1)] If every block of $A$ has rank at most $1$ then $Z_A$ is polynomial time computable.
 \item[(2)] If $A$ contains a block of rank at least $2$ then $Z_A$ is $\#\PP$-hard to compute.
\end{description}
\end{thm}
The fact that this result essentially identifies rank $1$ matrices as the 
polynomial time cases renders the result of Bulatov and Grohe widely applicable. 
Many results on partition functions rely on this theorem, as is the 
case of Bulatov's result \cite{bul08}.
A further example is the work of Dyer, Goldberg and Jerrum \cite{dyegoljer08} 
on \emph{hypergraph partition functions}. 
Dyer, Goldberg and Paterson \cite{dyegolpat07} give a dichotomy for partition functions $Z_A$ where $A$ is an unweighted directed acyclic graph. This has been extended to a dichotomy for all
directed $A$ with non-negative edge-weights by Cai and Chen \cite{caiche09}.

There are at least two further extensions 
of the result in \cite{bulgro05} to be mentioned. 
Goldberg, Grohe, Jerrum and Thurley \cite{golgrojerthu09} gave a dichotomy 
theorem for partition functions $Z_A$ on any symmetric real-valued matrix $A$,
that is, particularly negative entries in $A$ are allowed. 
Furthermore, Cai, Chen and Lu \cite{caichelu09} extended this dichotomy 
yet further to all symmetric complex valued matrices $A$.

\paragraph*{The main result of this paper.}
Here we will extend the work of \cite{golgrojerthu09} into another natural 
direction by studying partition functions on Hermitian matrices. Our main result
is a dichotomy for these partition functions. As a byproduct, our proof will give a clean account of the 
algebraic structure underlying the proof of \cite{golgrojerthu09}.

\begin{thm}\label{thm:hermitian_main}
Let $A \in \Calg^{m \times m}$ be a Hermitian matrix and 
$D \in \Ralg^{m \times m}$ a diagonal matrix of positive vertex weights. 
Then one of the following is true
\begin{description}\itemsep = 0pt
\item[(1)] The function $Z_{A,D}$ is $\#\PP$-hard to compute.
\item[(2)] The function $Z_{A,D}$ is polynomial time computable.
\end{description}
Furthermore, there is an algorithm which, given $(A,D)$ decides, whether (1) or (2) occurs.
\end{thm}
Since this work and \cite{caichelu09} are both generalizations of the same result
\cite{golgrojerthu09} it turns out that the proof structure and the polynomial 
time cases are rather similar from a high-level point of view.
However, the details are quite different in all three proofs.

After some preliminaries presented in the next section, 
we will give a detailed outline of the proof of this theorem in 
Section~\ref{sec:herm_prelim}.
The details of the proofs will then be given in the rest of the paper starting 
in Section~\ref{sec:technical_prelim}.

\section{Preliminaries}
For two counting problems $A$ and $B$ we write $A \Tle B$ if there is a polynomial time reduction from $A$ to $B$. If $A \Tle B$ and $B \Tle A$ holds, we write $A \Tequiv B$. 

\paragraph*{Computing with Complex Numbers.}
Recall that $\Calg$ denotes the algebraic numbers and $\Ralg = \Calg \cap \Real$.
For technical
reasons, we will always assume that numbers in $\Calg$ are given in
\emph{standard representation} in some algebraic extension field
$\Qu(\theta)$. 
That is, we consider numbers in $\Qu(\theta)$ as vectors in a
$d$-dimensional $\Qu$-vectorspace, where $d$ is the degree of $\Qu(\theta)$ over 
$\Qu$.
It is well-known that for any set of numbers from
$\Calg$ we can compute a $\theta$ which constitutes the corresponding
extension field (cf. \cite{coh93} p. 181). For further details see
also the treatment of this issue in \cite{dyegoljer08,thu09}.

\smallskip

\subsection{Basic Notation}\label{sec:notation}\marc{Section OK}

For natural numbers $a\le b$ we let $[a,b] := \{a, \ldots, b\}$ denote the set of natural numbers from $a$ to $b$ and we define $[b]:= [1,b]$.

\paragraph*{Graphs.} 
We will be mainly interested in \sdefi{directed graphs}{directed graph} 
$G = (V,E)$ which may have multiple edges. We denote edges by $uv \in E$, to explicate
that an edge $e$ has multiplicity $p$ we sometimes write $e^p$. 
We denote the out-degree (in-degree, resp.) of a vertex $v$ by $\outdeg_G(v)$ ($\indeg_G(v)$, resp.).
The \emph{degree} of $v$ is $\vdeg_G(v) = \indeg_G(v) + \outdeg_G(v)$ in contrast to
the \emph{grade} $\grade_G(v) = \outdeg_G(v) - \indeg_G(v)$. 
A \sdef{component} in a digraph $G=(V,E)$ will be what is usually 
known as \sdef{weakly connected component}. 

Let $G$ be a digraph and $p \in \Nat$. A graph $G'$ is obtained by \sdefi{$p$-thickening}{thickening}, if each edge $e \in E$ is replaced by $p$ many parallel copies of $e$. Similarly, $G'$ is obtained from $G$ by \sdefi{$p$-stretching}{stretching} if each edge is replaced by a path of length $p$ --- that is, a path on $p$ edges.

\paragraph*{Matrices and Vectors.} 

For an $m \times n$ matrix $A$ whenever convenient we separate indices by a comma \idxsymb{$A_{i,j}$} to increase readability.
Further, \idxsymb{$A\row i$} (\idxsymb{$A\col j$}, resp.) denote the $i$th row ($j$th column, resp.) of $A$.
For an $m\times n$ matrix $A$ and sets $I \subseteq [m],\, J \subseteq [n]$ the matrix $A_{IJ}$ is obtained from $A$ by deleting all rows with indices not occurring in $I$ and all columns not in $J$. If $I = [m]$ ($J = [n]$) we abbreviate this to $A\col J$ ($A \row I$, resp.).

For every $p \in \Qu$ we define the matrix $A^{(p)}$ by
$$
A^{(p)}_{ij} = \left\{ \begin{array}{l l}
                        (A_{ij})^p &, \text{ if } A_{ij} \neq 0 \\
                          0 &, \text{ otherwise.}
                       \end{array}\right.
$$

By \idxsymb{$\tr A$} we denote the trace of $A$ and $I_m$ is the $m \times m$ 
identity matrix. Let $A$ be an $m \times n$ matrix and $A'$ an $m' \times n'$ 
matrix. The \emph{tensor product} $A\otimes A'$ of $A$ and $A'$ is the matrix 
defined as follows. Let the indices be given as pairs $(i,k) \in [m]\times[m']$ for the row indices and $(j,l) \in [n]\times[n']$ for the column indices, then
$$
(A \otimes A')_{(i,k)(j,l)} = A_{ij} \cdot A_{kl}.
$$
With the bijections $\pi_\rw: [m]\times[m'] \rightarrow [m\cdot m']$ given by $\pi_\rw(i,k):= (i-1)\cdot m' + k$ and $\pi_\cl: [n]\times[n'] \rightarrow [n\cdot n']$ given by $\pi_\cl(j,l):= (j-1)\cdot n' + l$ the pairs translate to ordinary indices. But most often it will be convenient to explicitly refer to the indices as pairs. In this way we will be able to refer to the \sdefi{tiles}{tiles of a matrix}\sdefisub{}{matrix}{tiles of a} of $A \otimes A'$ which are the submatrices $(A\otimes A')_{(i,\absent)(j,\absent)} = A_{ij}\cdot A'$.

The \sdefis{direct sum}{direct sum of matrices}{$A \oplus B$}\sdefisub{}{matrix}{direct sum} of $A$ and $A'$ is the $(m+m')\times (n+n')$ matrix $A \oplus A'$ defined by
$$
(A \oplus A')_{ij} = \left\lbrace\begin{array}{l l}
                                   A_{ij}&, \text{ if } i \in [m],\; j \in [n] \\
                                   A'_{ij}&, \text{ if } i \in [m+1,m+m'],\; j \in [n+1,n+n'] \\
                                    0 &, \text{ otherwise} 
                                 \end{array}\right.
$$
Let $\vec a, \vec b \in \C^n$ be vectors. The \sdef{scalar product} 
of $\vec a$ and $\vec b$ is  
$\idxsymb{\scalp{\vec a,\vec b}} = a_1\cj b_1 + \ldots + a_n \cj b_n$. 
The \sdef{Hadamard product} is the vector 
$\idxsymbm{\vec a \circ \vec b} = (a_1b_1, \ldots, a_nb_n)$. The 
\sdef{conjugate Hadamard product} 
is $\idxsymbm{\vec a \bullet \vec b} = \vec a \circ \cj {\vec b}$.

We denote by \idxsymb{$\U$} the set of all complex numbers $a$ of
absolute value $1$. The algebraic numbers of absolute values $1$ are denoted by
$\Ualg$. Similarly, $\U_{\omega}$ denotes all $\omega$-th roots of unity.
A matrix $A \in \U^{m \times n}$ is called \sdef{normalized} if its first 
row and column are constantly $1$.
The expression $\cj A$ denotes the \sdefis{complex conjugate}{conjugate matrix}{$\cj A$} of $A$. 
A matrix $A \in \U^{m \times n}$ is called a \sdef{complex Hadamard matrix} 
if $A\cj A^T = nI_n$, that is all of its rows are pairwise orthogonal.
Note that this implies that the columns of $A$ are pairwise orthogonal as well,
which follows from elementary group theory. \marc{maybe somewhere where the 
$\U$ is used first?}

\paragraph*{Matrices, blocks and their connection to graphs.}

An $m \times n$ matrix $A$ is \sdefi{decomposable}{decomposable matrix}\sdefisub{}{matrix}{decomposable}, if there are non-empty index sets $I \subseteq [m]$, $J \subseteq [n]$ with $(I,J) \neq [m] \times [n]$ such that $A_{ij} = 0$ for all $(i,j) \in \bar I \times J$ and all $(i,j) \in I \times \bar J$. 
A matrix is \sdefi{indecomposable}{indecomposable matrix}\sdefisub{}{matrix}{indecomposable}  if it is not decomposable and a \sdefi{block}{block of a matrix}\sdefisub{}{matrix}{block of a} of $A$ is a maximal indecomposable submatrix.

Let $A$ be an $m \times m$ matrix. The \sdefi{graph underlying}{graph underlying a matrix} $A$ is the digraph defined by $G_A = ([m],E)$ with $ E:= \{ij \mid A_{ij} \neq 0\}$. For Hermitian matrices 
the digraph $G_A$ is symmetric in the sense that $ij \in E$ iff $ji \in E$. For considerations on the connectedness of this graph it is therefore sufficient to consider it as undirected.

We call a matrix $A$ \sdefisub{connected}{matrix}{connected}\sdefi{}{connected matrix} if $G_A$ is connected.
We call a submatrix $A_{II}$ a \sdefi{component}{component of a matrix} of $A$, if $G_A[I]$ is a component of $G_A$. If $G_A[I]$ is non-bipartite, then $A_{II}$ is a block of $A$. Note in particular that a connected non-bipartite matrix is is a block, and it is thus identical with its \emph{underlying block}. If otherwise $G_A[I]$ is bipartite then $A_{II}$ has an \sdef{underlying block} $B$ such that
$$
A = \left(\begin{array}{c c}
             0 & B \\
             \cj B^T & 0
          \end{array}\right).
$$

\subsection{Generalized Partition Functions}
Let $A \in \C^{m \times m}$ be a matrix and $D \in \C^{m \times m}$ a diagonal matrix. 
Let $G = (V,E)$ be some given digraph. A \sdef{pinning} of (vertices of) $G$ 
with respect to $A$ is a mapping $\vpin: W \rightarrow [m]$ for some subset $W \subseteq V$ of vertices of $G$.  
For a given graph $G = (V,E)$ and a pinning $\vpin$, we define the \sdef{partition function}
$$
Z_{A,D}(\vpin,G) = \sum_{\vpin \subseteq \vcfg: V \rightarrow [m]} \prod_{uv \in E} A_{\vcfg(u),\vcfg(v)} \prod_{v \in V \setminus{\df(\vpin)}} D_{\vcfg(v),\vcfg(v)}.
$$
Note in particular, that the sum is over all configurations $\vcfg: V \rightarrow [m]$ which extend the fixed given pinning $\vpin$ and, for technical reasons, the terms $D_{\vcfg(v),\vcfg(v)}$ for $v \in \df(\vpin)$ are excluded from the above expression. The \emph{weight} of the configuration $\vcfg$ is the term 
$\prod_{uv \in E} A_{\vcfg(u),\vcfg(v)} \prod_{v \in V \setminus \df(\vpin)} D_{\vcfg(v),\vcfg(v)}.
$
Whenever $\vpin$ is trivial, that is, $\df(\vpin) = \emptyset$ and $D$ is the identity we omit the terms $D$ ($\vpin$, respectively) in the expression. 
Let $\evalk(A,D)$ denote the problem of computing $Z_{A,D}(\vpin,G)$ on input 
$G,\vpin$. Analogously, $\eval(A,D)$ denotes the problem restricted to empty 
input pinnings.

\paragraph*{Congruential Partition Functions.}
We say that a matrix $A \in \Calg^{m \times m}$ is \sdef{$\omega$-algebraic}, 
if for every non-zero entry $A_{ij}$ there is some $\omega$-th root of unity 
$\zeta$ such that $A_{ij} = |A_{ij}|\cdot \zeta$.
Working with partition functions on $\denom$-algebraic matrices, it will be 
convenient to transition to a different formulation of such problems. 

Let, for a $d \in \Int$ the expression $\cngc{d}$ denote the modulus of $d$ after division by $\denom$. As $\denom$ is defined relative to $A$, the exact meaning of $\cngc{d}$ will always be clear from context.
A family $\mfam D = (D^{\cngc c})_{c \in \Int_{\denom}}$ of diagonal $m \times m$ matrices will be called a family of \sdefisub{congruential vertex weights}{congruential}{vertex weights}\sdefisub{}{vertex weights}{congruential}.  The \sdefisub{congruential partition function}{congruential}{partition function}\sdefisub{}{partition function}{congruential} on $A$, and $\mfam D = (D^{\cngc c})_{c \in \Int_{\denom}}$, for a digraph $G$ and a pinning $\vpin$ is defined by
$$
Z_{A, \mfam D}(\vpin, G) = \sum_{\vpin \subseteq \vcfg: V \rightarrow [m]} \prod_{uv \in E} A_{\vcfg(u),\vcfg(v)} \prod_{v \in V'} D^{\cngc{\grade(v)}}_{\vcfg(v),\vcfg(v)}
$$
where $V' = V \setminus \df(\vpin)$. Recall that $\grade(v) = \outdeg(v) - \indeg(v)$ denotes the grade of $v$ in $G$. The connection to ordinary partition functions easily follows 
from the definitions:
\begin{lemma}\label{lem:intro_cong}
Let $A \in \Calg^{m \times m}$ be a Hermitian $\denom$-algebraic matrix and $D \in \Ralg^{m \times m}$ a diagonal matrix of positive vertex weights. Let $\mfam D = (D^{\cngc c})_{c \in \Int_{\denom}}$ such that $D^{\cngc c} = D$ for all $c \in \Int_{\denom}$. Then 
$$
   Z_{A,D}(\vpin, G) = Z_{A,\mfam D}(\vpin, G) \quad \text{ for all digraphs $G$ and all pinnings $\vpin$}. 
$$
\end{lemma}
Correspondingly, $\evalk(A,\mfam D)$ denotes the evaluation problem associated with this partition function on input $(\phi,G)$.

\section{The Proof of the Main Theorem --- An Itinerary}\label{sec:herm_prelim}\marc{Section OK}

We will now give a detailed overview of the most important steps in the proof of Theorem~\ref{thm:hermitian_main}. Given an arbitrary Hermitian matrix $A$ and diagonal matrix $D$ it is rather routine to transition to the case of connected $A$ and the more general problems $\evalk(A,D)$.

Starting from this point, we will filter the abundance of problems $\evalk(A,D)$ until we are left with only the polynomial time computable ones. This process of filtering will be performed by transforming in several steps the problem $\evalk(A,D)$ into a problem satisfying certain conditions. In each step more and more such conditions will be accumulated and if any of these conditions fails for a certain problem it will be shown that this is due to the problem being $\#\PP$-hard.

The first step is to gather conditions such that the resulting problem can be transformed into a problem of the form $\evalk(A',\mfam D)$ where the block underlying the connected matrix $A'$ is a complex Hadamard matrix (cf. Lemma~\ref{lem:gen_red_to_Had}) --- we call such matrices \sdefi{Hadamard components}{Hadamard component}.
Once we have achieved this, we will be able to gather more conditions on these problems (cf. Theorem~\ref{thm:hadamard_reduction}). And eventually, we will have to solve the non-trivial task of providing a polynomial time algorithm for the remaining problems --- Theorem~\ref{thm:herm_ptime_partition_functions} yields this. 

Some words are in order for the additional statement of Theorem~\ref{thm:hermitian_main}: It is claimed that there is an algorithm deciding whether given matrices $A,D$ give rise to polynomial time computable partition functions or $\#\PP$-hard ones. This not very surprising fact will follow straightforwardly from the proof of Theorem~\ref{thm:hermitian_main}. We give details on the decision algorithm only if they do not follow straightforwardly from the details of the proofs given.

\subsection{From the General Case to Connected Matrices}

We start by transforming the original problem $\eval(A,D)$ into an equivalent 
problem $\evalk(A',\mfam D)$ with $A'$ an $\denom$-algebraic matrix.
The transition to $\denom$-algebraic matrices is captured by the following lemma.
Its proof can be found in Section~\ref{sec:algstrucpf}.
\begin{lemma}[The Arithmetical Structure Lemma]\label{lem:alg_to_w_alg_red}
Let $A \in \Calg^{m \times m}$ be a Hermitian matrix and $D \in \Ralg^{m \times m}$ a diagonal matrix of positive vertex weights. There is an $\denom \in \Nat$ and an $\denom$-algebraic matrix $A'$ whose entries satisfy $|A'_{ij}| \in \Nat$ and
$$
\eval(A',D) \Tequiv \eval(A,D).
$$
\end{lemma}
With the following lemma we then take the step which allows us to pin certain 
vertices of the input. The proof will be given in Section~\ref{sec:pin}. 
\begin{lemma}[Pinning Lemma]\label{lem:pinning}
Let $A\in \Calg^{m\times m}$ be a Hermitian matrix and $D \in \Ralg^{m\times m}$ a diagonal matrix of positive vertex weights.  Then
\[
  \evalk(A,D) \Tequiv \eval(A,D).
\]
\end{lemma}
The transition 
to congruential partition functions is a consequence of 
Lemma \ref{lem:intro_cong}:
\begin{cor}
We have $\evalk(A,D) \Tequiv \evalk(A,\mfam D)$.
\end{cor}

\subsection{From Connected Matrices to Hadamard Components}
We start under the following preconditions. We are given an $\denom$-algebraic Hermitian matrix $A \in \C^{m \times m}$ which satisfies that $|A_{ij}| \in \Nat$ for all $i,j \in [m]$ however this being in $\Nat$ is not essential. 
We say that a matrix $H$ and a family of diagonal matrices $\mfam D$ \emph{define} an \cond{H--STD} problem, if the following holds
\sdefi{}{\cond{H--STD}-problem}
\begin{condition}{(H--STD)}
There are $n,\denom \in \Nat$ and $\mfam D = (D^{\cngc c})_{c \in \Int_{\denom}}$ is a family of diagonal matrices such that
\begin{itemize}
 \item $H \in \U_{\denom}^{n \times n}$ is a normalized Hermitian Hadamard matrix..
 \item We have $D^{\cngc 0} = I_n$, $D^{\cngc c} \in (\{0\} \cup\Ualg)^{n \times n}$ and $D^{\cngc{-c}} = \cj{D^{\cngc{c}}}$ for all $c\in \Int_{\denom}$.
\end{itemize} 
\end{condition} 
Further, an evaluation problem $\evalk(A,\mfam D)$ is called a \cond{B--H--STD}-problem if it satisfies the following conditions.
\sdefi{}{\cond{B--H--STD}-problem}
\begin{condition}{(B--H--STD)}
There are $n,\denom \in \Nat$ such that $A$ is a connected Hermitian matrix with underlying block $H$ and $\mfam D = (D^{\cngc c})_{c \in \Int_{\denom}}$ is a family of diagonal matrices. Further,
\begin{itemize}
 \item $H \in \U_{\denom}^{n \times n}$ is a normalized complex Hadamard matrix.
 \item For all $c\in \Int_{\denom}$ there are diagonal matrices  $D^{\cngc c,\rw},D^{\cngc c,\cl} \in (\{0\} \cup\Ualg)^{n \times n}$ such that
             $$D^{\cngc c} = D^{\cngc c,\rw} \oplus D^{\cngc c,\cl}.$$
 \item $D^{\cngc 0} = I_{2n}$, and $D^{\cngc{-c}} = \cj{D^{\cngc{c}}}$ for all $c\in \Int_{\denom}$.
\end{itemize}
\end{condition}
The following lemma shows how we can reduce the complexity classification of evaluation problems on arbitrary connected Hermitian matrices to those on Hadamard components. Its proof will be given in Section~\ref{sec:herm_gen_case}.

\begin{lemma}\label{lem:gen_red_to_Had}
Let $A$ be a connected $\denom$-algebraic Hermitian matrix and $D$ a diagonal matrix of positive vertex weights. Then either $\evalk(A,D)$ is $\#\PP$-hard or the following holds.
\begin{itemize}
 \item[(1)] If $A$ is not bipartite then there is an $\denom$-algebraic Hermitian matrix $H$ and a family of diagonal matrices $\mfam D = (D^{\cngc c})_{c \in \Int_{\denom}}$ which define an \cond{H--STD} problem such that 
$$
\evalk(A,D) \Tequiv \evalk(H,\mfam D).
$$
 \item[(2)] If $A$ is bipartite then there is an $\denom$-algebraic Hermitian matrix $A'$ and a family of diagonal matrices $\mfam D = (D^{\cngc c})_{c \in \Int_{\denom}}$ which define a \cond{B--H--STD} problem such that 
$$
\evalk(A,D) \Tequiv \evalk(A',\mfam D).
$$
\end{itemize}
\end{lemma}

\subsection{Hadamard Components}

After applying Lemma~\ref{lem:gen_red_to_Had} we are left with the task of 
classifying the complexity of \cond{H--STD} and \cond{B--H--STD} problems. 
We will now develop further conditions on these problems which will eventually 
lead us to polynomial time computable problems. The definition of these conditions 
will be given in the following. We start with the statement of the theorem whose
proof can be found in Section~\ref{sec:herm_hadmard}.

\begin{thm}\label{thm:hadamard_reduction}
Let $A$ be an $\denom$-algebraic matrix and $\mfam D$ a family of diagonal matrices defining either an \cond{H--STD} problem or a \cond{B--H--STD} problem.
Then either $\evalk(A, \mfam D)$ is $\#\PP$-hard or the following holds.
\begin{itemize}
 \item[(1)] If $A, \mfam D$ define an \cond{H--STD} problem then they satisfy conditions \cond{GC}, \cond{R1} through \cond{R5} and the Affinity Condition \cond{AF}.
 \item[(2)] If $A, \mfam D$ define a \cond{B--H--STD} problem then they satisfy conditions \cond{GC}, \cond{B--R1} through \cond{B--R5} and the Affinity Condition \cond{B--AF}.
\end{itemize}
\end{thm}
The first condition \cond{GC} stipulates that the Hadamard matrix $H$ underlying every \cond{H--STD} and \cond{B--H--STD} problem has a certain group structure.

\paragraph*{The Group Structure of $H$.}
Let $R(H) : = \{ H\row i \mid i \in [n]\}$ denote the \sdefis{set of rows}{set of rows of a matrix}{$R(H)$} of $H$. The \sdefi{Group Condition}{Group Condition \cond{GC}} is defined as follows.
\begin{condition}{(GC)}
$H\row i \circ H\row j \in R(H)$ and $H^T\row i \circ H^T\row j \in R(H^T)$ for all $i,j \in [n]$. 
\end{condition}
The name ``Group Condition'' is justified as it guarantees that $R(H)$ and $R(H^T)$ are Abelian groups under the Hadamard product $\circ$. To see this, note that the definition of \cond{GC} is known as the \emph{subgroup criterion for finite groups}.

\bigskip

\noindent To introduce all further conditions, we will consider the non-bipartite case of \cond{H--STD} problems and the bipartite \cond{B--H--STD} problems separately. Let us begin with the non-bipartite case.

\paragraph*{A Group Theoretic Account of \cond{H--STD}-problems.}
When we are working with Abelian groups $\gG$, unless defined otherwise, 
the group operation will be denoted by $+$ and the neutral element will be $\gneut$. 
Let an $\denom$-algebraic matrix $H$ and a family $\mfam D = (D^{\cngc c})_{\Int_{\denom}}$ define an \cond{H--STD} problem.  Further, assume that $H$ satisfies the group condition \cond{GC}. 

Fix the canonical isomorphism $\caniso_{\denom}:\Int_{\denom} \rightarrow \U_{\denom}$ where $\U_\denom$ denotes the set of $\denom$-th roots of unity. That is, we have $\caniso_{\denom}(a) = \exp(\frac{2\pi \iu a}{\denom})$. By the group condition \cond{GC} the set $R(H)$ of rows of $H$ is an Abelian group under the Hadamard product. Further by the Fundamental Theorem of Finitely Generated Abelian Groups $R(H)$ is isomorphic to some Abelian group $\gG = \Int_{q_1} \oplus \cdots \oplus \Int_{q_z}$ with $q_i$ some prime powers.

Note first that the order of each element $g \in \gG$ divides $\denom$. To see this, recall that $g \in \gG$ corresponds to a row $H \row i$ by the above mentioned isomorphism. The order of $H \row i$ is the least common multiple of the orders of its entries $H_{i j}$. However, each of these entries is an $\denom$-th root of unity.

Let $f: R(H) \rightarrow \gG$ be an isomorphism. We see particularly that for all $i \in [n]$ there is a unique $a \in \gG$ such that $f(H\row i) = a$ and --- as $H$ is Hermitian --- $f(H\col i) = f(\cj{H\row i}) = - a$. 
We may therefore relabel the entries of $H$ by elements of $\gG$ such that $H\row a \circ H \row b = H\row{a+b}$ for all $a,b \in \gG$. In this way, if for $i, j \in [n]$ we let $a,b \in \gG$ be their corresponding group elements, then we have $H_{ij} = H_{a,-b}$. We thus obtain $H_{a,-b} = H_{ij} = \cj{H}_{ji} = \cj{H}_{b,-a}$.

We further define an operator $\bil{\absent, \absent}: \gG \times \gG \rightarrow \Int_{\denom}$ such that 
$$
H_{a,-b} = \caniso_{\denom}\left[\bil{a,b}\right] \text{ for all } a,b \in \gG.
$$
It thus follows directly from our considerations that
\begin{lemma}\label{lem:bil_bilinear}
The operator $\bil{\absent, \absent}$ has the following properties.
\begin{itemize}
 \item For all $g \in \gG$ the mappings $\bil{g, \absent}$ and $\bil{\absent, g}$ are group homomorphisms.
 \item For all $g,h \in \gG$ we have $\bil{g, h} = -\bil{h, g}$.
\end{itemize}
\end{lemma}
We summarize these properties by saying that the operator $\bil{\absent,\absent}$ is \sdef{skew-bilinear}. Extending this description, we would like to describe the non-zero entries of the diagonal matrices $D^{\cngc c}$ in a similar way, although our reasoning so far does \emph{not} allow us to stipulate that $D^{\cngc c}_{a,a} \neq 0$ implies $D^{\cngc c}_{a,a} \in \U_{\denom}$.

Therefore we extend $\Int_\denom$ to some set $\Omega$ and $\caniso_\denom$ to some mapping $\caniso: \Omega \rightarrow \U$ such that for all $c \in \Int_{\denom}$ and all $a \in \gG$, if $D^{\cngc c}_{a,a} \neq 0$ then there is some $\alpha \in \Omega$ such that $\caniso[\alpha] = D^{\cngc c}_{a,a}$.
We may further assume that $\Omega$ is itself an Abelian group and $\caniso$ is a homomorphism.
(Note that this is possible if we just define $\Omega$ as the real interval $[0,\denom)$ and $\caniso(a) = \exp(\frac{2\pi i}{\denom}\cdot a)$). 
Altogether, from conditions \cond{H--STD} and \cond{GC} we have derived the following.\label{pg:repres_criteria}
\sdefisub{}{Representability Conditions}{\cond{R1} -- \cond{R5}}
\begin{condition}{(R1)}
 There is an Abelian group $\Omega \supseteq \Int_{\denom}$ and a homomorphism $\caniso: \Omega \rightarrow \U$ such that $\caniso\upharpoonright_{\Int_{\denom}}$ is the canonical isomorphism between the additive group $\Int_{\denom}$ and the multiplicative group $\U_{\denom}$.
\end{condition}
\begin{condition}{(R2)}
$R(H)$ is isomorphic to some Abelian group $\gG = \Int_{q_1} \oplus \ldots \oplus \Int_{q_z}$ for $q_1, \ldots, q_z$ some prime powers. There is a skew-bilinear operator $\bil{\absent, \absent}: \gG \times \gG \rightarrow \Int_{\denom}$ such that
$$
H_{a,-b} = \caniso\left[\bil{a,b} \right] \text{ for all } a,b \in \gG.
$$ 
Furthermore, the order of each $g \in \gG$ divides $\denom$.
\end{condition}
We shall define three further conditions. The first of which will be shown to hold in the following. 
\begin{condition}{(R3)}
For every $c \in \Int_{\denom}$ let $\Lambda_c = \{ a \in \gG \mid D^{\cngc c}_{a,a} \neq 0\}$ be the \sdef{support} of $D^{\cngc c}$. Then there is some $\beta_c \in \gG$ and a subgroup $\gG_c$ of $\gG$ such that 
$$
\Lambda_c = \beta_c + \gG_c.
$$
\end{condition}
\begin{condition}{(R4)}
For every $c \in \Int_{\denom}$ with $D^{\cngc c} \neq 0$ there is a mapping $\rcwf_{c} : \gG_c \rightarrow \Omega$ such that
$$
    D^{\cngc c}_{a,a} = \caniso\left[\rcwf_{c}(a - \beta_c)\right] \text{ for all } a \in \Lambda_c.
$$
\end{condition}
\begin{condition}{(R5)}
For every $c \in \Int_{\denom}$  with $D^{\cngc c} \neq 0$ we have $\rcwf_{c}(\gneut) = \gneut$.
\end{condition}
Note that, by the above, condition \cond{R4} holds once we made sure that \cond{R3} is satisfied.
We will define a further criterion on the functions $\rcwf_{c}$ describing the non-zero vertex weights. We call this criterion the \sdefisub{Affinity Condition}{Affinity Condition}{\cond{AF}} for $D^{\cngc c}$:
\begin{condition}{\textup{(AF)}}
For all $c \in \Int_{\denom}$ there is a mapping $\gamma_c : \gG_c \rightarrow \gG$ such that following is true. We have
\begin{equation}
\rcwf_{c}(y + x) - \rcwf_{c}(x) - \rcwf_{c}(y) = \bil{\gamma_c(y), x}  \text{ for all } x,y \in \gG_c.
\end{equation}
\end{condition}

\bigskip 
\noindent\textsc{A Note on Decidability.} Apart from the conditions presented in other sections, the decidability of the conditions given here may, at this point, be a bit obscure to the reader. We shall give a bit of insight into this, now. Assume, we are given a matrix $H$ and a family $\mfam D$ of diagonal matrices defining an \cond{H--STD} problem and we want to decide whether this problem is polynomial time computable.
By Theorem~\ref{thm:hadamard_reduction} it turns out that $\evalk(H,\mfam D)$ is $\#\PP$-hard unless all of the conditions just described are satisfied. To be precise, as we have not given an algorithm which can compute the representation as given in \cond{R1} to \cond{R5}, for the cases of $\#\PP$-hardness Theorem~\ref{thm:hadamard_reduction} does only prove the existence of such a reduction.

Eventually however, if the evaluation problem is polynomial time computable, it will turn out that each of the matrices $D^{\cngc c}$ satisfies the following\footnote{The interested reader may find the proof of this in Lemma~\ref{lem:200109-2054}.}
\begin{equation}
\text{ For all $c \in \Int_\denom$ all non-zero entries of $D^{\cngc c}$ are in $\U_{2\denom}$.}
\end{equation}
In a first step we may thus check whether this is the case. If so it follows straightforwardly that the Abelian group $\Omega$ defined in condition \cond{R1} is finite and all of the conditions defined can be computed straightforwardly.

\paragraph*{A Group Theoretic Account of \cond{B--H--STD}-problems.}
Let $A$ be an $\denom$-algebraic matrix and $\mfam D$ a family of diagonal matrices define a \cond{B--H--STD}-problem. Assume further that the block $H$ underlying $A$ satisfies the group condition \cond{GC}. 

Fix the isomorphism $\caniso_{\denom}:\Int_{\denom} \rightarrow \U_{\denom}$ as before. By the group condition \cond{GC}, the set $R(H)$ of rows of $H$ and the set $R(H^T)$ of its columns are Abelian groups under the Hadamard product. Hence, by the Fundamental Theorem of Finitely Generated Abelian Groups $R(H)$ is isomorphic to some Abelian group $\gG_{\rw} = \Int_{q_1} \oplus \cdots \oplus \Int_{q_z}$ with $q_i$ some prime powers. Further, $R(H^T)$ is isomorphic to some Abelian group $\gG_{\cl} = \Int_{q'_1} \oplus \cdots \oplus \Int_{q'_{z'}}$ with $q'_i$ some prime powers.

Note first that the order of each element $g \in \gG_{\rw}$ ($g' \in \gG_{\cl}$ resp.) divides $\denom$. To see this recall that $g \in \gG_{\rw}$ corresponds to a row $H \row i$ by the above mentioned isomorphism. The order of $H \row i$ is the least common multiple of the orders of its entries $H_{i j}$ and each of these entries is a $\denom$-th root of unity.

Let $f_\rw: R(H) \rightarrow \gG_{\rw}$ and $f_\cl: R(H^T) \rightarrow \gG_{\cl}$ be group isomorphisms. We see particularly that for all $i \in [n]$ there is a unique $a \in \gG_\rw$ such that $f_\rw(H\row i) = a$ and the analogue of this holds for the columns. We may therefore relabel the entries of $H$ by elements of $\gG_\rw$ and $\gG_\cl$ such that $H\row a \circ H \row b = H\row{a+b}$ for all $a,b \in \gG_\rw$ and $H\col c \circ H \col d = H\col{c+d}$ for all $c,d \in \gG_\cl$.
We further define an operator $\bil{\absent, \absent}: \gG_\rw \times \gG_\cl \rightarrow \Int_{\denom}$ such that 
$$
H_{a,b} = \caniso_{\denom}\left[\bil{a,b}\right] \text{ for all } a \in \gG_\rw, \; b \in \gG_\cl.
$$
Be aware that, although we denote this operator in the same way as the corresponding one in the non-bipartite case, the two operators are different. However, it will always be unambiguously clear from the context, which of the two operators we are referring to.

We say that the operator $\bil{\absent, \absent}$ is \sdef{bilinear} if the following holds: For all $g \in \gG_\rw$ and all $g' \in \gG_{\cl}$ the mappings $\bil{g,\absent}$ and $\bil{\absent,g'}$ are group homomorphisms. It follows directly from the above considerations that
\begin{lemma}
The operator $\bil{\absent, \absent}$ is bilinear. 
\end{lemma}
We would like to describe the non-zero entries of the diagonal matrices $D^{\cngc c}$ in a similar way, although our reasoning so far does \emph{not} allow us to stipulate that $D^{\cngc c}_{a,a} \neq 0$ implies $D^{\cngc c}_{a,a} \in \U_{\denom}$.
Therefore we extend $\Int_\denom$ to some set $\Omega$ and $\caniso_\denom$ to some mapping $\caniso: \Omega \rightarrow \U$ such that for all $c \in \Int_{\denom}$ and all $a \in [2n]$, if $D^{\cngc c}_{a,a} \neq 0$ then there is some $\alpha \in \Omega$ such that $\caniso[\alpha] = D^{\cngc c}_{a,a}$. We further assume that $\Omega$ is itself an Abelian group and $\caniso$ is a homomorphism.

Altogether, from conditions \cond{B--H--STD} and \cond{GC} we have derived the following.\label{pg:bip_repres_criteria}
\sdefisub{}{Representability Conditions}{\cond{B--R1}--\cond{B--R5}}
\begin{condition}{(B--R1)}
 There is an Abelian group $\Omega$ and a homomorphism $\caniso: \Omega \rightarrow \U$ such that $\caniso\upharpoonright_{\Int_{\denom}}$ is the canonical isomorphism between the additive group $\Int_{\denom}$ and the multiplicative group $\U_{\denom}$.
\end{condition}
\begin{condition}{(B--R2)}
$R(H)$ is isomorphic to some Abelian group $\gG_\rw = \Int_{q_1} \oplus \ldots \oplus \Int_{q_z}$ for $q_1, \ldots, q_z$ some prime powers. $R(H^T)$ is isomorphic to some Abelian group $\gG_\cl = \Int_{q'_1} \oplus \ldots \oplus \Int_{q'_{z'}}$ for $q'_1, \ldots, q'_{z'}$ some prime powers.
There is a bilinear operator $\bil{\absent, \absent}: \gG_\rw \times \gG_\cl \rightarrow \Int_{\denom}$ such that
$$
H_{a,b} = \caniso\left[\bil{a,b} \right] \text{ for all } a \in \gG_\rw,\; b \in \gG_\cl.
$$ 
Furthermore, the order of each $g \in \gG_{\rw}$ and each $g' \in \gG_\cl$ divides $\denom$.
\end{condition}
We shall define three further conditions. The first of which will be shown to hold in the following. 
\begin{condition}{(B--R3)}
For every $c \in \Int_{\denom}$ let $\Lambda_{c,\rw} = \{ a \in \gG^\rw \mid D^{\cngc c,\rw}_{a,a} \neq 0\}$ be the \sdef{support} of $D^{\cngc c,\rw}$ and let $\Lambda_{c,\cl} = \{ a \in \gG^\cl \mid D^{\cngc c,\cl}_{a,a} \neq 0\}$ be the \sdef{support} of $D^{\cngc c,\cl}$. There are $\beta_{c,\rw} \in \gG_\rw$, $\beta_{c,\cl} \in \gG_\cl$ and  subgroups $\gG_{c,\rw}$ of $\gG_\rw$ and $\gG_{c,\cl}$ of $\gG^\cl$ such that 
$$
\Lambda_{c,\rw} = \beta_{c,\rw} + \gG_{c,\rw} \quad \text{ and } \quad \Lambda_{c,\cl} = \beta_{c,\cl} + \gG_{c,\cl}.
$$
\end{condition}
\begin{condition}{(B--R4)}
For every $c \in \Int_{\denom}$ there are mappings $\rcwf_{c,\rw} : \gG_{c,\rw} \rightarrow \Omega$ and $\rcwf_{c,\cl} : \gG_{c,\cl} \rightarrow \Omega$ such that
$$
    D^{\cngc c,\rw}_{a,a} = \caniso\left[\rcwf_{c,\rw}(a - \beta_{c,\rw})\right] \text{ for all } a \in \Lambda_{c,\rw}.
$$
$$
D^{\cngc c,\cl}_{a,a} = \caniso\left[\rcwf_{c,\cl}(a - \beta_{c,\cl})\right] \text{ for all } a \in \Lambda_{c,\cl}.
$$
\end{condition}

\begin{condition}{(B--R5)}
For every $c \in \Int_{\denom}$ the following holds.
If $D^{\cngc c,\rw} \neq 0$ then $\rcwf_{c,\rw}(\gneut) = \gneut$ and if 
$D^{\cngc c,\cl} \neq 0$ then $\rcwf_{c,\cl}(\gneut) = \gneut$.
\end{condition}
Note that, by the above, condition \cond{B--R4} holds once we made sure that \cond{B--R3} is satisfied.
As in the description of the \cond{H--STD} problems we will give a further condition on the mappings $\rcwf_{c,\rw},\rcwf_{c,\cl}$ describing the non-zero vertex weights. Again, we call this the \sdefisub{Affinity Condition}{Affinity Condition}{\cond{B--AF}} for $D^{\cngc c}$, which reads
\begin{condition}{\textup{(B--AF)}}
For all $c \in \Int_{\denom}$ there are mappings $\gamma_{c,\rw} : \gG_{c,\rw} \rightarrow \gG_{\cl}$ and $\gamma_{c,\cl} : \gG_{c,\cl} \rightarrow \gG_{\rw}$ such that the following is true. 
\begin{eqnarray}
\rcwf_{c,\rw}(y + x) - \rcwf_{c,\rw}(x) -  \rcwf_{c,\rw}(y) &=& \bil{x,\gamma_{c,\rw}(y)}  \text{ for all } x,y \in \gG_{c,\rw}\\
\rcwf_{c,\cl}(y + x) - \rcwf_{c,\cl}(x) - \rcwf_{c,\cl}(y) &=& \bil{\gamma_{c,\cl}(y), x}  \;\text{ for all } x,y \in \gG_{c,\cl}.
\end{eqnarray}
\end{condition}

\bigskip

\noindent\textsc{Decidability.} Note that the question whether a given  \cond{B--H--STD} problem is polynomial time computable or $\#\PP$-hard can be solved analogously to the non-bipartite case. Recall the discussion given there.

\subsection{The Polynomial Time Case}

By now we have arrived at \cond{H--STD} problems and \cond{B--H--STD} problems satisfying all conditions we have defined so far.
It remains to show that this implies polynomial time computability of the corresponding problems.

\begin{thm}\label{thm:herm_ptime_partition_functions}
Let $A$ be an $\denom$-algebraic matrix and $\mfam D$ a family of diagonal matrices. The problem $\evalk(A, \mfam D)$ is polynomial time computable if the following holds.
\begin{itemize}
 \item[(1)] If $A, \mfam D$ define an \cond{H--STD} problem then they satisfy conditions \cond{GC}, \cond{R1} through \cond{R5} and the Affinity Condition \cond{AF}.
 \item[(2)] If $A, \mfam D$ define a \cond{B--H--STD} problem then they satisfy conditions \cond{GC}, \cond{B--R1} through \cond{B--R5} and the Affinity Condition \cond{B--AF}.
\end{itemize}
\end{thm}

This theorem will follow by reducing the partition functions satisfying the 
above conditions to a problem which has been shown to be polynomial time 
computable in \cite{caichelu09}. The proof details will be given in Section~\ref{sec:ptime}.

\subsection{The Proof of Theorem~\ref{thm:hermitian_main}}

With the high-level results we have derived in the previous sections we are now able to prove the Main Theorem~\ref{thm:hermitian_main}. In a first step, let us show how we can easily derive the result for problems $\evalk(A, D)$ on connected matrices $A$.

\begin{lemma}\label{lem:Herm_general_case_conn}
Let $A \in \Calg^{m \times m}$ be a connected $\denom$-algebraic Hermitian matrix and $D \in \Ralg^{m \times m}$ a diagonal matrix of positive vertex weights. Then either $\evalk(A, D)$ is $\#\PP$-hard or polynomial time computable.
\end{lemma}
\begin{proof}
Let us consider the case first that $A$ is not bipartite 
-- the bipartite case follows by analogous reasoning. Assume that $\evalk(A, D)$ is not $\#\PP$-hard. Then Lemma~\ref{lem:gen_red_to_Had} implies that there is an $\denom$-algebraic Hermitian matrix $H$ and a family $\mfam D = (D^{\cngc c})_{c \in \Int_{\denom}}$ of diagonal matrices which define an \cond{H--STD} problem such that 
\begin{equation}\label{eq:2605091342}
\evalk(A,D) \Tequiv \evalk(H,\mfam D). 
\end{equation}
Therefore $\evalk(H, \mfam D)$ cannot be $\#\PP$-hard. 
By Theorem~\ref{thm:hadamard_reduction} the matrix $H$ and the family $\mfam D$ thus satisfy conditions \cond{GC},\cond{R1} through \cond{R5} and the Affinity Condition \cond{AF}. Hence by Theorem~\ref{thm:herm_ptime_partition_functions} the problem $\evalk(H, \mfam D)$ is polynomial time computable.
Therefore, equation \eqref{eq:2605091342} implies that $\evalk(A,D)$ is also polynomial time computable.
\end{proof}

\begin{proof}[of Theorem \ref{thm:hermitian_main}]
Let $A \in \Calg^{m \times m}$ be a Hermitian matrix and 
$D \in \Ralg^{m \times m}$ a diagonal matrix of positive vertex weights. 
Assume first that $A$ is not $\denom$-algebraic for some $\denom \in \Nat$, 
then the Arithmetical Structure Lemma~\ref{lem:alg_to_w_alg_red} implies that 
there is some $\denom \in \Nat$ and an $\denom$-algebraic matrix $A'$ such that 
$\eval(A,D) \Tequiv \eval(A',D)$. We therefore assume for simplicity that $A$ is
$\denom$-algebraic. We have
\begin{equation}\label{eq:2605091419}
\evalk(A,D) \Tequiv \eval(A,D)
\end{equation}
by the Pinning Lemma~\ref{lem:pinning}.
Let $\m I$ be a partition of $[m]$ into subsets $I \in \m I$ such that 
each $A_{II}$ is a component of $A$. The following is true for each $I \in \m I$.
\begin{claim}\label{cl:2605091416}
If $\evalk(A_{II}, D_{II})$ is $\#\PP$-hard then $\eval(A,D)$ is $\#\PP$-hard.
\end{claim}
To see this, let $G,\vpin$ be the input to $\evalk(A_{II}, D_{II})$ 
for some $I \in \m I$. We may assume that $G$ is connected and 
$\df(\vpin) \neq \emptyset$.  Therefore, we have 
$Z_{A_{II}, D_{II}}(\vpin,G) = Z_{A,D}(\vpin,G)$ and thus 
$\evalk(A_{II}, D_{II}) \Tle \evalk(A,D)$. By equation \eqref{eq:2605091419} 
this proves Claim~\ref{cl:2605091416}.
We further have, for all digraphs $G$,
$$
Z_{A,D}(G) = \sum_{I \in \m I} Z_{A_{II},D_{II}}(G).
$$
This proves that $\eval(A,D)$ is polynomial time computable if for all $I \in \m I$ the problem $\eval(A_{II},D_{II})$ is.
By Lemma~\ref{lem:Herm_general_case_conn} we see that 
for each component $A_{II}$ of $A$ either the preconditions of 
Claim~\ref{cl:2605091416} are given or $\evalk(A_{II},D_{II})$ is polynomial 
time computable. This finishes the proof.
\end{proof}

\section{Basic Technical Results}\label{sec:technical_prelim}\marc{Section OK}
In this section we will give proofs of many technical results we will need later
on. Most importantly, we will prove the 
Arithmetical Structure Lemma~\ref{lem:alg_to_w_alg_red} in 
Section~\ref{sec:algstrucpf} and the Pinning Lemma~\ref{lem:pinning} in 
Section~\ref{sec:pin}.
The proofs of these results require a considerable amount of other tools which 
we will develop in the following.

\subsection{General Principles} \label{sec:gen_princip}\label{sec:gen_princip_cng}

Most reductions which we will present here have to deal with arbitrary 
input digraphs $G$ and pinnings $\vpin$. To avoid tedious reasoning on trivial
or vacuous cases we will 
introduce some \emph{general principles} which we will apply whenever necessary.

\begin{lemma}[Connectedness Principle]\label{lem:principle_conn}\label{lem:principle_conn_cng}
Fix one of $\evalk(A,D)$, $\evalk(A, \mfam D)$ or $\eval(A,D)$. 
If we know how to solve the problem in polynomial time given that the input is restricted to connected digraphs, then the general problem can be solved in polynomial time. 
\end{lemma}
\begin{proof}
We give the proof for $\evalk(A,D)$, the other cases are analogous.
Let $G=(V,E)$ be a given digraph with components $G_1, \ldots, G_c$.
Each pinning $\vpin$ of vertices of $G$ to entries of $A$ can be partitioned into pinnings (possibly empty) $\vpin_1,\ldots,\vpin_c$ corresponding to the vertices of the components of $G$. Then
$$
Z_{A,D}(\vpin, G) = \prod_{i=1}^c Z_{A,D}(\vpin_i,G_i).
$$
Since we know how to compute the values $Z_{A,D}(\vpin_i,G_i)$ in polynomial time, the value $Z_{A,D}(\vpin, G)$ can be computed by the above equation.
\end{proof}

\begin{lemma}[The Pinning Principle] \label{lem:principle_pin} \label{lem:principle_pin_cng}
Fix one of $\evalk(A,D)$ or $\evalk(A, \mfam D)$.
Assume that this problem can be solved in polynomial time if the input allows only non-trivial pinnings.
Then the general problem can be solved in polynomial time. 
\end{lemma}
\begin{proof}
We give the proof for $\evalk(A,D)$, the other cases are analogous.
Assume that $A$ is an $m \times m$ matrix and let $G$, $\vpin$ be an instance of $\evalk(A,D)$. If $\df(\vpin) = \emptyset$, fix an arbitrary vertex $v \in V$ and let $\vpin_i$ for every $i \in [m]$ be the mapping defined by $v \mapsto i$. We have
$$
Z_{A,D}(G) = \sum_{i=1}^{m} D_{ii} \cdot Z_{A,D}(\vpin_i,G).
$$
Each value $Z_{A,D}(\vpin_i,G)$ can be computed in polynomial time.
\end{proof}
For $A$ an $m \times m$ matrix and $\pi : [m] \rightarrow [m]$ a permutation, 
define \idxsymb{$A_{\pi\pi}$} by
$(A_{\pi\pi})_{ij} = A_{\pi(i)\pi(j)} \text{ for all } i,j \in [m]$.
It is straightforward to see that simultaneous permutation of the rows and 
columns of $A$ do not alter the partition function:
\begin{lemma}[Permutability Principle] \label{lem:principle_permute}
  \label{lem:principle_permute_cng}
\marc{true?}
Let $A,D \in \Calg^{m \times m}$ and $\pi : [m] \rightarrow [m]$ a permutation. Then $\evalk(A,D) \Tequiv \evalk(A_{\pi\pi},D_{\pi\pi})$.
This holds analogously for congruential partition functions.
\end{lemma}

\subsection{Technical Lemmas}
\begin{lemma}\label{lem:coeff_zero_inf}
Let $0 < x_1 < \ldots < x_n$ be positive in $\Ralg$. $c_1,\ldots,c_n \in \Calg$.
There is a $p_0 \in \Nat$ such that for all $p \ge p_0$, the following equation holds if, and only if, all coefficients $c_i$ are zero.
\begin{equation}\label{eq:coeff_zero_inf}
0 =  \sum_{i = 1}^n c_i x_i^p.
\end{equation}
Further $p_0$ is computable from input $x_1,\ldots, x_n,c_1,\ldots,c_n$.
\end{lemma}
\begin{proof}
Note that, if $0 = c_1 = \ldots = c_n$, then equation \eqref{eq:coeff_zero_inf} is satisfied for all $p\ge 0$.
Assume therefore that there is at least one non-zero $c_i$. We perform induction on the maximum index $m \le n$ such that $c_m \neq 0$.
If $m =1$ then $\sum_{i = 1}^n c_i x_i^p. = c_1 x_1^p$ and the proof follows for $p_0 = 0$. For $m > 1$ note that
\begin{eqnarray*}
\left\vert \sum_{i = 1}^n c_ix_i^p \right\vert &\ge& \left\vert |c_mx_m^p| - \left| \sum_{i = 1}^{m-1} c_ix_i^p \right| \right\vert
\end{eqnarray*}
and the right hand side is larger than zero, if
$|c_m|x_m^p > \sum_{i = 1}^{m-1} |c_i|x_i^p$.
If all $c_1,\ldots, c_{m-1}$ are zero, we are done with $p_0 = 0$ and let $x = \max\{ x_i \mid c_i \neq 0,\; i \in [m-1]\}$ otherwise. We see that equation \eqref{eq:coeff_zero_inf} is not satisfied if
\begin{eqnarray*}
\left(\dfrac{x_m}{x}\right)^p &>&  \sum_{i = 1}^{m-1} \dfrac{|c_i|}{|c_m|}.
\end{eqnarray*}
As $x_m > x$ this inequality is satisfied for large enough $p$. We define $p_0$ such that this holds. 
\end{proof}

\subsubsection{Twin Reduction.}\label{sec:twin_resolution}
Let $A \in \C^{m\times m}$ be Hermitian. We say that two rows $A \row i$ 
and $A \row j$ are \sdefi{twins}{twin}, if $A \row i = A \row j$. 
A matrix $A$ is \sdef{twin-free} if it has no twins.
Twins induce an equivalence relation on the rows of $A$. 
Let $I_1,\ldots ,I_k$ be the equivalence classes of this relation. 
Since $A$ is Hermitian these classes are also the equivalence classes of the 
twin relation on the columns of $A$.
The \sdef{twin resolvent} of $A$ is the $k \times k$ matrix $\twres{A}$, defined by
$$
\twres{A}_{i,j} = A_{\mu,\nu} \text{ for some } \mu \in I_{i} \text{ and } \nu \in I_{j}.
$$
We say that $\twres A$ is obtained from $A$ by \sdef{twin reduction}. 
The \sdef{twin resolution mapping} $\tau: [m] \rightarrow
[k]$ of $A$ is defined such that $\mu \in I_{\tau(\mu)}$ for all $\mu \in [m]$. 
Hence $ \twres A_{\tau(i),\tau(j)} = A_{i,j} \text{ for all } i,j \in [m]$.

To use twin reductions in the context of partition functions, we need to consider
their effect on the 
diagonal matrices $D$ associated to $A$. We will see that the following diagonal 
$k \times k$ matrix $D^{\twres A}$ captures this effect
 \begin{equation}
     D^{\twres A}_{i,i} = \sum_{\nu \in I_i} D_{\nu,\nu} \text{ for all } i \in [k].
 \end{equation}
Analogously, for a family $\mfam D = (D^{\cngc c})_{c \in \Int_{\denom}}$ 
of congruential vertex weights, we define a family 
$\mfam D^{\twres A} = (D^{\twres A ,{\cngc c}})_{c \in \Int_{\denom}}$ 
of diagonal $k \times k$ matrices by
\[
D^{\twres A, {\cngc c}}_{i,i} = \sum_{\nu \in I_i} D^{\cngc c}_{\nu,\nu} \text{ for all } i \in [k] \text{ and all } c \in \Int_{\denom}.
\]

\begin{lemma}[Twin Reduction Lemma]\label{lem:cng_twin_red}
Let$\, A \in \C^{m \times m}$ be Hermitian.
Let $I_1,\ldots, I_k$ be the equivalence classes of the twin relation on $A$ and
$\tau$ be the twin resolution mapping of $A$. The following is true 
for all digraphs $G$ and pinnings $\vpin$:
\begin{description}
 \item[(1)] Let $\mfam D = (D^{\cngc c})_{c \in \Int_{\denom}}$ a family of 
       congruential vertex weights, then       
  \[
  Z_{A, \mfam D}(\vpin,G) = Z_{\twres A,\mfam D^{\twres A}}(\tau\circ \vpin, G).
  \]
\item[(2)] Let $D$ be a diagonal $m \times m$ matrix of vertex weights. Then
\[
  Z_{A,D}(\phi,G) = Z_{\twres A,D^{\twres A}}(\tau\circ \phi, G).
\]  
\end{description}
\end{lemma}
\begin{proof}
Let us start by proving (1).
For a digraph $G = (V,E)$ and a pinning $\vpin$, let $V' = V\setminus \df(\vpin)$. We have
\begin{eqnarray*}
Z_{A,\mfam D}(\vpin,G) &=& \sum_{\vpin \subseteq \vcfg:V \rightarrow [m]} \prod_{uv \in E} A_{\vcfg(u),\vcfg(v)} \prod_{v \in V'} D^{\cngc{\grade{(v)}}}_{\vcfg(v),\vcfg(v)} \\
&=& \sum_{\vpin \subseteq \vcfg:V \rightarrow [m]} \prod_{uv \in E} \twres A_{\tau\circ\vcfg(u),\tau \circ \vcfg(v)} \prod_{v \in V'} D^{\cngc{\grade{(v)}}}_{\vcfg(v),\vcfg(v)}
\end{eqnarray*}
where the second equality follows from the definition of $\tau$. As for all configurations $\vcfg: V \rightarrow [m]$ we have $\tau \circ \vcfg : V \rightarrow [k]$, we can partition the $\vcfg$ into classes according to their images under concatenation with $\tau$ and obtain:
\begin{eqnarray*}
Z_{A,\mfam D}(\vpin,G) &=& \sum_{\vcfg':V \rightarrow [k]}\sum_{\substack{\vpin \subseteq \vcfg:V \rightarrow [m] \\ \tau \circ \vcfg = \vcfg'}} \prod_{uv \in E} \twres A_{\vcfg'(u), \vcfg'(v)} \prod_{v \in V'} D^{\cngc{\grade{(v)}}}_{\vcfg(v), \vcfg(v)} \\
           &=& \sum_{\tau\circ \vpin \subseteq \vcfg':V \rightarrow [k]} \prod_{uv \in E} \twres A_{\vcfg'(u), \vcfg'(v)} \left(\sum_{\substack{\vpin \subseteq \vcfg:V \rightarrow [m] \\ \tau \circ \vcfg = \vcfg'}} \prod_{v \in V'} D^{\cngc{\grade{(v)}}}_{\vcfg(v), \vcfg(v)}\right)
\end{eqnarray*}
Fix some $\vcfg': V \rightarrow [k]$. For $\vcfg: V \rightarrow [m]$ we have $\tau
\circ \vcfg = \vcfg'$ if and only if ${\vcfg'}^{-1}(\{i\}) =
\vcfg^{-1}(I_i)$ for all $i \in
[k]$. Define for all $i \in [k]$ the set $V_i := {\vcfg'}^{-1}(\{i\})$ and the mapping $\vpin_i := \vpin\upharpoonright_{\df(\vpin) \cap V_i}$, then
\begin{eqnarray*}
\sum_{\substack{\vpin \subseteq \vcfg:V \rightarrow [m] \\ \tau \circ \vcfg = \vcfg'}} \prod_{v \in V'} D^{\cngc{\grade{(v)}}}_{\vcfg(v), \vcfg(v)}
&=&\sum_{\substack{\vpin \subseteq \vcfg: V \rightarrow [m] \\ \forall \; i \in [k]:\;\vcfg(V_i) \subseteq I_i }} \prod_{v \in V'} D^{\cngc{\grade{(v)}}}_{\vcfg(v),\vcfg(v)} \\
&=& \prod_{i=1}^k \sum_{\vpin_i \subseteq \vcfg_i:V_i \rightarrow I_i} \prod_{v \in V_i\setminus\df(\vpin_i)}  D^{\cngc{\grade{(v)}}}_{\vcfg_i(v), \vcfg_i(v)}\\
 &=&\prod_{i=1}^k \prod_{v \in V_i\setminus\df(\vpin_i)} \sum_{\nu \in I_i}   D^{\cngc{\grade{(v)}}}_{\nu, \nu} \\
&=&\prod_{v \in V'} D^{\twres A,\cngc{\grade{(v)}}}_{\vcfg'(v), \vcfg'(v)}
\end{eqnarray*}
Hence 
\[
Z_{A,\mfam D}(\phi, G) = \sum_{\tau\circ \phi \subseteq \vcfg':V \rightarrow [k]} \prod_{uv \in E} \twres A_{\vcfg'(u), \vcfg'(v)} \prod_{v \in V'} D^{\twres A,\cngc{\grade{(v)}}}_{\vcfg'(v), \vcfg'(v)}
 = Z_{\twres A,\mfam D^{\twres A}}(\tau \circ \phi, G).
\]
\medskip
For the proof of (2) notice that $D^{\cngc{c}} = D$ implies
$D^{\twres{A},\cngc{c}} = D^{\twres A}$ for every $c \in \Int_\denom$. The claim
hence follows by Lemma~\ref{lem:intro_cong}.
\end{proof}

\begin{cor}
We have $\evalk(A,\mfam D) \Tequiv  \evalk(\twres A,\mfam D^{\twres{A}})$.
\end{cor}

\subsubsection{Root Of Unity Transfer}
Lemma~\ref{lem:unity_transfer} explains a way to transfer certain values from 
the Hermitian matrix $A \in \C^{m \times m}$ into the vertex weights 
of the partition function.

\begin{lemma}[Root Of Unity Transfer Lemma]\label{lem:unity_transfer}
Let $A \in \C^{m \times m}$ be Hermitian and $\denom$-algebraic. Let $\mfam D = (D^{\cngc c})_{c \in \Int_{\denom}}$ be a family of congruential vertex weights and let $\Pi$ be a diagonal $m \times m$ matrix whose diagonal entries are $\denom$-th roots of unity. 
Define $A' = \Pi A \cj \Pi$ and a family of diagonal matrices $\mfam D' = ({D'}^{\cngc c})_{c \in \Int_{\denom}}$ by $(D')^{\cngc c} = \Pi^c D^{\cngc c}$.

There is a polynomial time computable function $f_\Pi$  such that for every digraph $G$ and pinning $\vpin$ we have.
$$
Z_{A, \mfam D}(\vpin, G) = f_\Pi(\vpin) \cdot Z_{A', \mfam D'}(\vpin, G).
$$
Furthermore, $f_\Pi$ is non-zero for all $\vpin$.
\end{lemma}
\begin{proof}\marc{proof OK}
Let $G = (V,E)$ be a given digraph and $\vpin$ a pinning.
With $V' = V \setminus \df(\phi)$ we have
\begin{eqnarray*}
Z_{A, \mfam D}(\vpin, G)  &=& \sum_{\substack{\vcfg:V\to [m]\\ \vcfg \supseteq \vpin}} \prod_{uv \in E} A_{\vcfg(u)\vcfg(v)} \prod_{v \in V'} D^{\cngc{\grade(v)}}_{\vcfg(v)\vcfg(v)} \\
&=& \sum_{\substack{\vcfg:V\to [m]\\ \vcfg \supseteq \vpin}} \prod_{uv \in E} \Pi_{\vcfg(u)\vcfg(u)} A'_{\vcfg(u)\vcfg(v)} \cj \Pi_{\vcfg(v)\vcfg(v)} \prod_{v \in V'} D^{\cngc{\grade(v)}}_{\vcfg(v)\vcfg(v)} \\
   &=& \left(\prod_{v \in \df(\vpin)} \Pi^{\grade(v)}_{\vpin(v)\vpin(v)}\right) \sum_{\substack{\vcfg:V\to [m]\\ \vcfg \supseteq \vpin}} \prod_{uv \in E} A'_{\vcfg(u)\vcfg(v)} \prod_{v \in V'} \Pi^{\grade(v)}_{\vcfg(v)\vcfg(v)}D^{\cngc{\grade(v)}}_{\vcfg(v)\vcfg(v)} \\  
\end{eqnarray*}
Now $\Pi^{\grade(v)} = \Pi^{\cngc{\grade(v)}}$ as all diagonal entries of $\Pi$ are $\denom$-th roots of unity. 
Therefore,
\begin{eqnarray*}
Z_{A, \mfam D}(\vpin, G) &=& \left(\prod_{v \in \df(\vpin)} \Pi^{\grade(v)}_{\vpin(v)\vpin(v)}\right) \sum_{\substack{\vcfg:V\to [m]\\ \vcfg \supseteq \vpin}} \prod_{uv \in E} A'_{\vcfg(u)\vcfg(v)} \prod_{v \in V'} \Pi^{\cngc{\grade(v)}}_{\vcfg(v)\vcfg(v)}D^{\cngc{\grade(v)}}_{\vcfg(v)\vcfg(v)} \\
&=& \left(\prod_{v \in \df(\vpin)} \Pi^{\grade(v)}_{\vpin(v)\vpin(v)}\right) \sum_{\substack{\vcfg:V\to [m]\\ \vcfg \supseteq \vpin}} \prod_{uv \in E} A'_{\vcfg(u)\vcfg(v)} \prod_{v \in V'} (D')^{\cngc{\grade(v)}}_{\vcfg(v)\vcfg(v)} \\
&=& \left(\prod_{v \in \df(\vpin)} \Pi^{\grade(v)}_{\vpin(v)\vpin(v)}\right) Z_{A', \mfam D'}(\vpin, G)
\end{eqnarray*}
Note that the factor $\left(\prod_{v \in \df(\vpin)} \Pi^{\grade(v)}_{\vpin(v),\vpin(v)}\right)$
is nonzero and polynomial time computable. In particular, this factor equals $1$ if $\df(\vpin) = \emptyset$.
\end{proof}
The above lemma directly yields,
\begin{cor}
We have
$\evalk(A, \mfam D) \Tequiv \evalk(A',\mfam D')$.
\end{cor}

\subsection{The Arithmetical Structure Lemma}\label{sec:algstrucpf}

In this section we will give the proof of Lemma~\ref{lem:alg_to_w_alg_red}. 
We need some preparation.

\paragraph*{$\eval(A,D)$ and $\cnt(A,D)$.} 
\newcommand{\svcfg}{\textup{C}}
Let $A \in \Calg^{m \times m}$ be a matrix and $G=(V,E)$ a digraph.
We define a set of \sdefis{potential weights}{potential weights}{$\wset_A(G)$}
\begin{equation}\label{eq:define_wset}
 \wset_{A}(G) := \left\lbrace \prod_{i,j \in [m]} A^{m_{ij}}_{ij} \,\mid \, \sum_{i,j \in [m]} m_{ij} = \vert E \vert,\;  \text{ and } m_{ij} \ge 0, \text{ for all } i,j \in [m]  \right\rbrace.
\end{equation}
For some $w \in \Calg$ denote the set of configurations with weight $w$ by
$$
\svcfg_{A}(G,w) := \left\lbrace\vcfg: V \rightarrow [m] \, \vert \, w = \prod_{uv \in E} A_{\vcfg(u)\vcfg(v)}\right\rbrace.
$$
Define the value
$$
N_{A,D}(G,w) := \sum_{\vcfg \in \svcfg_{A}(G,w)} \prod_{v \in V}D_{\vcfg(v)\vcfg(v)}
$$ 
%
$\cnt(A,D)$ denotes the problem of computing $N_{A,D}(G,w)$ for some
given digraph $G=(V,E)$ and a weight $w \in \Calg$.
We use an interpolation technique from \cite{dyegre00}:
\begin{lemma}\label{lem:interpolate}
Let, for some fixed $\theta \in \Calg$ let $x_1, \ldots, x_n \in \Q(\theta)$ be pairwise different and non-zero. Let $b_1,\ldots,b_n \in \Q(\theta)$ be arbitrary such that
$$
 b_j = \sum_{i=1}^n c_i x^j_i \text{ for all } j \in [n].
$$
Then the coefficients $c_1,\ldots, c_n$ are uniquely determined and can be computed in polynomial time.
\end{lemma}

\begin{lemma}\label{lem:cnt_eq_eval}
For every matrix $A \in \Calg^{m\times m}$ and $D \in \Calg^{m \times m}$ we have 
$$\eval(A,D) \Tequiv \cnt(A,D).
$$ 
\end{lemma}
\begin{proof}
Let $G = (V,E)$ be a digraph. We have
\[
Z_{A,D}(G) = \sum_{w \in \wset_{A}(G)} w \sum_{\vcfg \in \svcfg_{A}(G,w)} \prod_{v \in V}D_{\vcfg(v)\vcfg(v)}\\
= \sum_{w \in \wset_{A}(G)} w \cdot N_{A,D}(G,w). 
\]
As the cardinality of $\wset_{A}(G)$ is polynomial in the size of $G$ this proves $
\eval(A,D) \Tle \cnt(A,D)$. For the backward direction let $G^{(t)}$ denote the digraph obtained from $G$ by replacing each edge with $t$ copies of it. We have
$$
Z_{A,D}(G^{(t)}) =  \sum_{\vcfg: V \rightarrow [m]}\left(\prod_{uv \in E} 
A_{\vcfg(u)\vcfg(v)}\right)^t\prod_{v \in V} D_{\vcfg(v)\vcfg(v)} \\
 = \sum_{w \in \wset_{A}(G)} w^t \cdot N_{A,D}(G,w).
$$
Using an $\eval(A,D)$ oracle, we can evaluate this for $t = 1, \ldots, |\wset_{A}(G)|$. 
Therefore the values $N_{A,D}(G,w)$ can be recovered in polynomial time by Lemma~\ref{lem:interpolate}.
\end{proof}

\subsubsection{The Proof of Lemma~\ref{lem:alg_to_w_alg_red}}

To measure the time needed to perform operations on algebraic numbers, 
we need to introduce their \emph{description length}. 
For an integer $a \in \Int$ let its \sdef{description length} 
be $\dlen a := \lceil \log_2 |a_i|\rceil + 2$. For a rational number 
$q = \frac{a}{b}$ with $a,b \in \Int$ we define $\dlen q := \dlen a + \dlen b$.
For a vector $\vec a \in \Q^n$ we define 
$\dlen{\vec a} = \dlen{a_1} + \ldots + \dlen{a_n}$. 
This defines the description lengths of algebraic numbers in $\Q(\theta)$ 
in standard representation. The description length $\dlen A$ of a matrix 
$A \in \Q^{n \times n}$ is the sum of the description lengths of it entries.

Let $B = \{b_1, \ldots, b_n\} \subseteq \Calg$ be a set of algebraic numbers. By $\mspan{B}$ we denote the \sdef{multiplicative span} of $B$ that is $\mspan B = \{ \zeta \cdot \prod_{b \in B} b^{\lambda_b} \mid \lambda_b \in \Int, \text{ for all } b \in B, \; \zeta \text{ is a root of unity}\}$. 
The set $B$ is called \sdef{multiplicatively independent}, if for all $\lambda_1, \ldots, \lambda_n \in \Int$ the following holds: if $b_1^{\lambda_1} \cdots b_n^{\lambda_n}$ is a root of unity then $\lambda_1 = \ldots = \lambda_n = 0$.
In all other cases we say that $B$ is \sdef{multiplicatively dependent}. 
We say that a set $S$ is \sdef{effectively representable} in terms of $B$, if for given $x \in S$ we can compute $\lambda_1, \ldots ,\lambda_n \in \Int$ such that
$x \cdot b_1^{\lambda_1} \cdots b_n^{\lambda_n}$ is a root of unity.
A set $B$ is an \sdef{effective representation system} for a set $S$, if $S$ is effectively representable in terms of $B$ and $B$ is multiplicatively independent.
We need a result from \cite{ric01}:
\begin{lemma}[Theorem 2 in \cite{ric01}]\label{lem:Richardson}
Let $a_1,\dots, a_n \in \Q(\theta)$ given in standard representation, each of description length at most $s$. There is a matrix $A \in \Int^{n \times n}$ such that, for vectors $ \vec \lambda \in \Int^n$ we have
\begin{equation}\label{eq:0506091450}
 \prod_{i=1}^n a_i^{\lambda_i} \text{ is a root of unity, if, and only if, } A\cdot \vec \lambda = 0.
\end{equation}
The description length of $A$ is bounded by a computable function in $n$ and $s$.
\end{lemma}
This theorem has a straightforward algorithmic consequence.
\begin{cor}\label{cor:05061625}
Let $a_1,\dots, a_n \in \Q(\theta)$ be given in standard representation. There is an algorithm which decides if there is a non-zero vector $\vec \lambda = (\lambda_1,\ldots, \lambda_n) \in \Int^n$ such that
\begin{equation}\label{eq:0506091456}
\prod_{i=1}^n a_i^{\lambda_i} \text{ is a root of unity}. 
\end{equation}
Furthermore, if it exists, the algorithm computes such a vector $\vec \lambda$.
\end{cor}

\begin{lemma}\label{lem:04061619}
Let $B =\{b_1, \ldots, b_n\} \subseteq \Calg$ be a multiplicatively dependent set of algebraic numbers. Then there is a set $B'$ such that $|B'| < |B|$ and $B$ is effectively representable by $B'$.
\end{lemma}
\begin{proof}
Since $B$ is multiplicatively dependent, Corollary~\ref{cor:05061625} implies that we can compute a non-zero vector $\vec \lambda \in \Int^n$ such that 
$b_1^{\lambda_1}\cdots b_n^{\lambda_n}$ is a root of unity.
We can easily make sure that at least one of the $\lambda_i$ is larger than zero. Assume therefore w.l.o.g. that $\lambda_1 > 0$. Fix a set $B'=\{b'_2,\ldots,b'_{n}\}$ where each $b'_i$ is a $\lambda_1$-th root of $b_i$, that is $(b'_i)^{\lambda_1} = b_i$. Then
$$
b_1^{\lambda_1} \cdot \left(\prod_{i = 2}^{n} (b'_i)^{\lambda_i} \right)^{\lambda_1} \text{ is a root of unity and hence }
b_n \cdot \prod_{i = 0}^{n-1} (b'_i)^{\lambda_i} \text{ is a root of unity}.
$$
All operations are computable and effective representation of $B$ by $B'$ follows. 
\end{proof}

\begin{lemma}\label{lem:04061632}
Let $S \subseteq \Calg$ be a set of algebraic numbers. There is an effective representation system $B$ for $S$ which can be computed effectively from $S$. 
\end{lemma}
\begin{proof}
Let $B' = S$ and apply Lemma~\ref{lem:04061619} recursively on $B$. Since the empty set is multiplicatively independent, after at most finitely many steps, we find an effective representation system $B$ for $S$.
\end{proof}

\begin{proof}[of Lemma \ref{lem:alg_to_w_alg_red}]
Let $S$ be the set of non-zero entries of $A$. By Lemma~\ref{lem:04061632} we can compute an effective representation system $B$ for $S$. However, with respect to our model of computation we need to be a bit careful, here: assume that $S \subseteq \Q(\theta)$ for some primitive element $\theta$. The application of Lemma~\ref{lem:04061632} does not allow us to stipulate that $B  \subseteq \Q(\theta)$. But in another step of precomputation, we can compute another primitive element $\theta'$ for the elements of $B$ such that $B \subseteq \Q(\theta')$. Then we may consider all computations as taking place in $\Q(\theta')$.

Assume that $B=\{b_1,\ldots, b_n\}$, then every non-zero entry of $A$ has a unique computable representation 
$$
A_{ij} = \zeta_{ij} \cdot \prod_{\nu = 1}^n b_{\nu}^{\lambda_{ij\nu}} \quad \text{ for $\zeta_{ij}$ a root of unity.}
$$
Let $p_1,\ldots,p_\beta$ be $\beta = |B|$ distinct prime numbers and define $A'$ as the matrix obtained from $A$ by replacing in each non-zero entry $A_{ij}$ the powers of $b \in B$ by the corresponding powers of primes, that is,
$$
A'_{ij} = \zeta_{ij} \cdot \prod_{\nu = 1}^n p_\nu^{\lambda_{ij \nu}}.
$$
Recall the definition of $\wset_A(G)$ in equation \eqref{eq:define_wset}. For each $w \in \wset_A(G)$ we can, in polynomial time compute a representation $w = \prod_{i,j} A_{ij}^{m_{ij}}$ as powers of elements in $S$. 
The effective representation of $S$ in terms of $B$ extends to $\wset_A(G)$ being effectively representable by $B$. Moreover, as $S$ depends only on $A$, the representation of each $w \in \wset_A(G)$ is even polynomial time computable.
We have
$$
Z_{A,D}(G) = \sum_{w \in \wset_{A}(G)} w \cdot N_{A,D}(G,w)
$$
In particular, for each $w \in \wset_A(G)$, we can compute unique $\lambda_{w,1},\ldots, \lambda_{w,n} \in \Int$ such that $w  \cdot  b_1^{\lambda_{w,1}} \cdots b_n^{\lambda_{w,n}}$ is a root of unity.
Define functions $f$ and $g$ such that for every $w \in \wset_A(G)$ we have 
$$
f(w) =  \prod_{\nu = 1}^n p_\nu^{\lambda_{w,\nu}} \quad \text { and } \quad
g(w) =  \prod_{\nu = 1}^n b_\nu^{\lambda_{w,\nu}}.
$$
Thus we obtain
$$
Z_{A',D}(G) = \sum_{w \in \wset_{A}(G)} w \cdot \dfrac{f(w)}{g(w)} \cdot N_{A,D}(G,w).
$$
This yields a reduction for $\eval(A',D) \Tle \eval(A,D)$. The other direction follows by
\[
Z_{A,D}(G) = \sum_{w' \in \wset_{A'}(G)} w' \cdot \dfrac{g(w)}{f(w)} \cdot N_{A',D}(G,w').
\]
\end{proof}

\subsection{Pinning Vertices}\label{sec:pin}
In this section we will prove the Pinning Lemma~\ref{lem:pinning}.
Before we can do this we need to establish an important property 
of partition functions $Z_{A,D}$: they allow 
to \emph{reconstruct} the pair $(A,D)$, in the sense that
the values $Z_{A,D}(G)$ for all digraphs $G$ determine the pair $(A,D)$ 
up to isomorphism. 

The notion of \sdef{isomorphism} employed here is 
given by considering $A$ as (the adjacency matrix of) a weighted graph. 
Let $A,D \in \C^{m \times m}$ and $A',D' \in \C^{m' \times m'}$ be matrices 
such that $D$ and $D'$ are diagonal. Then
$A$ and $A'$ are \sdef{isomorphic}, 
if $m=m'$ and they admit a
bijection $\alpha: [m] \rightarrow [m']$ such that for all $i,j \in [m]$ 
we have $A_{ij} = A'_{\alpha(i),\alpha(j)}$. We call $\alpha$ an \emph{isomorphism}.
If further $D_{ii} = D'_{\alpha(i),\alpha(i)}$ for all $i \in [m]$ then
the pairs $(A,D)$ and $(A',D')$ are \sdef{isomorphic}.

\subsubsection{The Reconstruction Lemma}
It will be convenient in the following to fix pinnings and consider digraphs 
compatible with these. To define this, fix $\vpin:[k] \rightarrow [m]$ to denote
our pinning. A \sdef{$k$-labeled digraph} $G$ is then a digraph with $k$ 
distinct vertices labeled $1, \ldots, k$, in this way $\vpin$ is compatible with
every $k$-labeled digraph.
The main technical result of this section, is the following.
\begin{lemma}[The Reconstruction Lemma]\label{lem:reconstruct}
Let $A \in \C^{m \times m}$ and $A' \in \C^{m' \times m'}$ be twin-free, $D \in \Real^{m \times m}$ and $D' \in \Real^{m' \times m'}$ diagonal matrices of positive vertex weights and $\vpin:[k] \rightarrow [m]$ and $\vpin':[k] \rightarrow [m']$ pinnings.
Assume that 
$$Z_{A,D}(\vpin,G) = Z_{A',D'}(\vpin',G) \text{ for all $k$-labeled digraphs $G$.}$$
Then there is an isomorphism $\alpha$ between $(A,D)$ and $(A',D')$ such that $\vpin' = \vpin\circ \alpha$.
\end{lemma}
Similar results been given first by Lov\'{a}sz \cite{lov67} for the case that $A$ is the adjacency matrix of a graph and in \cite{lov06} for symmetric real-valued $A$. Schrijver \cite{sch08} 
gave a proof for symmetric complex-valued matrices.

We will prove Lemma~\ref{lem:reconstruct} 
by extending the proof in \cite{lov06}. We will therefore fix $m \ge m'$, twin-free matrices $A \in \C^{m \times m}$ and $A' \in \C^{m' \times m'}$ and diagonal matrices $D \in \Real^{m \times m}$ and $D' \in \Real^{m' \times m'}$ of positive vertex weights.
For convenience we will consider $k$-tuples $\vec{x} \in [m]^k$ as pinnings $\vec{x} : [k] \rightarrow [m]$ and for any $x \in [m]$ the expression $\vec x x$ then denotes a $(k+1)$-tuple with last component $x$.
We define an equivalence relation between tuples $\vec x \in [m]^k$ and $\vec y \in [m']^k$
by 
$$
\vec x \sim \vec y \;\text{ iff }\; Z_{A,D}(\vec x,G) = Z_{A',D'}(\vec y,G) \text{ for all $k$-labeled digraphs } G.
$$
The product $G_1G_2$ of two $k$-labeled digraphs is the digraph obtained from $G_1$ and $G_2$ by taking their disjoint union and identifying vertices with the same label. For $k$-labeled digraphs $G_1$ and $G_2$ and every $\vec{x} \in [m]^k$ we have
\begin{equation}\label{eq:left_product_identity}
Z_{A,D}(\vec x,G_1G_2) = Z_{A,D}(\vec x, G_1) \cdot Z_{A,D}(\vec x, G_2).
\end{equation}

\begin{lemma}\label{lem:extension}
Let $\vec x\in [m]^k$ and $\vec y  \in [m']^k$ such that $\vec x \sim \vec y$.
Then for every and $x \in [m]$ there is a $y \in [m']$ such that $\vec xx \sim \vec yy$.
\end{lemma}
\begin{proof}\marc{proof OK}
Note first that
$$
Z_{A,D}(\vec x,G) = \sum_{x=1}^m D_{xx}Z_{A,D}(\vec xx,G).
$$
The condition that $\vec x \sim \vec y$ therefore implies that, for every $(k+1)$--labeled digraph $G$
\begin{equation}\label{eq:x_equivalent_y}
\sum_{x=1}^m D_{x x}Z_{A,D}(\vec x x, G) = \sum_{y=1}^{m'} D'_{y y} Z_{A',D'}(\vec y y, G).
\end{equation}
Define pairs $(X_1,Y_1), \ldots, (X_s,Y_s)$ of sets in the following way. The sets $X_1,\ldots, X_s$ partition $[m]$ with possibly some empty parts and $Y_1, \ldots, Y_s$ are partitions of $[m']$ with possibly empty parts, as well. The partitions satisfy, for every $i \in [s]$, that $x,x'\in X_i$ if, and only if, 
$Z_{A,D}(\vec x x,G) = Z_{A,D}(\vec x x',G)$ for all $k+1$-labeled graphs $G$. The analogue holds for the $y,y' \in Y_i$ with $Z_{A',D'}(\vec y y,G) = Z_{A',D'}(\vec y y',G)$ for all $k+1$-labeled graphs $G$.
Further for all $i \in [s]$, the set $X_i \cup Y_i$ is non-empty and for all $x \in X_i$ and $y \in Y_i$ we have 
\begin{equation}\label{eq:2503091600}
Z_{A,D}(\vec x x,G) = Z_{A',D'}(\vec y y,G) \text{ for all $k+1$-labeled digraphs } G. 
\end{equation}
Define for all $i \in [s]$ a $(k+1)$-tuple $\vec z_i$ and a function $Z(\vec z_i, G)$ such that $\vec z_i := \vec x x$ for some $x\in X_i$  and $Z(\vec z_i, G) := Z_{A,D}(\vec x x, G)$ if $X_i$ is non-empty and $\vec z_i := \vec yy$ for $y \in Y_i$ with $Z(\vec z_i, G) := Z_{A',D'}(\vec y y, G)$ otherwise. Let furthermore 
$$
c_i := \sum_{x \in X_i} D_{xx} - \sum_{y \in Y_i} D'_{yy}
$$
Equation \eqref{eq:x_equivalent_y} now simplifies to
$$
0 = \sum_{i=1}^s c_i Z(\vec z_i, G).
$$
We claim that $c_i = 0 \text{ for all } i \in [s]$.
If this holds the proof of the lemma follows as by the non-negativity of the vertex weights $c_i=0$ implies $X_i \neq \emptyset$ and $Y_i \neq \emptyset$ which, by equation \eqref{eq:2503091600}, provides us with the desired extensions of $\vec x$ and $\vec y$.
We will show, more generally, that for all sets $I \subseteq [s]$ the equation
\begin{equation} \label{eq:2403091837}
0 = \sum_{i\in I} c_i Z(\vec z_i, G) \quad \text{ for all digraphs } G.
\end{equation}
implies that $c_i = 0$ for all $i \in I$. We will prove this by induction on the cardinality of $I$.

For $I= \{i\}$ let $E_{k+1}$ be the $k+1$-labeled digraph on $k+1$ vertices which contains no edges. If $\vec z_i = \vec x x$, we have $Z(\vec z_i, E_{k+1}) = Z_{A,D}(\vec z_i, E_{k+1}) = 1$ and thus equation \eqref{eq:2403091837} implies $c_i =0$. An analogous argument proves the case that $\vec z_i = \vec y y$.

Let $|I| > 1$. By definition, for all $i \neq j \in I$ there is a digraph $G_{ij}$ such that $Z(\vec z_i, G_{ij}) \neq Z(\vec z_j, G_{ij})$.
Partition $I$ into equivalence classes $J_0, \ldots, J_t$ such that $a,b \in J_\nu$ iff $Z(\vec z_a, G_{ij}) = Z(\vec z_b, G_{ij})$. Define, for each $i \in [0,t]$ the tuple $\vec z'_{i}$ such that $\vec z'_i = \vec z_a$ for some $a \in J_i$.
Let $H^p$ denote the $k+1$-labelled graph product of $H$ with itself taken $p$ times, then by equation \eqref{eq:left_product_identity} and \eqref{eq:2403091837} we have
$$
0 = \sum_{\nu = 0}^t \left(\sum_{\mu \in J_{\nu}} c_\mu \right) Z(\vec z'_{\nu}, G^p_{ij}G) = \sum_{\nu = 0}^t Z(\vec z'_{\nu}, G_{ij})^p
\left(\sum_{\mu \in J_{\nu}} c_{\mu}Z(\vec z_{\mu}, G) \right).
$$
By definition, the values $Z(\vec z'_{\nu}, G_{ij})$ are pairwise different. It may be the case that one of these is zero. Assume w.l.o.g. that $0 = Z(\vec z'_{0}, G_{ij})$.
For $p=1,\ldots, t$, this gives rise to a homogeneous system of linear equations with a Vandermonde determinant: 
$$
0 = \sum_{\nu = 1}^t Z(\vec z'_{\nu}, G_{ij})^p
\left(\sum_{\mu \in J_{\nu}} c_{\mu}Z(\vec z_{\mu}, G) \right).
$$
Solving this system yields $0 = \sum_{\mu \in J_{\nu}} c_{\mu}Z(\vec z_{\mu}, G)$ for all $\nu \ge 1$. 
Since $|J_{\nu}| < n$, the induction hypothesis implies that we have $c_i = 0$ for all $i \in J_{\nu}$ and every $\nu \ge 1$. Therefore, by equation \eqref{eq:2403091837}, we further obtain
$$
0 = \sum_{j \in J_0} c_j Z(\vec z_j, G).
$$
which, again by the induction hypothesis, implies $c_j = 0$ for all $j \in J_0$.
\end{proof}

\begin{lemma}\label{lem:inj_twin_free}
Every map $\psi:[m] \rightarrow [m']$ such that $A_{ij} = A'_{\psi(i)\psi(j)}$ for all $i,j \in [m]$ is injective.
\end{lemma}
\begin{proof}\marc{proof OK}
Let $J$ be the image of $\psi$ and define a mapping $\psi': J \rightarrow [m]$ such that for each $j \in J$ we fix an $i \in [m]$ with $\psi(i) = j$ and define $\psi'(j) = i$. Define $\phi: [m] \rightarrow [m]$ by $\phi = \psi' \circ \psi$.

The precondition of the lemma implies that $A_{ij} = A_{\phi(i)\phi(j)}$ for all $i,j \in [m]$. We will show first that $\phi$ has a power $\phi^s$ which is \emph{idempotent}, that is $\phi^s = \phi^{2s}$. To see this, note that by the finiteness of $[m]$ not all powers of $\phi$ can be different. Hence, there are $l,k$ such that $\phi^l = \phi^{l+k} = \phi^{l+ik}$ for all $i \in \Nat$. Multiplying from the left with $\phi^j$ we have $\phi^{l+j} = \phi^{l+j+ik}$ for all $i,j \in \Nat$. Let $i$ be large enough such that $ik - l > 0$ then with $j = ik -l$ we have $2(l+j) = l+j+ik$. Hence $\phi^{l+j} = \phi^{l+j + ik} = \phi^{2(l+j)}$.

Now $A_{ij} = A_{\phi(i)\phi(j)}$ implies $A_{ij} = A_{\phi^s(i)\phi^s(j)}$ for all $s$. Hence
for $s$ such that $\phi^s = \phi^{2s}$ we have
$$
A_{ij} = A_{\phi^s(i)\phi^s(j)} = A_{\phi^{2s}(i)\phi^s(j)} = A_{\phi^s(i)j} \text{ for all } j \in [m].
$$
Recall that $A$ is twin-free. Thus $\phi^s$ is the identity and therefore $\phi$ must be bijective. Further by the definition of $\phi$ this implies that $\psi$ is injective.
\end{proof}

\begin{lemma}\label{lem:2503091143}
Let $\vec x \in [m]^m$ and $\vec y \in [m']^m$ such that $\vec x$ is bijective and $\vec x \sim \vec y$. Then there is an isomorphism $\alpha$ between $(A,D)$ and $(A',D')$ such that $\vec y= \alpha \circ \vec x$. 
\end{lemma}
\begin{proof}\marc{proof OK}
Define $G_{ij}$ as the $m$-labeled digraph on $m$ vertices with the single edge $ij$. Assume w.l.o.g. that $x_i = i$ for all $i \in [m]$.
By $\vec x \sim \vec y$ we have
\begin{equation}\label{eq:A_automorphism}
A_{ij} = Z_{A,D}(\vec x, G_{ij}) = Z_{A',D'}(\vec y, G_{ij}) = A'_{y_iy_j}. 
\end{equation}
Since $A'$ is twin-free Lemma~\ref{lem:inj_twin_free} implies that $\vec y$ is injective and by the condition that $m \ge m'$ it is thus bijective and $m= m'$. This gives rise to an isomorphism $\alpha:[m] \rightarrow [m]$ between $A$ and $A'$ by defining $\alpha(x_i) = \alpha(i) =: y_i$ for all $i \in [m]$.

It remains to show that $D_{ii} = D'_{\alpha(i),\alpha(i)}$ holds for all $i \in [m]$. Let $Q = \{i \in [m] \mid A\row i \neq 0\}$ and $P = \{j \in [m] \mid A\col j \neq 0\}$ be the set of indices corresponding to non-zero rows and columns of $A$. If one of these sets is empty, then $Q=P = \emptyset$ and $A =(0)$ since $A$ is twin-free and then the following claim implies $D_{11} = D'_{11}$.

\begin{claim}\label{cl:2906091500}
We have
$$
\sum_{j = 1}^m D_{jj} = \sum_{j = 1}^m D'_{\alpha(j)\alpha(j)}.
$$
\end{claim}
\begin{clproof}
Let $G_0$ be the $m$-labeled digraph on vertex set $[m] \dot\cup \{v\}$ without edges. By $\vec x \sim \vec y$ we have 
\[
\sum_{j = 1}^m D_{jj} = Z_{A,D}(\vec x, G_0) = Z_{A',D'}(\vec y, G_0) = \sum_{j = 1}^m D'_{\alpha(j)\alpha(j)}.
\]
\end{clproof}
Assume therefore that $Q,P \neq \emptyset$. Let $p : Q \to \Nat$ be an mapping, we say that it is \emph{non-vanishing} if there is an $i \in Q$ such that $p(i) > 0$. Define $G^{[p]}$ as the $m$-labeled digraph on vertex set $[m] \dot\cup \{v\}$ with edges $E = \{(iv)^{p(i)} \mid i \in Q\}$, that is for each $i \in Q$ there is an edge with multiplicity $p(i)$. Note in particular that the edge $iv$ does not exist if $p(i)=0$.

For every non-vanishing $p: Q \to \Nat$, we have
$$
Z_{A,D}(\vec x, G^{[p]}) = \sum_{j \in P} D_{jj}\prod_{i \in Q} (A_{ij})^{p_i}.
$$
By $\vec x \sim \vec y$ we have $Z_{A,D}(\vec x, G^{[p]}) = Z_{A',D'}(\vec y, G^{[p]})$ and therefore, 
\begin{eqnarray*}
 \sum_{j \in P} D_{jj}\prod_{i \in Q} (A_{ij})^{p(i)} &=&
 \sum_{j \in P} D'_{\alpha(j)\alpha(j)}\prod_{i \in Q} (A'_{\alpha(i)\alpha(j)})^{p(i)}
\end{eqnarray*}
Now $\alpha$ is an isomorphism between $A$ and $A'$, that is $A_{ij} = A'_{\alpha(i)\alpha(j)}$ and thus the above simplifies to
\begin{equation}\label{eq:2502091210}
0 = \sum_{j \in P} \left(D_{jj} - D'_{\alpha(j)\alpha(j)}\right)\prod_{i \in Q} (A_{ij})^{p(i)} \quad \text{ for all non-vanishing } p: Q \to \Nat. 
\end{equation}
We shall show the following
\begin{claim}\label{cl:2906091444}
Let $J \subseteq P$ of cardinality at least $2$ and assume that
\begin{equation}\label{eq:2906091447}
0 = \sum_{j \in J} \left(D_{jj} - D'_{\alpha(j)\alpha(j)}\right)\prod_{i \in Q} (A_{ij})^{p(i)} \quad \text{ for all non-vanishing } p: Q \to \Nat. 
\end{equation}
Then for every $j \in J$, there is a set $J' \subset J$ with $j \in J'$ such that
\begin{equation}\label{eq:2906091448}
0 = \sum_{j \in J'} \left(D_{jj} - D'_{\alpha(j)\alpha(j)}\right)\prod_{i \in Q} (A_{ij})^{p(i)} \quad \text{ for all non-vanishing } p: Q \to \Nat. 
\end{equation}
\end{claim}
Before we prove this claim, let us see how it finishes the proof of the lemma. Its statement will be particularly true for $J = P$ and the condition of which is given by equation \eqref{eq:2502091210}. Thus, for every $j \in P$ iterative application of this claim eventually implies
\begin{equation*}
0 = \left(D_{jj} - D'_{\alpha(j)\alpha(j)}\right)\prod_{i \in Q} (A_{ij})^{p(i)} \quad \text{ for all non-vanishing } p: Q \to \Nat. 
\end{equation*}
Since $j \in P$ there is an $i \in Q$ such that $A_{ij} \neq 0$. Fix a $p:Q\to \Nat$ which satisfies $p(i) = 1$ and $p(i') = 0$ for $i'\neq i$. Then we have $0 = \left(D_{jj} - D'_{\alpha(j)\alpha(j)}\right)A_{ij}$ and thus $D_{jj} = D'_{\alpha(j)\alpha(j)}$.

Therefore we obtain $D_{jj} = D'_{\alpha(j)\alpha(j)}$ for all $j \in P$. If $A$ contains no zero columns, we are done as then $P=[m]$. Otherwise, since $A$ is twin-free, there is exactly one $k$ such that $A\col k = 0$ and $P \cup \{k\} = [m]$. Thus Claim~\ref{cl:2906091500} further implies $D_{kk} = D'_{\alpha(k)\alpha(k)}$.

\paragraph*{The Proof of Claim~\ref{cl:2906091444}.}
As $J$ is of cardinality at least $2$, we may fix distinct $j,j' \in J$. By $A$ being twin-free we may fix an $i^* \in Q$ such that $A_{i^* j} \neq A_{i^* j'}$. Let $p: Q \to \Nat$ be a non-vanishing mapping. Define a non-vanishing $p_q: Q \to \Nat$ by $p_q(i^*) = p(i^*) + q$ and $p_q(i) = p(i)$ for all $i \neq i^*$.
The condition of the claim implies that
\begin{eqnarray*}
0 &=& \sum_{j \in J} \left(D_{jj} - D'_{\alpha(j)\alpha(j)}\right)\prod_{i \in Q} (A_{ij})^{p_q(i)} \\
  &=& \sum_{j \in J} (A_{i^* j})^{q}\left(D_{j j} - D'_{\alpha(j)\alpha(j)}\right)\prod_{i \in Q} (A_{i j})^{p(i)}
\end{eqnarray*}
Partition $J$ into equivalence classes $J_0,\ldots,J_t$ such that $A_{i^* j} = A_{i^* j'}$ iff $j, j' \in J_k$ for some $k \in [t]$. Define, for all $k \in [0,t]$ values $a_{k} := A_{i^* \nu}$ for some $\nu \in J_k$. Then
$$
0 = \sum_{k=0}^t (a_{k})^{q}\left(\sum_{j \in J_k} \left(D_{jj} - D'_{\alpha(j)\alpha(j)}\right)\prod_{i \in Q} (A_{i j})^{p(i)}\right)
$$
It may be the case that one of the $a_k$ values is zero. Assume w.l.o.g. that $a_0 = 0$. By condition, the above holds for all $q$. In particular, for $q = 1, \ldots, t$ this gives rise to a homogeneous system of linear equations
$$
0 = \sum_{k=1}^t (a_{k})^{q}\left(\sum_{j \in J_k} \left(D_{jj} - D'_{\alpha(j)\alpha(j)}\right)\prod_{i \in Q} (A_{i j})^{p(i)}\right).
$$
Since all the values $a_1, \ldots, a_k$ are pairwise distinct and non-zero, this system has a Vandermonde determinant and is thus solvable, which implies, for all $k \in [t]$,
$$
0 = \sum_{j \in J_k} \left(D_{jj} - D'_{\alpha(j)\alpha(j)}\right)\prod_{i \in Q} (A_{i j})^{p(i)} 
\quad \text{ for all non-vanishing } p: Q \to \Nat. 
$$
In combination with the condition \eqref{eq:2906091447} of the claim we also obtain
\[
0 = \sum_{j \in J_0} \left(D_{jj} - D'_{\alpha(j)\alpha(j)}\right) \prod_{i \in Q} (A_{ij})^{p(i)}
\quad \text{ for all non-vanishing } p: Q \to \Nat. 
\]
Since the $J_0,\ldots,J_t$ are a partition of $J$ and each of them has cardinality strictly smaller than $J$, the claim follows.
\end{proof}
Now we are able to prove the Reconstruction Lemma.

\begin{proof}[of Lemma \ref{lem:reconstruct}]\marc{proof OK}
Let $A,D$ be $m \times m$ matrices and $A',D'$ be $m'\times m'$ matrices as given in the lemma and $\vpin, \vpin'$ pinnings. In accordance with the notation of this section we define $\vec x : = \vpin$ and $\vec y := \vpin'$ and we assume w.l.o.g. that $m \ge m'$. The precondition that $Z_{A,D}(\vpin,G) = Z_{A',D'}(\vpin',G)$ for all digraphs $G$ is tantamount to $\vec x \sim \vec y$.

It will be easier to work with surjective pinnings. Extend $\vec x$ to a surjective $\vec x' := \vec x \vec z \in [m]^l$ with $l \ge m$. Lemma~\ref{lem:extension} implies the existence of a $\vec z'$ such that for $\vec y' = \vec y \vec z'$ we have $\vec x' \sim \vec y'$. Observe that if we find an isomorphism $\alpha$ between $(A,D)$ and $(A',D')$ such that $\vec y' = \alpha \circ \vec x'$ then the proof follows as this implies $\vec y = \alpha \circ \vec x$.

We shall prove the existence of such an isomorphism. Assume w.l.o.g. that $x'_i = i$ for all $i \in [m]$ --- we can achieve this by permuting the labels appropriately. Define $\vec x'|_{I} = (x'_i)_{i \in I}$ for a set of indices $I \subseteq [l]$.

\begin{claim}\label{cl:2503091343}
If $\vec x'' x \sim \vec y'' y$ for some $x \in [m]$ and $y  \in [m']$ then $\vec x'' \sim \vec y''$.
\end{claim}
\begin{clproof}
To see that this holds, assume the contrary. That is, for some $k$-labeled digraph $G$ we have $Z_{A,D}(\vec x'',G) \neq Z_{A',D'}(\vec y'', G)$. Let $G'$ be the digraph obtained from $G$ by adding a single isolated vertex labeled $k+1$, then
$$
Z_{A,D}(\vec x''x,G') = Z_{A,D}(\vec x'', G) \neq Z_{A',D'}(\vec y'', G) =  Z_{A',D'}(\vec y''y,G')
$$
contradicting the assumption.
\end{clproof}
This claim implies that $\vec x'|_{[m]} \sim \vec y'|_{[m]}$. The $m$-tuple $\vec x'|_{[m]}$ is bijective by definition. Lemma~\ref{lem:2503091143} thus implies $\vec y'|_{[m]} = \alpha \circ \vec x'|_{[m]}$ for some isomorphism $\alpha$ between $(A,D)$ and $(A',D')$.

It remains to show that for all $j \in [m+1,l]$ we have $y'_j = \alpha(x'_j)$. The bijectivity of $\vec x'|_{[m]}$ implies $x'_j = x'_r$ for an appropriate $r \in [m]$. Define $I = \{1, \ldots, r-1,r+1, \ldots, m,j\}$, then $\vec x'|_I$ is bijective. Therefore, Lemma~\ref{lem:2503091143} implies the existence of an isomorphism $\beta$ between $(A,D)$ and $(A',D')$ such that $\vec y'|_I = \beta \circ \vec x'|_I$, particularly implying bijectivity of $\vec y'|_I$. It follows thus for our original automorphism $\alpha$ that $y'_j = y'_r = \alpha(x'_r) = \alpha(x'_j)$.
\end{proof}

\subsubsection{The Proof of The Pinning Lemma.}
Applying the Reconstruction Lemma~\ref{lem:reconstruct}, we will now prove the
Pinning Lemma.
\begin{lemma}\label{lem:june10_1}
Let $A\in \Calg^{m\times m}$ be a Hermitian matrix and $D \in \Ralg^{m\times m}$ a diagonal matrix of positive vertex weights. Let $\twres{A}$ be the twin resolvent of $A$. Then
\[
\evalk(\twres{A},D^{\twres{A}}) \Tequiv \evalk(A,D)
\;\text{ and }\;
\eval(\twres{A},D^{\twres{A}}) \Tequiv \eval(A,D).
\]
\end{lemma}
\begin{proof}
Let $\tau$ be the twin resolution mapping on $A$. By the Twin Reduction Lemma~\ref{lem:cng_twin_red}(2) we have 
\[
  Z_{A,D}(\phi,G) = Z_{\twres A,D^{\twres A}}(\tau\circ \phi, G), \text{ for every digraph } G \text{ and pinning } \vpin.
  \]
This yields $\evalk(A,D) \Tle \evalk(\twres{A},D^{\twres A})$. On the other hand, for some instance $G$ and $\vpin'$ of $\evalk(\twres{A},D^{\twres A})$ let $\vpin$ be a pinning on $A$ such that $\tau \circ \vpin = \vpin'$, then the above also yields the backward direction. The result for $\eval(A,D)$ follows as the above holds likewise for empty pinnings.
\end{proof}
\begin{lemma}\label{lem:pinning_twin_free}
Let $A\in \Calg^{m\times m}$ be a Hermitian and twin-free matrix and $D \in \Ralg^{m\times m}$ a diagonal matrix of positive vertex weights.  Then
$$
  \evalk(A,D) \Tequiv \eval(A,D).
$$
\end{lemma}
\begin{proof}
It suffices to show $\evalk(A,D) \Tle \eval(A,D)$ as the other direction holds trivially.

Let $G=(V,E)$ and a pinning $\vpin$ be an instance of $\evalk(A,D)$. 
By appropriate permutation of the rows/columns of $A$ and $D$ (cf. Lemma~\ref{lem:principle_permute}) we may assume that $[k] = \textup{img } \vpin \subseteq [m]$ for some $k \le m$.
Let $\hat{G} = (\hat V, \hat E)$ be the graph obtained from $G$ by collapsing the the sets $\vpin^{-1}(i)$ for all $i \in [k]$. Formally, define a map
$$
\gamma(v) = \left\lbrace \begin{array}{l l}
                          i &, v \in \vpin^{-1}(i) \text{ for some } i \in [k] \\
                          v &, \text{ otherwise}
                         \end{array}\right.
$$
Then $\hat G$ is a multi-digraph (with possibly some self-loops) defined by 
\begin{eqnarray*}
\hat{V} & = & [k] \;\dot\cup\; (V \setminus \text{def}(\vpin))\\
\hat{E} & = & \{ \gamma(u)\gamma(v) \mid uv \in E\}.
\end{eqnarray*}
Hence $Z_{A,D}(\vpin,G) = Z_{A,D}(\idfunc_{[k]},\hat G)$ where $\idfunc_{[k]}$ denotes the identity map on $[k]$.
Call two mappings $\chi,\psi: [k] \rightarrow [m]$ \emph{equivalent} if there is an automorphism $\alpha$ of $(A,D)$ such that $\chi = \alpha\circ \psi$. Partition the mappings $\psi:[k]\rightarrow [m]$ into equivalence classes $I_1, \ldots, I_c$ according to this definition and for all $i \in [c]$ fix some $\psi_i \in I_i$.
Assume furthermore, that $\psi_1 = \idfunc_{[k]}$.
Clearly for any two $\chi, \psi$ from the same equivalence class, we have $Z_{A,D}(\chi,F) = Z_{A,D}(\psi,F)$ for every graph $F$. Therefore, for every graph $G'$,
\begin{equation}\label{eq:std_sum_fixation}
Z_{A,D}(G') = \sum_{i = 1}^c Z_{A,D}(\psi_i,G')\cdot \left(\sum_{\psi \in I_i} \prod_{v \in \df(\psi)} D_{\psi(v)\psi(v)}\right) 
\end{equation}
Define, for each $i \in [c]$ the value $c_i = \left(\sum_{\psi \in I_i} \prod_{v \in \df(\psi)} D_{\psi(v)\psi(v)} \right)$.
We claim the following
\begin{claim}\label{cl:2605091505}
Let $I \subseteq [c]$ be a set of cardinality at least $2$ such that $1 \in I$. Assume that we can, for every $k$-labeled digraph $G'$, compute the value
\begin{equation}\label{eq:260320091543}
\sum_{i \in I} c_i\cdot Z_{A,D}(\psi_i,G').
\end{equation}
Then there is a proper subset $I' \subset I$ which contains $1$ such that we can compute, for every $k$-labeled digraph $G''$, the value
\begin{equation}\label{eq:2605090815}
\sum_{i \in I'} c_i\cdot Z_{A,D}(\psi_i,G'').
\end{equation}
\end{claim}
\bigskip
This claim will allow us to finish the proof. To see this, note first that by equation \eqref{eq:std_sum_fixation} we can compute the value \eqref{eq:260320091543} for $I = [c]$ and $G' = \hat G$. Thus after at most $c$ iterations of Claim~\ref{cl:2605091505} we arrive at $c_1\cdot Z_{A,D}(\psi_1,\hat G)$. Further, $c_1$ is effectively computable in time depending only on $D$ and therefore we can compute $Z_{A,D}(\idfunc_{[k]},\hat G) = Z_{A,D}(\vpin,G)$. This proves the reducibility $\evalk(A,D) \Tle \eval(A,D)$. 

\paragraph*{Proof Of Claim~\ref{cl:2605091505}.} Assume that we can compute the value given in \eqref{eq:260320091543}.
Lemma~\ref{lem:reconstruct} implies that for every pair $i\neq j \in I$ there is a $k$-labeled digraph $\Gamma$ such that 
\begin{equation}\label{eq:distinguish_pf}
Z_{A,D}(\psi_i,\Gamma) \neq Z_{A,D}(\psi_j,\Gamma).
\end{equation}
Fix such a pair $i\neq j \in I$ and a graph $\Gamma$ satisfying this equation. Note that this graph can be computed effectively in time depending only on $A,D$ and $\psi_i,\psi_j$. Let $G^s$ denote the digraph obtained from $G$ by iterating $s$ times the $k$-labeled product of $G$ with itself. 
We can thus compute
\begin{equation}\label{eq:2603091557}
\sum_{i \in I} c_i\cdot Z_{A,D}(\psi_i,G'\Gamma^s) = \sum_{i \in I}^c c_i Z_{A,D}(\psi_i,G')\cdot Z_{A,D}(\psi_i,\Gamma)^s. 
\end{equation}
Partition $I$ into classes $J_0,\ldots ,J_z$ such that for every $\nu \in [0,z]$ we have $i',j' \in J_\nu$ if, and only if, $Z_{A,D}(\psi_{i'},\Gamma) = Z_{A,D}(\psi_{j'},\Gamma)$.
Since one of these sets $J_\nu$ contains $1$ and all of these are proper subsets of $I$, it remains to show that we can compute, for each $\nu \in [0,z]$, the value 
$$
\sum_{i' \in J_\nu} c_i Z_{A,D}(\psi_{i'},G').
$$
To prove this, define $x_\nu : = Z_{A,D}(\psi_{i'},\Gamma)$ for each $\nu \in [z]$ and an $i' \in J_{\nu}$. Equation \eqref{eq:2603091557} implies that we can compute
$$
\sum_{\nu = 0}^z x_\nu^s \left(\sum_{i' \in J_\nu} c_{i'} Z_{A,D}(\psi_{i'},G')\right). 
$$
One of the values $x_\nu$ might be zero. Assume therefore w.l.o.g. that $x_0 = 0$, then evaluating the above for $s= 1, \ldots, z$ yields a system of linear equations, which by Lemma~\ref{lem:interpolate} can be solved in polynomial time such that we can recover the values $\sum_{i' \in J_\nu} c_{i'} Z_{A,D}(\psi_{i'},G')$ for each $\nu \ge 1$. 
Using equation \eqref{eq:260320091543} we can thus also compute the value
\[
\left(\sum_{i \in I} c_i\cdot Z_{A,D}(\psi_i,G')\right) - \sum_{\nu = 1}^z \left(\sum_{i' \in J_\nu} c_{i'} Z_{A,D}(\psi_{i'},G')\right) = \sum_{i' \in J_0} c_{i'} Z_{A,D}(\psi_{i'},G'). 
\]
\end{proof}
\begin{proof}[of the Pinning Lemma \ref{lem:pinning}]
Fix the twin resolvent $A' = \twres A$ of $A$ and let $D' = D^{\twres A}$. 
By Lemma~\ref{lem:pinning_twin_free} we have $\evalk(A',D') \Tequiv \eval(A',D')$.
Hence together with Lemma~\ref{lem:june10_1} we obtain the chain of reductions
\[\evalk(A,D) \Tequiv \evalk(A',D') \Tequiv \eval(A',D') \Tequiv \eval(A,D).\]
\end{proof}

\subsection{Basic Complexity Results for Congruential Partition Functions}

\paragraph*{A Basic Polynomial Time Case.}
We provide the following basic tractability result which is a straightforward extension of Theorem 6 in \cite{bulgro05}.
\begin{lemma}\label{lem:ptime_rank1_cng}
Let $A \in \C^{m\times m}$ be a Hermitian matrix and $\mfam D$ a family of diagonal matrices of vertex-weights.
If $A$ has rank $1$ then $\evalk(A,\mfam D)$ is polynomial time computable.
\end{lemma}
\begin{proof}
Let $G = (V,E)$ be a given digraph and $\vpin$ a pinning. Let $V' = V \setminus \df(\vpin)$ then
\[
Z_{A, \mfam D}(\vpin, G) = \sum_{\vpin \subseteq \vcfg: V \rightarrow [m]} \prod_{uv \in E} A_{\vcfg(u),\vcfg(v)} \prod_{v \in V'} D^{\cngc{\grade(v)}}_{\vcfg(v),\vcfg(v)}
\]
As $\rank A = 1$, there are vectors $\vec  a, \vec b \in \Calg^m$ such that $A = \vec a \vec b^T$.
Then, for every configuration $\vcfg: V \rightarrow [m]$,
$$
\prod_{uv \in E} A_{\vcfg(u),\vcfg(v)} = \prod_{uv \in E} a_{\vcfg(u)} b_{\vcfg(v)} = \prod_{v \in V} a^{\outdeg(v)}_{\vcfg(v)} b^{\indeg(v)}_{\vcfg(v)}.
$$
Therefore,
\begin{eqnarray*}
Z_{A,D}(\vpin,G) &=& \sum_{\vpin \subseteq \vcfg : V \rightarrow [m]} \prod_{uv \in E} A_{\vcfg(u),\vcfg(v)} \prod_{v \in V'} D^{\cngc{\grade(v)}}_{\vcfg(v),\vcfg(v)} \\
           &=& \left(\prod_{v \in \df(\vpin)} a^{\outdeg(v)}_{\vpin(v)} b^{\indeg(v)}_{\vpin(v)} \right) \sum_{\vcfg : V' \rightarrow [m]} \prod_{v \in V'} a^{\outdeg(v)}_{\vcfg(v)} b^{\indeg(v)}_{\vcfg(v)} D^{\cngc{\grade(v)}}_{\vcfg(v),\vcfg(v)}\\
           &=& \prod_{v \in \df(\vpin)} a^{\outdeg(v)}_{\vpin(v)} b^{\indeg(v)}_{\vpin(v)} \prod_{v \in V'} \sum_{i = 1}^m  a^{\outdeg(v)}_{i} b^{\indeg(v)}_{i} D^{\cngc{\grade(v)}}_{i,i} 
\end{eqnarray*}
And the term in the last line can be evaluated in polynomial time.
\end{proof}

\paragraph*{Basic $\#\PP$-hardness Results.}
We need the following extension of the $\#\PP$-hardness criterion of Theorem~\ref{thm:BG}.
\begin{lemma}\label{lem:BG_extended_hardness}
Let $A \in \Ralg^{m \times m}$ be a non-negative symmetric matrix and $D \in \Ralg^{m \times m}$ a diagonal matrix of positive vertex weights. 
If $A$ contains a block of rank at least $2$ then $\eval(A,D)$ is $\#\PP$-hard.
\end{lemma}
We will give the proof of this Lemma below.
From this result we derive a basic $\#\PP$-hardness criterion 
for congruential partition functions.
\begin{lemma}\label{lem:cng_block2_hard}
Let $A$ be a Hermitian matrix and $\mfam D = (D^{\cngc c})_{c \in \Int_{\denom}}$ such that $D^{\cngc 0}$ is a diagonal matrix of positive vertex weights.
If $\abs{A}$ contains a block of row rank at least $2$ then 
$\evalk(A, \mfam D)$ is \#\PP-hard.
\end{lemma}
\begin{proof}
Note first, that we have $\evalk(\abs {A}^{(2)}, D^{\cngc 0}) \Tle \evalk(A, \mfam D)$.
To see this, let a graph $G =(V,E)$ and a pinning $\vpin$ be an instance of $\evalk(\abs A, D^{\cngc 0})$.
Let $G'$ be the graph obtained from $G$ by adding for each existing arc $uv$ an arc $vu$, we have $Z_{\abs{A}^{(2)},D^{\cngc 0}}(\vpin, G) =  Z_{A,\mfam D}(\vpin, G')$.

As $\abs A$ contains a block of row rank at least $2$ the matrix $\abs{A}^{(2)}$ does so as well. By Lemma~\ref{lem:BG_extended_hardness} the problem $\evalk(\abs{A}^{(2)}, D^{\cngc 0})$ 
is $\#\PP$-hard. 
\end{proof}

\paragraph*{The proof of Lemma~\ref{lem:BG_extended_hardness}}
The proof follows from Theorem~\ref{thm:BG} and the following Lemma which is a
straightforward extension of Theorem 3.2 in \cite{dyegre00}: 
\begin{lemma}\label{lem:dg_omit_vertexweights}
Let $A \in \Ralg^{m \times m}$ be a symmetric matrix with non-negative entries such that every pair of rows in $A$ is linearly independent. Let $D \in \Ralg^{m \times m}$ be a diagonal matrix of positive vertex weights. Then $$\eval(A) \Tle \eval(A,D).$$ 
\end{lemma} 
The proof adapts the one in \cite{dyegre00}; we 
present it for completeness.
The following lemmas are restatements of those in \cite{dyegre00} 
(see Lemma~3.4, 3.6, 3.7 and Theorem~3.1).
\begin{lemma}\label{lem:DG__eig_interpolate}
Let $A\in \Ralg^{m \times m}$ be symmetric and non-singular, $G =(V,E)$ a graph and $F \subseteq E$.
If we know the values
\begin{equation}\label{eq:3012091445}
f_r(G) = \sum_{\vcfg: V \rightarrow [m]} c_A(\sigma) \prod_{uv \in F} A^{r}_{\vcfg(u) \vcfg(v)}
\end{equation}
for all $r \in [(|F| + 1)^{m^2}]$, where $c_A$ is a function depending on $A$ but not on $r$. Then we can evaluate
\begin{equation}\label{eq:204101135}
\sum_{\vcfg: V \rightarrow [m]} c_A(\sigma) \prod_{uv \in F} (I_m)_{\vcfg(u) \vcfg(v)} 
\end{equation}
in polynomial time.
\end{lemma}
\begin{proof}
As $A$ is symmetric and non-singular, there is an orthogonal matrix $P$ such that $P^T A P =: D$ is a diagonal matrix with non-zero diagonal.
Every entry of $A^r$ satisfies
$A^r_{ij} = (P D^{r} P^T)_{ij} = \sum_{\mu = 1}^m P_{\vcfg(u)\mu} P_{ \vcfg(v) \mu} (D_{\mu\mu})^r$.
By equation \eqref{eq:3012091445}
\begin{eqnarray*}
f_r(G) 
       &=& \sum_{\vcfg: V \rightarrow [m]} c_A(\sigma) \prod_{uv \in F} \sum_{\mu = 1}^m P_{\vcfg(u)\mu} P_{ \vcfg(v) \mu} (D_{\mu\mu})^r 
\end{eqnarray*}
Define
$
\mathcal{W} = \left\{ \prod_{i=1}^m (D_{ii})^{\alpha_i} \mid 0 \le \alpha_i \text{ for all } i \in [m], \; \sum_{i=1}^m \alpha_i = |F|\right\}
$
which can be constructed in polynomial time. We rewrite
$$
f_r(G) = \sum_{w \in \mathcal{W}} c_w w^r.
$$
for unknown coefficients $c_w$. By interpolation (cf. Lemma~\ref{lem:interpolate}), we can recover these coefficients in polynomial time and can thus calculate $f_0(G) = \sum_{w \in \mathcal{W}} c_w$. We have
\begin{eqnarray*}
f_0(G) &=& \sum_{\vcfg: V \rightarrow [m]} c_A(\sigma) \prod_{uv \in F} \sum_{\mu = 1}^m P_{\vcfg(u)\mu} P_{ \vcfg(v) \mu} (D_{\mu\mu})^0 \\
\end{eqnarray*}
which is equal to \eqref{eq:204101135} by inspection.
\end{proof}
\begin{lemma}\label{lem:a_to_ada}
Let $A \in \Real^{m\times m}$ be a symmetric matrix in which every pair of distinct rows is linearly independent. Let $D \in \Real^{m \times m}$ be a diagonal matrix of non-negative vertex weights. Then every pair of rows in $ADA$ is linearly independent.
Furthermore there is an $0 < \epsilon < 1$ such that
$$
|(ADA)_{ij}| \le \epsilon \sqrt{(ADA)_{ii} (ADA)_{jj}}
$$
\end{lemma}
\begin{proof}
Define $Q = AD^{(1/2)}$. We have $ADA = AD^{(1/2)} D^{(1/2)}A^T = QQ^T$. That is $(ADA)_{ij} = \scalp{Q\row i, Q\row j}$ every pair of rows in $Q$ is linearly independent as it is linearly independent in $A$. 
By Cauchy-Schwarz $\scalp{Q\row i, Q\row j} < \sqrt{\scalp{Q\row i, Q\row i}\scalp{Q\row j, Q\row j}}$,
which implies that the corresponding $2\times 2$ submatrix of $ADA$ defined by $i$ and $j$ has non-zero determinant.
The existence of $\epsilon$ follows.
\end{proof}

\begin{lemma}\label{lem:make_non-sing}
Let $A\in \Real^{m \times m}$ be a symmetric non-negative matrix in which every pair of distinct rows is linearly independent. Let $D\in \Real^{m \times m}$ be a diagonal matrix of positive vertex weights. Then there is a $p \in \Nat$ such that the matrix $(ADA)^{(p)}$ is non-singular.
\end{lemma}
\begin{proof}
Let $A' = ADA$ and consider the determinant
$$
\det(A') = \sum_{\pi \in S_m} \pm \prod_{i=1}^m A'_{i\pi(i)}.
$$
Let $S_m$ be the set of permutations $\pi: [m] \rightarrow [m]$. For some $\pi \in S_m$ define $t(\pi) = |\{i \mid \pi(i) \neq i\}|$. Let $\epsilon$ be as in Lemma~\ref{lem:a_to_ada}. Then
\begin{equation}\label{eq:2606091840}
\prod_{i=1}^m| A'_{i\pi(i)}| \le \epsilon^{t(\pi)} \prod_{i=1}^m \sqrt{A'_{ii}}  \prod_{i=1}^m \sqrt{ A'_{\pi(i)\pi(i)}} = \epsilon^{t(\pi)} \prod_{i=1}^m A'_{ii}. 
\end{equation}
Let $\idfunc$ denote the trivial permutation. Then
\begin{eqnarray*}
\det((A')^{(p)}) &\ge& \left(\prod_{i=1}^m A'_{ii}\right)^p - \sum_{\pi \in S_m\setminus\{\idfunc\}} \left(\prod_{i=1}^m A'_{i\pi(i)}\right)^p.
\end{eqnarray*}
By equation \eqref{eq:2606091840}, we have
\begin{eqnarray*}
  m!\epsilon^p\left(\prod_{i=1}^m A'_{ii}\right)^p &\ge& \sum_{\pi \in S_m\setminus\{\idfunc\}} \left(\prod_{i=1}^m A'_{i\pi(i)}\right)^p
\end{eqnarray*}
and hence, as $0 < \epsilon < 1$, the matrix $(ADA)^{(p)}$ is non-singular for large enough $p$.
\end{proof}

\begin{proof}[of Lemma \ref{lem:dg_omit_vertexweights}]
Fix a a $p \in \Nat$ such that $(ADA)^{(p)}$ is non-singular by Lemma~\ref{lem:make_non-sing}.
Let a graph $G=(V,E)$ be an instance of $\eval(A)$. Define a graph $G' = (V',E')$ 
by $V'  = \{ v_0,\ldots,v_{d-1} \mid v\in V, d= \vdeg_G(v)\}$ and 
$E' = E'' \cup E'''$ for sets $E'',E'''$ to be defined next.
Let $E'''$ be the set which contains, for each $v\in V$ 
the cycle on $\{v_0,\ldots, v_{\vdeg_G(v)-1}\}$, 
i.e. $\{v_0v_1, \ldots, v_{\vdeg_G(v)-1}v_0\} \subseteq
E'''$. Let $E''$ be the set of edges, such that each edge from $E$ incident with
$v$ is connected to exactly one of the $v_i$. 

We further construct a graph $G'_{p,r}$ from $G'$ as follows.
Let $T_p$ be a graph which consists 
$p$ many length $2$ paths connecting terminals $a$ and $b$. 
Let $T_{p,r}$ be the series composition of $r$ many copies of $T_p$. 
$G'_{p,r}$ is obtained from $G'$ by replacing each edge in $E'''$ 
with a distinct copy of $T_{p,r}$.
We call $a$ and $b$ the "start" and "end" vertex of $T_p$ and $T_{p,r}$ has start
and end vertices induced by the start and 
end vertices of of the series composition of $T_p$. 

For a graph $H$ with designated ``start'' and ``end'' vertex let
$Z_{A,D}(i,j; H) := Z_{A,D}(\phi, H)$ such that $\phi$ pins the start vertex to $i$ 
and the end vertex to $j$. 

\begin{claim}
Let $C = (ADA)^{(p)}$, then with $X = D^{(1/2)}$, we have for all $i,j \in [m]$ and $r \in \Nat$,
\begin{equation}\label{eq:2303091755}
Z_{A,D}(i,j; T_{p,r}) = (X_{ii}X_{jj})^{-1}(XCX)^r_{ij}.
\end{equation}
\end{claim}
\begin{clproof}
Straightforwardly,
\begin{eqnarray*}
Z_{A,D}(i,j; T_p) &=& \left(\sum_{k=1}^m A_{ik}A_{kj}D_{kk}\right)^p = (ADA)^{(p)}_{ij} = C_{ij}
\end{eqnarray*}
and therefore
\begin{eqnarray*}
Z_{A,D}(i,j; T_{p,r}) &=& \sum_{\substack{\vcfg:[r+1] \rightarrow [m] \\ \vcfg(1)=i,\,\vcfg(r+1)=j}}
\prod_{k=1}^r Z_{A,D}(\vcfg(k),\vcfg(k+1);T_p) \prod_{k=2}^r D_{\vcfg(k),\vcfg(k)} \\
&=& (X_{ii}X_{jj})^{-1} \sum_{\substack{\vcfg:[r+1] \rightarrow [m] \\ \vcfg(1)=i,\,\vcfg(r+1)=j}} \prod_{k=1}^r X_{\vcfg(k),\vcfg(k)} C_{\vcfg(k),\vcfg(k+1)} X_{\vcfg(k+1),\vcfg(k+1)} 
\end{eqnarray*}
By inspection, the last line equals the right hand side of equation \eqref{eq:2303091755} --- as claimed.
\end{clproof}
%
%
For convenience, we count the indices of 
the vertices $v_{0}, \ldots, v_{\vdeg_G(v) - 1}$ modulo $\vdeg_G(v)$. 
Particularly, $v_{\vdeg_G(v)} = v_0$ and we have, for all $r \in \Nat$,
\begin{eqnarray*}
Z_{A,D}(G'_{p,r}) = \sum_{\vcfg:V' \rightarrow [m]} \prod_{uv \in E''} A_{\vcfg(u)\vcfg(v)} \prod_{v\in V'} D_{\vcfg(v),\vcfg(v)} \prod_{v\in V} \prod_{i=0}^{\vdeg_G(v) - 1} Z_{A,D}(\sigma(v_i),\sigma(v_{i+1});T_{p,r}).
\end{eqnarray*}
As the vertices in $V'$ are grouped according to the vertices in $V$ we have
$$
\prod_{v\in V'} D_{\vcfg(v),\vcfg(v)} = \prod_{v\in V}\prod_{i=0}^{\vdeg_G(v)-1} D_{\vcfg(v_i),\vcfg(v_i)}.
$$
By \eqref{eq:2303091755}, the expression
$\prod_{v\in V'} D_{\vcfg(v),\vcfg(v)} \prod_{v\in V} \prod_{i=0}^{\vdeg_G(v) - 1} Z_{A,D}(\vcfg(v_i),\vcfg(v_{i+1});T_{p,r})$
equals
\begin{eqnarray*}
\prod_{v\in V} \prod_{i=0}^{\vdeg_G(v) - 1} D_{\vcfg(v_i),\vcfg(v_i)} \dfrac{(XCX)^r_{\vcfg(v_i),\vcfg(v_{i+1})}}{X_{\vcfg(v_i),\vcfg(v_i)}X_{\vcfg(v_{i+1}),\vcfg(v_{i+1})}}
&=&
\prod_{v\in V} \prod_{i=0}^{\vdeg_G(v) - 1} (XCX)^r_{\vcfg(v_i),\vcfg(v_{i+1})}
\end{eqnarray*}
Thus
\begin{eqnarray*}
Z_{A,D}(G'_{p,r})
&=& \sum_{\vcfg:V' \rightarrow [m]} \prod_{uv \in E''} A_{\vcfg(u)\vcfg(v)} \prod_{uv\in E'''} (XCX)^r_{\sigma(u)\sigma(v)}
\end{eqnarray*}
Given the $\eval(A,D)$ oracle, we can, in polynomial time, evaluate this expression for every $r$ which is polynomial in the size of $G$. Lemma~\ref{lem:DG__eig_interpolate} implies that we can compute the value
\begin{eqnarray*}
Z &=&  \sum_{\vcfg:V' \rightarrow [m]} \prod_{uv \in E''} A_{\vcfg(u)\vcfg(v)} \prod_{uv\in E'''} I_{\sigma(u)\sigma(v)} 
\end{eqnarray*}
It remains to show that $Z = Z_{A}(G)$. To see this, note that, for every 
configuration $\vcfg:V' \rightarrow [m]$ the weight 
$\prod_{uv \in E''} A_{\vcfg(u)\vcfg(v)} \prod_{uv\in E'''} I_{\sigma(u)\sigma(v)}$ 
in the above expression is zero unless the following holds:
For all $v \in V$ we have $\vcfg(v_0) = \ldots = \vcfg(v_{d-1})$ for $d = d_G(v)$.
For such a configuration, define $\vcfg' : V \rightarrow [m]$ by 
$\vcfg(v) = \vcfg(v_0)$ for every $v \in V$.
Then
$$
\prod_{uv \in E''} A_{\vcfg(u),\vcfg(v)} = \prod_{uv \in E} A_{\vcfg'(u),\vcfg'(v)}.
$$
Since every configuration $\vcfg':V \rightarrow [m]$ arises this way 
this finishes the proof.
\end{proof}

\section{Connected Hermitian Matrices}\label{sec:herm_gen_case}

\noindent In this section we will prove Lemma~\ref{lem:gen_red_to_Had}. To do this we will split the proof into two parts --- one for the non-bipartite case and one for the bipartite case. The following two lemmas provide these two parts.

\begin{lemma}[The Non-Bipartite Case]\label{lem:gen_red_to_Had_non-bip}
Let $A$ be a connected non-bipartite $\denom$-algebraic and Hermitian matrix and $D$ a diagonal matrix of positive vertex weights. Then either the problem $\evalk(A,D)$ is $\#\PP$-hard or the following holds.

There is an $\denom$-algebraic Hermitian matrix $H$ and a family $\mfam U = (U^{\cngc c})_{c \in \Int_{\denom}}$ of diagonal matrices which define an \cond{H--STD} problem such that 
$$
\evalk(A,D) \Tequiv \evalk(H,\mfam U).
$$
\end{lemma}

\begin{lemma}[The Bipartite Case]\label{lem:gen_red_to_Had_bip}
Let $A$ be a connected bipartite $\denom$-algebraic Hermitian matrix and $D$ a diagonal matrix of positive vertex weights. Then either $\evalk(A,D)$ is $\#\PP$-hard or the following holds.

There is an $\denom$-algebraic Hermitian matrix $A'$ and a family $\mfam U = (U^{\cngc c})_{c \in \Int_{\denom}}$ of diagonal matrices which define a \cond{B--H--STD} problem such that 
$$
\evalk(A,D) \Tequiv \evalk(A',\mfam U).
$$
\end{lemma}
The proof of Lemma~\ref{lem:gen_red_to_Had} follows directly from these 
two lemmas. In the remainder of this section we will give the proof of the Non-Bipartite Case Lemma~\ref{lem:gen_red_to_Had_non-bip}. The proof of the 
Bipartite Case Lemma~\ref{lem:gen_red_to_Had_bip} is conceptually very similar,
therefore we omit it, the full proof can be found in \cite{thu09}.

\paragraph*{The proof of Lemma~\ref{lem:gen_red_to_Had_non-bip}.}
The strategy is to filter the problems $\evalk(A,D)$ until we are left with 
only those for which we can find a polynomial time equivalent \cond{H--STD} 
problem $\evalk(H, \mfam U)$. This process of filtering will be performed by 
transforming in several steps the problem $\evalk(A,D)$ into a problem 
$\evalk(C, \mfam D)$ where the matrix $C$ and the family $\mfam D$ satisfy 
certain conditions, which will be given below. In each step, if a transformation
is not possible, it will be shown that this implies $\#\PP$-hardness of the corresponding problem.

A bit of preparation is necessary. We say that a family $\mfam D = (D^{\cngc c})_{c \in \Int_{\denom}}$ \sdefi{fits}{fit} a matrix $C$ if the dimensions of each $D^{\cngc c}$ match those of $C$. The matrix $C$ under consideration will have a tensor product structure $X \otimes H$. Recall that we index the entries of such a matrix by pairs $(\mu,i),(\nu,j)$ such that $C_{(\mu,i),(\nu,j)} = X_{\mu\nu} \cdot H_{i,j}$. The diagonal matrices $(D^{\cngc c})_{c \in \Int_{\denom}}$ fit $C$ and therefore we index these in the same way. For convenience, we will alter the notation of the indices of these diagonal matrices a bit and let
$$
D^{\cngc {c};\kappa}_{k,k} := D^{\cngc c}_{(\kappa,k),(\kappa,k)} \text{ for all $c \in \Int_{\denom}$}.
$$

\paragraph*{Shape Conditions.} 
Let $C$ be an $\denom$-algebraic Hermitian matrix and $\mfam D = (D^{\cngc c})_{c \in \Int_{\denom}}$ a family of diagonal matrices of vertex weights that fits $C$.
We define conditions on the shape of $C$ and $\mfam D$, two of which determine the shape of $C$.
\sdefisub{}{Shape Conditions}{\cond{C1},\cond{C2}}
\begin{condition}{(C1)}
There are $r,m \in \Nat$, $\varsigma \in \{-1,1\}$, a non-singular normalized matrix $H \in \U_{\denom}^{r \times r}$ and positive vectors $v,w \in \Ralg^m$ such that $v_1 < \ldots < v_m$ and $w$ depends linearly on $v$ and 
       $$
          C = \varsigma \cdot vw^T \otimes H = \left(\begin{array}{c c c}
               \varsigma \cdot v_1w_1 H & \ldots & \varsigma \cdot v_1w_m H\\
               \vdots  & \ddots & \vdots \\
               \varsigma \cdot v_mw_1 H & \ldots & \varsigma \cdot v_mw_m H 
            \end{array}\right).
       $$
The submatrices $C_{(\mu,\absent)(\nu,\absent)} = \varsigma \cdot v_\mu w_\nu H$ are the \sdefi{tiles}{tile} of $C$. 
\end{condition}
\begin{condition}{(C2)}
The matrix $H$ is a normalized Hermitian complex Hadamard matrix.
\end{condition}
Three conditions give structure of the diagonal matrices in $\mfam D$.
\sdefisub{}{Shape Conditions}{\cond{D1},\cond{D2},\cond{D3}}
\begin{condition}{(D1)}
$D^{\cngc 0}$ has positive diagonal and for all $c \in \Int_{\denom}$ we have $D^{\cngc{-c}} = \cj{D^{\cngc c}}$.
\end{condition}
\begin{condition}{(D2)}
There is an $m \times m$ diagonal matrix $\Delta^{\cngc 0}$ with positive diagonal such that $D^{\cngc 0} = \Delta^{\cngc 0} \otimes I_{r}$.
\end{condition}
\begin{condition}{(D3)}
For each $c \in \Int_{\denom}$ there is a diagonal matrix $U^{\cngc c} \in (\{0\} \cup \U)^{r \times r}$ and a diagonal matrix $\Delta^{\cngc c}$ such that   $D^{\cngc c} = \Delta^{\cngc c} \otimes U^{\cngc c}$. 
Further, if $D^{\cngc c} = 0$ then $U^{\cngc c} = 0$.
\end{condition}
These conditions will play a central role throughout the whole section.
Once we make sure that they are satisfied, the problem $\evalk(C,\mfam D)$ 
has all properties necessary to find a polynomial time equivalent 
\cond{H--STD} problem. Then the following lemma provides us with the 
last part of the proof of Lemma~\ref{lem:gen_red_to_Had_non-bip}.

\begin{lemma}\label{lem:cng_non-bip_end_decomp}
Let $C,\mfam D$ satisfy conditions \cond{C1},\cond{C2} and \cond{D1}--\cond{D3}. Let a matrix $H$ and a family $\mfam U = (U^{\cngc c})_{c \in \Int_{\denom}}$ be defined as in conditions \cond{C1},\cond{C2} and \cond{D1}--\cond{D3}, respectively. Then
$$
\evalk(C,\mfam D) \Tequiv \evalk(H,\mfam U).
$$
\end{lemma}
Before we give the proof of the Lemma, let us first develop a small technical result which will play an important role in that proof.

\begin{lemma}\label{lem:cng_k_tensor_decomp}
Let $A \in \C^{m \times m}$ and $A' \in \C^{m' \times m'}$ be $\denom$-algebraic. Let $\mfam D = (D^{\cngc c})_{c \in \Int_{\denom}}$ be a family of $m \times m$ diagonal matrices and $\mfam D' = (D'^{\cngc c})_{c \in \Int_{\denom}}$ a family of $m' \times m'$ diagonal matrices. 
Let $A'' = A \otimes A'$ and $\mfam D'' = (D^{\cngc c} \otimes D'^{\cngc c})_{c \in \Int_{\denom}}$.
There are projections $\tau$ and $\tau'$ such that for every digraph $G$ and pinning $\vpin$ we have
$$
Z_{A'',\mfam D''}(\vpin, G) =
Z_{A,\mfam D}(\tau \circ \vpin, G) \cdot Z_{A',\mfam D'}(\tau' \circ \vpin, G).$$
\end{lemma}
\begin{proof}\marc{proof OK}
For convenience, we write the indices of $A''$ as pairs according to the tensor product structure. That is 
$$
A''_{(i,k),(j,l)} = A_{ij} \cdot A'_{kl} \text{ for all $i,j \in [m]$ and $k,l \in [m']$}.
$$
This holds analogously for the elements of $\mfam D''$. We define $\tau: [mm'] \rightarrow [m]$ and $\tau': [mm'] \rightarrow [m']$ by $\tau(i,k) = i$ and $\tau'(i,k) = k$ for all $(i,k) \in [m]\times [m']$.
 
Let $G = (V,E)$ be a digraph and $\vpin$ a pinning w.r.t. $A''$. Define $V' = V\setminus \df(\vpin)$.
\begin{eqnarray*}
Z_{A'',\mfam D''}(\vpin, G) &=&
\sum_{\vpin \subseteq \vcfg: V \rightarrow [m] \times [m']} \prod_{uv \in E} A''_{\vcfg(u),\vcfg(v)} \prod_{v \in V'} (D^{\cngc {\grade(v)}} \otimes D'^{\cngc {\grade(v)}})_{\vcfg(v),\vcfg(v)} \\
&=&
\sum_{\vpin \subseteq \vcfg: V \rightarrow [m] \times [m']} \prod_{uv \in E} A_{\tau \circ \vcfg(u), \tau \circ \vcfg(v)} A'_{\tau' \circ \vcfg(u), \tau' \circ \vcfg(v)}\\
& &
\phantom{\sum_{\vpin \subseteq \vcfg: V \rightarrow [m] \times [m']}}
\cdot \prod_{v \in V'} D^{\cngc{\grade (v)}}_{\tau \circ \vcfg(v), \tau \circ\vcfg(v)} \cdot  D'^{\cngc{\grade (v)}}_{\tau' \circ \vcfg(v), \tau' \circ\vcfg(v)}
\end{eqnarray*}
For each configuration $\vcfg \supseteq \vpin$ we can form two independent maps $\vcfg^*$ and $\vcfg'$ such that $\tau \circ \vpin \subseteq \vcfg^* = \tau\circ \vcfg$ and $\tau' \circ \vpin \subseteq \vcfg' = \tau'\circ \vcfg$.
Therefore we obtain
\begin{eqnarray*}
Z_{A'',\mfam D''}(\vpin, G) &=&
\left(\sum_{\tau \circ \vpin \subseteq \vcfg^*: V \rightarrow [m]} \prod_{uv \in E} A_{\vcfg^*(u), \vcfg^*(v)} \prod_{v \in V'} D^{\cngc{\grade (v)}}_{\vcfg^*(v), \vcfg^*(v)}\right)\\
& &
\cdot 
\left(\sum_{\tau' \circ \vpin \subseteq \vcfg': V \rightarrow [m']} \prod_{uv \in E} A'_{\vcfg'(u), \vcfg'(v)}\prod_{v \in V'} D'^{\cngc{\grade (v)}}_{\vcfg'(v), \vcfg'(v)}\right) 
\end{eqnarray*}
In other words, $Z_{A'',\mfam D''}(\vpin, G) = Z_{A,\mfam D}(\tau \circ \vpin, G) \cdot Z_{A',\mfam D'}(\tau' \circ \vpin, G)$.
\end{proof}

\begin{proof}[of Lemma \ref{lem:cng_non-bip_end_decomp}]\marc{proof OK}
Define $M = \varsigma \cdot vw^T$ and let $\mfam \Delta = (\Delta^{\cngc c})_{c \in \Int_{\denom}}$ be a family of $m \times m$ diagonal matrices defined as in conditions \cond{D2} and \cond {D3}.
In particular, $C = M \otimes H$ and for each $c \in \Int_\denom$ we have $D^{\cngc c} = \Delta^{\cngc c} \otimes U^{\cngc c}$. Then the following claim follows from Lemma~\ref{lem:cng_k_tensor_decomp}.
\begin{claim}\label{cl:0304091524}
There are mappings $\tau$ and $\tau'$ such that for every digraph $G = (V,E)$ and every pinning $\vpin$, we have
$$
Z_{C,\mfam D}(\vpin,G) = Z_{M,\mfam \Delta}(\tau\circ\vpin,G) \cdot Z_{H,\mfam U}(\tau'\circ\vpin,G).
$$
\end{claim}
By Lemma~\ref{lem:ptime_rank1_cng} the function $\evalk(M,\mfam \Delta)$ is polynomial time computable and therefore Claim~\ref{cl:0304091524} implies the existence of a reduction witnessing 
$\evalk(C,\mfam D) \Tle \evalk(H,\mfam U)$.

\bigskip 

\noindent 
It remains to prove the reducibility $\evalk(H,\mfam U) \Tle \evalk(C,\mfam D)$,
the proof of which is slightly more complicated. The complications arise from the fact that there may be graphs $G$ for which $Z_{M,\mfam \Delta}(\tau\circ\vpin,G) = 0$ implying $Z_{C,\mfam D}(\vpin,G) = 0$ although $Z_{H,\mfam U}(\tau'\circ\vpin,G)$ might be non-zero.

We will tackle this problem by $(p\cdot \denom + 1)$-thickening. Let $G^{(p\denom + 1)}$ be the graph obtained from $G$ by this operation. Note that 
$(p\cdot \denom + 1)\grade(v) \equiv \grade(v) \mmod{\denom}$ for all vertices $v$ whence it thus follows by inspection that
$$
Z_{H,\mfam U}(\tau'\circ\vpin,G^{(p\denom + 1)}) = Z_{H^{(p\cdot \omega + 1)},\mfam U}(\tau'\circ\vpin,G).
$$
Since every entry of $H$ is an $\denom$-th root of unity, we have $H^{(p\cdot \omega + 1)} = H$. Therefore,
$$
Z_{C,\mfam D}(\vpin,G^{(p\denom + 1)}) = Z_{M,\mfam \Delta}(\tau\circ\vpin,G^{(p\denom + 1)}) \cdot Z_{H,\mfam U}(\tau'\circ\vpin,G).
$$
We will show the following claim.
\begin{claim}\label{cl:0704091735}
There is a computable $p \in \Nat$ which depends only on $M$ and $\mfam \Delta$ such that $Z_{H,\mfam U}(\tau'\circ\vpin,G) \neq 0$ implies $Z_{M,\mfam \Delta}(\tau\circ\vpin,G^{(p\denom + 1)}) \neq 0$.
\end{claim}
Before we give the proof let us see how this finishes the proof of the lemma. 
Given this $p$ we can compute $Z_{M,\mfam \Delta}(\tau\circ\vpin,G^{(p\denom + 1)})$ in polynomial time. If it is zero then we know that $Z_{H,\mfam U}(\tau'\circ\vpin,G)$ is and we are done.
Otherwise, we can compute
$$
\left(Z_{M,\mfam \Delta}(\tau\circ\vpin,G^{(p\denom + 1)})\right)^{-1} \cdot Z_{C,\mfam D}(\vpin,G^{(p\denom + 1)}) = Z_{H,\mfam U}(\tau'\circ\vpin,G).
$$
This yields $\evalk(H,\mfam U) \Tle \evalk(C,\mfam D)$.

\subparagraph*{Proof of Claim~\ref{cl:0704091735}.}
Recall that $Z_{M,\mfam \Delta}(\tau\circ\vpin,G) = \varsigma^{|E|}\cdot Z_{vw^T,\mfam \Delta}(\tau\circ\vpin,G)$. We will therefore focus on $Z_{vw^T,\mfam \Delta}(\tau\circ\vpin,G)$. Define $V' = V \setminus \df(\tau \circ \vpin)$ and recall that by \cond{C1} we have $w = \lambda \cdot v$ for some positive $\lambda$. Therefore
\begin{eqnarray*}
Z_{vw^T,\mfam \Delta}(\tau\circ\vpin,G) &=& \sum_{\tau\circ\vpin \subseteq \vcfg: V \rightarrow [m]} \prod_{ab \in E} v_{\vcfg(a)}w_{\vcfg(b)} \prod_{a \in V'} \Delta^{\cngc{\grade(a)}}_{\vcfg(a)\vcfg(a)} \\
&=& \lambda^{|E|} \sum_{\tau\circ\vpin \subseteq \vcfg: V \rightarrow [m]} \prod_{a \in \df(\vpin)}v^{\vdeg(a)}_{\vcfg(a)} \prod_{a \in V'}v^{\vdeg(a)}_{\vcfg(a)} \Delta^{\cngc{\grade(a)}}_{\vcfg(a)\vcfg(a)} \\
&=& \lambda^{|E|} \prod_{a \in \df(\vpin)}v^{\vdeg(a)}_{\tau\circ\vpin(a)}  \sum_{\vcfg: V' \rightarrow [m]} \prod_{a \in V'}v^{\vdeg(a)}_{\vcfg(a)} \Delta^{\cngc{\grade(a)}}_{\vcfg(a)\vcfg(a)}\\
\end{eqnarray*}
It thus follows that
\begin{eqnarray*}
Z_{vw^T,\mfam \Delta}(\tau\circ\vpin,G^{(p\denom + 1)}) 
&=& \lambda^{|E|} \prod_{a \in \df(\vpin)}v^{(p\denom + 1)\cdot\vdeg_G(a)}_{\tau\circ\vpin(a)}  \sum_{\vcfg: V' \rightarrow [m]} \prod_{a \in V'}v^{(p\denom + 1)\cdot\vdeg_G(a)}_{\vcfg(a)} \Delta^{\cngc{(p\denom + 1)\cdot\grade_G(a)}}_{\vcfg(a)\vcfg(a)}\\
&=& \lambda^{|E|} \prod_{a \in \df(\vpin)}v^{(p\denom + 1)\cdot\vdeg_G(a)}_{\tau\circ\vpin(a)}  \prod_{a \in V'}\sum_{i=1}^m v^{(p\denom + 1)\cdot\vdeg_G(a)}_{i} \Delta^{\cngc{\grade_G(a)}}_{ii}
\end{eqnarray*}
By this calculation,  $Z_{vw^T,\mfam \Delta}(\tau\circ\vpin,G^{(p\denom + 1)})$ is non-zero for some $p \in \Nat$, if for each $a \in V'$,
\begin{equation}\label{eq:2903091629}
0 \neq \sum_{i=1}^m v^{(p\denom + 1)\cdot\vdeg_G(a)}_{i} \Delta^{\cngc{\grade_G(a)}}_{ii}
\end{equation}
To finish the proof it suffices to prove, for each $a \in V'$, the existence of a $p_a$ such that with $p\ge p_a$ the inequality in \eqref{eq:2903091629} is satisfied.
If further each $p_a$ depends only on $M$ and $\mfam \Delta$, then with $p$ being the maximum of all these $p_a$ the proof of the claim follows.

The assumption $Z_{H,\mfam U}(\tau'\circ\vpin,G) \neq 0$ implies that for each $a \in V'$ the matrix $U^{\cngc{\grade_G(a)}} \neq 0$ and thus by condition \cond{D3} the matrix $D^{\cngc{\grade_G(a)}}$ contains a non-zero diagonal entry.
By the General Principles of Section~\ref{sec:gen_princip_cng} we may assume that $G$ is connected. Further the reduction is trivial if $G$ contains no edge, therefore we may assume further that $\vdeg_G(a) \ge 1$ for all $a \in V$.

Fix an $a \in V'$ and let $i_a$ be the maximum index $i \in[m]$ such that $\Delta^{\cngc{\grade_G(a)}}_{ii}\neq 0$. If there is only one index $i \in [m]$ such that $\Delta^{\cngc{\grade_G(a)}}_{ii} \neq 0$, then the inequality in \eqref{eq:2903091629} holds for all $p \ge p_a := 0$.
Otherwise, let $z \in [m]$, with $z \neq i_a$ be maximal such that $\Delta^{\cngc{\grade_G(a)}}_{zz} \neq 0$.
The inequality \eqref{eq:2903091629} is satisfied if
\begin{equation}
\left\vert v^{(p\denom + 1)\cdot\vdeg_G(a)}_{i_a} \Delta^{\cngc{\grade_G(a)}}_{i_ai_a}
\right\vert
>
\left\vert\sum_{i\neq i_a} v^{(p\denom + 1)\cdot\vdeg_G(a)}_{i} \Delta^{\cngc{\grade_G(a)}}_{ii}\right\vert
\end{equation}
By the definition of $z$ this is the case if
\begin{equation}\label{eq:2806091058}
\left(\dfrac{v_{i_a}}
{v_{z}}\right)^{(p\denom + 1)\cdot\vdeg_G(a)}
>
\left\vert\sum_{i\neq i_a} \Delta^{\cngc{\grade_G(a)}}_{ii}\right\vert
\left\vert  \Delta^{\cngc{\grade_G(a)}}_{i_ai_a}
\right\vert^{-1}
\end{equation}
As $\vdeg_G(a) \ge 1$, $\denom \ge 1$ and $0< v_z < v_{i_a}$, there is a $p_a$ such that 
for all $p \ge p_a$ the inequality \eqref{eq:2806091058} (and hence \eqref{eq:2903091629}) is satisfied. To finish the proof, observe that $p_a$ depends only on $\mfam \Delta$ and $v_1,\ldots, v_m$.
\end{proof}

\subsection{Satisfying Shape Conditions \cond{C1} and \cond{D1}}

We will now show, in a sequence of several small steps, how to transform the 
initial problem $\evalk(A,D)$ into a problem satisfying the shape conditions. 
In each step, if transformation is not possible we will argue that this is due 
to the problem at hand being $\#\PP$-hard.

The first step will provide us with the transformation of $\evalk(A,D)$ 
into a problem satisfying \cond{C1} and \cond{D1}. 

\begin{lemma}\label{lem:cond_C1_D1}
Let $A \in \Calg^{n \times n}$ be a Hermitian $\denom$-algebraic matrix which is connected and non-bipartite and $D \in \Ralg^{n \times n}$ a diagonal matrix of positive vertex weights.
Then either $\evalk(A,D)$ is $\#\PP$-hard or the following holds.

There is a matrix $C$ satisfying condition \cond{C1} and a family $\mfam D$ of diagonal matrices which satisfies condition \cond{D1} such that
$$
\evalk(A,D) \Tequiv \evalk(C, \mfam D).
$$
Further, there are two technical results following from the proof.
\begin{itemize}
 \item[(1)] $H(D^{\cngc 0;\mu})\cj H^T = \tr{D^{\cngc 0;\mu}} I_r$ for all $\mu \in [m]$.
 \item[(2)] There are constants $d_1, \ldots,d_m$ such that $D^{\cngc 0;\mu} = d_\mu D^{\cngc 0;1}$ for all $\mu \in [m]$.
\end{itemize}
\end{lemma}
\medskip

\noindent
We split the proof into two parts. The first of which is given in the following lemma.

\begin{lemma}\label{lem:130109-1}
Let $A \in \Calg^{n \times n}$ be a Hermitian $\denom$-algebraic matrix which is connected and non-bipartite and $D \in \Ralg^{n \times n}$ a diagonal matrix of positive vertex weights.
Either $\evalk(A,D)$ is $\#\PP$-hard or the following holds.

There is a Hermitian $\denom$-algebraic and twin-free matrix $C$ and a family 
$\mfam D = (D^{\cngc c })_{c \in \Int_{\denom}}$ of diagonal matrices which 
satisfies condition \cond{D1} such that
$$
\evalk(A,D) \Tequiv \evalk(C, \mfam D).
$$
The matrix $C$ further has the following properties. There is a $\varsigma \in \{-1,1\}$ and positive vectors $v, w \in \Ralg^m$ such that $v_1 < \ldots < v_m$ and $w = \lambda v$ for some positive $\lambda$ such that
$$
C =\left(\begin{array}{c c c}
               v_1w_1 P^{1 1} & \ldots &  v_1w_{m} P^{1 m}\\
               \vdots  & \ddots & \vdots \\
                v_mw_1 P^{m 1} & \ldots & v_mw_m P^{m m} 
            \end{array}\right).
$$
For appropriate matrices $P^{\mu\nu}$ of $\denom$-th roots of unity such that $P^{\mu 1}\col 1$ and $P^{1\mu}\row 1$ are constantly $\varsigma$ for all $\mu \in [m]$.
\end{lemma}
\begin{proof}\marc{proof OK}
For convenience we will assume in the following that
\begin{equation}
|A_{ii}| \le |A_{jj}| \text{ for all } i \le j \in [n].
\end{equation}
This is possible w.l.o.g. by the Permutability Principle~\ref{lem:principle_permute_cng}, since every matrix $A$ satisfies this condition up to symmetric permutations of rows and columns.
Let $\mfam D' = (D'^{\cngc c})_{c \in \Int_{\denom}}$ with $D'^{\cngc c} = D$ for all $c \in \Int_{\denom}$. By Lemma~\ref{lem:intro_cong} we have 
\begin{equation}\label{eq:1305091704}
Z_{A,D}(\vpin, G) = Z_{A,\mfam D'}(\vpin, G) \quad \text{ for all digraphs $G$ and all pinnings $\vpin$}
\end{equation}
and therefore, $\evalk(A,D) \Tequiv \evalk(A, \mfam D')$. 
By Lemma~\ref{lem:cng_block2_hard} the problem $\evalk(A, \mfam D')$ 
is $\#\PP$-hard, if $\abs A$ has rank at least $2$.
Assume therefore, that $\abs A$ has rank $1$ that is, 
$\abs A = xy^T$ for non-negative real vectors $x$ and $y$.
The vectors $x$ and $y$ are positive. To see this, note that $\abs A$ is a block,
because $A$ is. On the other hand a value $x_i = 0$ would contradict this fact,
as then $\abs A\row i = x_iy^T = 0$ implying the decomposability of $\abs A$. 
Therefore for appropriate $\denom$-th roots of unity $\zeta_{ij}$ we have
\begin{equation}
A_{ij} = x_iy_j\cdot \zeta_{ij} \text{ for all } i,j \in [n].
\end{equation}
By the Hermitianicity of $A$ we have $\zeta_{11} \in \{-1,1\}$. Define $\varsigma := \zeta_{11}$ and a diagonal $n \times n$ matrix $\Pi$ by $\Pi_{ii} = \varsigma\cdot \cj{\zeta_{i1}}$ for all $i \in [n]$. Let $A' = \Pi A \cj \Pi$. The first column of this matrix satisfies 
$A'_{i1} = \Pi_{ii} A_{i1} \cj\Pi_{11} = \varsigma\cj{\zeta_{i1}} \zeta_{i1} x_iy_1 \cj{\zeta_{11}}\zeta_{11} = \varsigma x_iy_1$ for all $i \in [n]$.
As this follows similarly for the first row of $A'$ we have
\begin{equation}\label{eq:150109-1}
A'_{i1} = \varsigma|A'_{i1}| = \varsigma|A'_{1i}| = A'_{1i} \text{ for all } i \in [n]. 
\end{equation}
With
$\mfam D'' = (D''^{\cngc c})_{c \in \Int_{\denom}}$ defined by $D''^{\cngc c} = \Pi^cD'^{\cngc c}$ the Root Of Unity Transfer Lemma~\ref{lem:unity_transfer} yields
\begin{equation}\label{eq:1305091711}
Z_{A, \mfam D'}(\vpin, G) = f_\Pi(\vpin) \cdot Z_{A', \mfam D''}(\vpin, G) \quad \text{ for all digraphs $G$ and all pinnings $\vpin$}.
\end{equation}
The function $f_\Pi$ is polynomial time computable and non-zero for every $\vpin$. By definition, all diagonal entries of $\Pi$ are $\denom$-th roots of unity. Hence $\Pi^{\denom-c} = \cj \Pi^{c}$ for all $c \in \Int_{\denom}$, that is,  $\mfam D''$ satisfies condition \cond{D1} .

We will now perform twin reduction. Let $I_1, \ldots, I_{n'}$ be the equivalence classes of the twin relation on $A'$ and let $\tau$ be the twin resolution mapping. 
By Lemma~\ref{lem:cng_twin_red} we have
\begin{equation}\label{eq:1305091717}
Z_{A', \mfam D''}(\vpin,G) = Z_{\twres{ A'},\mfam D^{\twres {A'}}}(\tau\circ \vpin, G) \text{ for all digraphs } G \text { and all pinnings } \vpin.
\end{equation}
Define $C = \twres{A'}$ and $\mfam D = \mfam D^{\twres {A'}} = (D^{\twres {A'},\cngc c})_{c \in \Int_{\denom}}$. The family $\mfam D$ therefore satisfies $D^{\cngc c}_{ii} = \sum_{\nu \in I_i} D''^{\cngc c}_{\nu,\nu}$ for all $i \in [n']$ and all $c \in \Int_{\denom}$.
It follows from the fact that $\mfam D''$ satisfies condition \cond{D1} that this is also true for $\mfam D$.  Since $C$ is the twin resolvent of $A'$ we have
$$
C_{ij} = A'_{\mu\nu} \text{ for some } \mu\in I_i,\,\nu \in I_j.
$$
We may assume w.l.o.g. that $1 \in I_1$ and that for all $i,j \in [n]$ and all $\mu \in I_i,\nu\in I_j$ we have $|A'_{\mu\mu}| \le |A'_{\nu\nu}|$ iff $i \le j$. This is possible if we assume an appropriate ordering of the $I_1, \ldots, I_{n'}$.
We thus obtain, from equation \eqref{eq:150109-1},
\begin{equation}\label{eq:150109-a}
C_{i1} = \varsigma|C_{i1}| = \varsigma|C_{1i}| = C_{1i} \text{ for all } i \in [n']
\end{equation}
and, as this was true for $A$ and $A'$,
\begin{equation}\label{eq:130109-1}
|C_{ii}| \le |C_{jj}| \text{ for all } i \le j \in [n'].
\end{equation}
With $\rank \abs A = 1$ we have $\rank \abs C = 1$ and still there are no zero entries in this matrix. There are thus positive vectors $u$ and $z$ such that $\abs C = uz^T$. By the symmetry of $\abs C$ we have $u_iz_j = u_jz_i$ for all $i,j \in [n']$ which implies that $u$ and $z$ are linearly dependent with $z_i = \dfrac{z_1}{u_1}\cdot u_i$. Let $\lambda = z_1 \cdot u_1^{-1}$.
This rephrases equation \eqref{eq:130109-1} to $u^{2}_i\lambda \le u^{2}_j\lambda \text{ for all  } i \le j \in [n']$ and therefore the entries of $u$ are ordered increasingly. Let $v$ be the vector of increasingly ordered distinct entries of $u$ and $w$ the analogon for $z$. We have $w = \lambda v$. Let $m$ be the length of $w$. The definition of $C$ can thus be rephrased as
$$
C =\left(\begin{array}{c c c}
               v_1w_1 P^{1 1} & \ldots & v_1w_{m} P^{1 m}\\
               \vdots  & \ddots & \vdots \\
               v_mw_1 P^{m 1} & \ldots & v_mw_m P^{m m} 
            \end{array}\right)
$$
for appropriate matrices $P^{\mu\nu}$ of $\denom$-th roots of unity.
From \eqref{eq:150109-a} it follows that $P^{\mu1}\col 1$ and $P^{1\mu}\row 1$ are constant for all $\mu \in [m]$ --- each entry equals $\varsigma$.
Combining equations \eqref{eq:1305091704}, \eqref{eq:1305091711} and \eqref{eq:1305091717} yields,
$$
Z_{A, D}(\vpin,G) = f_\Pi(\vpin)\cdot Z_{C,\mfam D}(\tau\circ \vpin, G) \text{ for all digraphs } G \text { and all pinnings } \vpin.
$$
This finishes the proof.
\end{proof}

\begin{proof}[of Lemma~\ref{lem:cond_C1_D1}]\marc{proof OK}
Assume that $\evalk(A, D)$ is not $\#\PP$-hard. 
By Lemma~\ref{lem:130109-1} we have $\evalk(A, D) \Tequiv \evalk(C,\mfam D)$ 
for a family $\mfam D$ of diagonal matrices satisfying \cond{D1} and a matrix 
$C$ which has the following properties. There are vectors $v,w \in \Ralg^m$ 
satisfying $0 < v_1 < \ldots < v_m$ and $w = \lambda v$ for some $\lambda > 0$ such that
$$
C =\left(\begin{array}{c c c}
               v_1w_1 P^{1 1} & \ldots & v_1w_{m} P^{1 m}\\
               \vdots  & \ddots & \vdots \\
               v_mw_1 P^{m 1} & \ldots & v_mw_m P^{m m} 
            \end{array}\right)
$$
for appropriate matrices $P^{\mu\nu}$ of $\denom$-th roots of unity.
By the Hermitianicity of $C$, there are numbers $m_1,\ldots, m_m$ such that $P^{\mu\nu}$ is an $m_\mu \times m_{\nu}$ matrix and $P^{\mu\nu} = \cj{P^{\nu\mu}} $ for all $\mu,\nu \in [m]$.
We index $C$ so as to explicate this structure. That is for $\mu,\nu \in [m]$, $i \in [m_{\mu}]$ and $j \in [m_\nu]$ we let $C_{(\mu,i)(\nu,j)} = v_{\mu}w_{\nu} P^{\mu\nu}_{ij}$. Index the matrices $D^{\cngc c}$ analogously and write, for convenience, $D^{\cngc c; \mu}_{ii} := D^{\cngc c}_{(\mu,i)(\mu,i)}$ for all $i \in [m_\mu]$.
\begin{claim}\label{cl:150109-1}
For all $\mu,\nu,\kappa \in [m]$ and $i \in [m_{\mu}],\, j \in [m_{\nu}]$ we have
$
\left| \scalp{ D^{\cngc 0; \kappa}, P^{\nu\kappa}\row j \bullet P^{\mu\kappa}\row i} \right| \le \tr{D^{\cngc 0;\kappa}}$.
Furthermore, equality is assumed only if $P^{\nu\kappa}\row j$ and $P^{\nu\kappa}\row i$ are linearly dependent.
\end{claim}
\begin{clproof}
This is straightforward, we present the calculations for completeness. Application of the triangle inequality yields
\[\begin{aligned}
\left| \scalp{ D^{\cngc 0; \kappa}, P^{\nu\kappa}\row j \bullet P^{\mu\kappa}\row i} \right| 
&=& \left| \sum_{k = 1}^{m_\kappa} D^{\cngc 0; \kappa}_{k,k}\cdot  \cj P^{\nu\kappa}_{j,k} \cdot P^{\mu\kappa}_{i,k} \right|
&\le&  \sum_{k = 1}^{m_\kappa} \left|D^{\cngc 0; \kappa}_{k,k}\cdot  \cj P^{\nu\kappa}_{j,k} \cdot P^{\mu\kappa}_{i,k} \right|
&=& \tr{D^{\cngc 0;\kappa}}.
\end{aligned}\]
Here, equality can be assumed only if all of the terms $\cj P^{\nu\kappa}_{j,k} \cdot P^{\mu\kappa}_{i,k}$ are equal.
\end{clproof}

\begin{claim}\label{cl:150109-5}
$\evalk(C,\mfam D)$ is $\#\PP$-hard unless the following is true.
For all $i \in [m_{\mu}]$ and all $j \in [m_{\nu}]$ we have either 
\begin{equation}
\scalp{D^{\cngc 0;\kappa}, P^{\nu\kappa}\row j \bullet P^{\mu\kappa}\row i} = 0 \text{ for all } \kappa \in [m] \text{ or }
\scalp{D^{\cngc 0;\kappa}, P^{\nu\kappa}\row j \bullet P^{\mu\kappa}\row i} = \tr{ D^{\cngc{0};\kappa}} \text{ for all } \kappa \in [m].
\end{equation}
In particular the latter case occurs only if $P^{\mu\kappa}\row i =  P^{\nu\kappa}\row j$.
\end{claim}
\begin{clproof}
Define, for every $q \in \Nat$ the value $p = q\cdot \denom + 1$ and a matrix $C' = C'(p) = C^{(p)}D^{\cngc 0}C^{(p)}$. For a given digraph $G = (V,E)$, let $G_{p}$ be the result of $2$-stretching followed by $p$-thickening. Note that $p\cdot \grade(v) \equiv \grade(v) \mmod{\denom}$ for all $v\in V$. Thus
we obtain for every pinning $\vpin$
\begin{equation}\label{eq:1305091750}
Z_{C'(p),\mfam D}(\vpin, G) = Z_{C,\mfam D}(\vpin, G_p).
\end{equation}
The matrix $C'$ satisfies
\begin{eqnarray*}
C'_{(\mu,i)(\nu,j)} &=& \sum_{\kappa=1}^m\sum_{k = 1}^{m_{\kappa}} C^{(p)}_{(\mu,i)(\kappa,k)}\cj{C^{(p)}_{(\nu,j)(\kappa,k)}}D^{\cngc 0}_{(\kappa,k)(\kappa,k)} \\
&=& v^p_{\mu}v^p_{\nu} \sum_{\kappa=1}^m\sum_{k = 1}^{m_{\kappa}} w^{2p}_{\kappa}(P^{\mu\kappa}_{ik}\cj{P^{\nu\kappa}_{jk}})^p D^{\cngc 0;\kappa}_{k,k} \\
&=& v^p_{\mu}v^p_{\nu} \sum_{\kappa=1}^m w^{2p}_{\kappa}\scalp{D^{\cngc 0;\kappa}, (P^{\nu\kappa}\row j)^{(p)} \bullet (P^{\mu\kappa}\row i)^{(p)}}.
\end{eqnarray*}
The definition of $p$ further implies $(P^{\mu\nu}_{ij})^p = P^{\mu\nu}_{ij}$ for all $i,j$ and all $\mu,\nu$. Hence
\begin{equation}\label{eq:150109-2}
C'_{(\mu,i)(\nu,j)} = (v_{\mu}v_{\nu})^{q\cdot \denom + 1} \sum_{\kappa=1}^m w^{2(q\cdot \denom + 1)}_{\kappa}\scalp{D^{\cngc 0;\kappa}, P^{\nu\kappa}\row j \bullet P^{\mu\kappa}\row i}
\end{equation}
In particular all diagonal entries of $C'$ are positive, as for all $\mu \in [m],\, i \in [m_{\mu}]$ we have
$$
C'_{(\mu,i)(\mu,i)} 
= v^{2(q\cdot \denom + 1)}_{\mu}\sum_{\kappa=1}^m w^{2(q\cdot \denom + 1)}_{\kappa}\tr{D^{\cngc 0;\kappa}}.
$$
If there is a $q$ such that the matrix $\abs{C'}$ contains a block of rank at 
least $2$, then by Lemma~\ref{lem:cng_block2_hard} the problem 
$\evalk(C',\mfam D)$ is $\#\PP$-hard. Then further, by equation 
\eqref{eq:1305091750}, $\evalk(C,\mfam D)$ will be $\#\PP$-hard.
Assume therefore that, for all $q$, all blocks of $\abs{C'}$ have rank $1$. 
We shall show that under this assumption $C$ satisfies \cond{C1}.

Consider some $2\times 2$ submatrix of $\abs{C'}$ of the form
$$
\left(\begin{array}{c c}
 |C'_{(\mu,i)(\mu,i)}| & |C'_{(\mu,i)(\nu,j)}| \\
 |C'_{(\nu,j)(\mu,i)}| & |C'_{(\nu,j)(\nu,j)}| \\
\end{array}\right)
$$
By our assumption that for all $q$ this submatrix is not a witness for the existence of a block of rank at least $2$ in $\abs{C'}$ we see that either its determinant is zero or $C'_{(\mu,i)(\nu,j)} = 0$.
More precisely we arrive at the assumption that, for all $\mu,\nu \in [m],\, i\in [m_{\mu}],j \in [m_{\nu}]$ and
\begin{equation}
\text{For all } q \in \Nat \text{ either } C'_{(\mu,i)(\nu,j)} = 0 \text{ or }  |C'_{(\nu,j)(\nu,j)}||C'_{(\mu,i)(\mu,i)}| = |C'_{(\mu,i)(\nu,j)}|^2.
\end{equation}
If $C'_{(\mu,i)(\nu,j)} = 0$ for infinitely many $q$ then by the definition of $C'$ in equation \eqref{eq:150109-2}, we have
$$
0 = \sum_{\kappa=1}^m (w^{2\denom}_{\kappa})^q w^{2}_{\kappa}\scalp{D^{\cngc 0;\kappa}, P^{\nu\kappa}\row j \bullet P^{\mu\kappa}\row i} \quad \text{ for all $q$}
$$
which by Lemma~\ref{lem:coeff_zero_inf} implies $0 = \scalp{D^{\cngc 0;\kappa}, P^{\nu\kappa}\row j \bullet P^{\mu\kappa}\row i}$ for all $\kappa \in [m]$ --- as claimed.
Assume therefore that $C'_{(\mu,i)(\nu,j)} = 0$ holds only for a finite number of $q \in \Nat$. Then clearly,
$|C'_{(\nu,j)(\nu,j)}||C'_{(\mu,i)(\mu,i)}| = |C'_{(\mu,i)(\nu,j)}|^2$ must be true for infinitely many such $q$.
By inspection of equation \eqref{eq:150109-2}, this implies
\begin{equation*}
\left|\sum_{\kappa=1}^m w^{2(q\cdot \denom + 1)}_{\kappa}\tr{D^{\cngc 0;\kappa}}\right|^2=
\left|\sum_{\kappa=1}^m w^{2(q\cdot \denom + 1)}_{\kappa}\scalp{D^{\cngc 0;\kappa}, P^{\nu\kappa}\row j \bullet P^{\mu\kappa}\row i}\right|^2
\end{equation*}
and thus
\begin{equation}\label{eq:150109-4}
 \sum_{\kappa=1}^m w^{2p}_{\kappa}\tr{D^{\cngc 0;\kappa}} =
\left|\sum_{\kappa=1}^m w^{2p}_{\kappa}\scalp{D^{\cngc 0;\kappa}, P^{\nu\kappa}\row j \bullet P^{\mu\kappa}\row i} \right|
\end{equation}
However, application of the triangle inequality yields
\[\begin{aligned}
\left|\sum_{\kappa=1}^m w^{2p}_{\kappa}\scalp{D^{\cngc 0;\kappa}, P^{\nu\kappa}\row j \bullet P^{\mu\kappa}\row i} \right| &\le & \sum_{\kappa=1}^m w^{2p}_{\kappa}\left|\scalp{D^{\cngc 0;\kappa}, P^{\nu\kappa}\row j \bullet P^{\mu\kappa}\row i} \right| &\le & \sum_{\kappa=1}^m w^{2p}_{\kappa}\tr{D^{\cngc 0;\kappa}}.
\end{aligned}\]
By Claim~\ref{cl:150109-1} we have $\left|\scalp{D^{\cngc 0;\kappa}, P^{\nu\kappa}\row j \bullet P^{\mu\kappa}\row i} \right| \le \tr{D^{\cngc 0;\kappa}}$ for each of the $\kappa \in [m]$.
Equation \eqref{eq:150109-4} therefore implies the existence of some $\alpha \in \U$ such that
$$
 0 = \sum_{\kappa=1}^m w^{2p}_{\kappa}\left( \tr{D^{\cngc 0;\kappa}} - \alpha\scalp{D^{\cngc 0;\kappa}, P^{\nu\kappa}\row j \bullet P^{\mu\kappa}\row i}\right). 
$$
As we may choose $p$ arbitrarily large, depending only on $D^{\cngc 0}$ and $C$, Lemma~\ref{lem:coeff_zero_inf} implies that
$$
\alpha \scalp{D^{\cngc 0;\kappa}, P^{\nu\kappa}\row j \bullet P^{\mu\kappa}\row i} = \tr{D^{\cngc 0;\kappa}} \text{ for all } \kappa \in [m].
$$
That is, $P^{\nu\kappa}\row j = \alpha P^{\mu\kappa}\row i$ for all $\kappa$, in particular
$P^{\nu 1}_{j,1} = \alpha P^{\mu 1}_{i,1}$ which by our condition from Lemma~\ref{lem:130109-1} that $P^{\nu 1}_{j,1} = P^{\mu 1}_{i,1}$ implies $\alpha = 1$.
\end{clproof}
As $C$ is twin-free, we see that, for every $\mu \in [m]$ and all $i \neq j \in [m_\mu]$ there is some $\kappa \in [m]$ such that $P^{\mu\kappa}\row j \bullet P^{\mu\kappa}\row i$ is not constant.
Thus, by the above
\begin{equation}\label{eq:31031902}
\text{For all } \mu \in [m] \text{ and all } i\neq j \in [m_{\mu}] \text{ we have } \scalp{D^{\cngc 0;\kappa}, P^{\mu\kappa}\row j \bullet P^{\mu\kappa}\row i} = 0 \text{ for all } \kappa \in [m]. 
\end{equation}
Fix some $\kappa$ and define $\Delta^{\kappa} := (D^{\cngc 0;\kappa})^{(1/2)}$. Let $T^{\mu \kappa}:= P^{\mu\kappa} \Delta^{\kappa}$, then
$$
0 = \scalp{D^{\cngc 0;\kappa}, P^{\mu\kappa}\row j \bullet P^{\mu\kappa}\row i} = \scalp{T^{\mu \kappa}\row i, T^{\mu \kappa}\row j} \text{ for all } i \neq j \in [m_{\mu}].
$$
That is, $T^{\mu \kappa}$ has full row rank for all $\mu,\kappa \in [m]$ and by the definition of $T^{\mu \kappa}$ it has the same rank as $P^{\mu\kappa}$. Thus each $P^{\mu\kappa}$ has full row rank and by $P^{\mu\kappa} = \cj{P^{\kappa\mu}}$ it has full column rank, as well.
Therefore each $P^{\mu\nu}$ is non-singular. Let $r = m_1$ be the rank of $P^{11}$ therefore $P^{1\kappa}$ being a non-singular $r \times m_\kappa$ matrix for every $\kappa \in [m]$ we see that $m_1 =\ldots = m_m = r$. Altogether
\begin{equation}
\text{ For all } \mu, \nu \in [m], P^{\mu\nu} \text{ is a non-singular } r\times r \text{ matrix}.
\end{equation}

\begin{claim}
There are mappings $\tau_1,\ldots \tau_{m} :[r] \rightarrow [r]$ such that
for all $\mu,\nu \in [m]$ we have 
$$
P^{11}_{ij} = P^{\mu\nu}_{\tau_\mu(i),\tau_{\nu}(j)} \text{ for all } i,j \in [r].
$$
\end{claim}
\begin{clproof}
Consider the matrix
$$
T^{\absent 1} = \left( \begin{array}{c}
                        T^{11} \\ \vdots \\ T^{m1} 
                       \end{array}\right)
              = \left( \begin{array}{c}
                        P^{11}\Delta^{1} \\ \vdots \\ P^{m1} \Delta^{1}
                       \end{array}\right).
$$
This matrix has rank $r$ and by the above the first $r$ rows of it form a set of independent vectors. In particular, for each $\mu \in [m]$ and each $j \in [r]$ there is an $i \in [r]$ such that
$$
\scalp{T^{11}\row i, T^{\mu 1}\row j} \neq 0.
$$
By $\scalp{T^{11}\row i, T^{\mu 1}\row j} = \scalp{D^{\cngc 0;1}, P^{\mu 1}\row j \bullet P^{11}\row i}$ and Claim~\ref{cl:150109-5} we see that
$P^{\mu 1}\row j = P^{11}\row i$ and as $C$ is twin-free we further see that for each $\mu \in [m]$ this is unique.
Thus define $\tau_{\mu}(i) : = j$ and by Hermitianicity the claim follows.
\end{clproof}
Define a permutation $\pi :[m]\times[r] \rightarrow [m]\times[r]$ by
$\pi(\mu,i) := (\mu,\tau^{-1}_\mu(i))$.
By the above we have $C_{(\mu,i)(\nu,j)} = v_\mu w_\nu P^{\mu\nu}_{ij} = v_\mu w_\nu P^{11}_{\tau_\mu(i)\tau_\nu(j)}$.
Thus the matrix $C_{\pi\pi}$ satisfies
$$
(C_{\pi\pi})_{(\mu,i),(\nu,j)} = C_{\pi(\mu,i), \pi(\nu,j)} = C_{\pi(\mu,\tau^{-1}_\mu(i)), \pi(\nu,\tau^{-1}_\nu(j))}
=  v_\mu w_\nu P^{11}_{\tau_\mu(\tau^{-1}_\mu(i))\tau_\nu(\tau^{-1}_\nu(j))}
$$
That is 
$$
(C_{\pi\pi})_{(\mu,i),(\nu,j)}  = v_\mu w_\nu P^{11}_{ij}
$$
Define $H := \varsigma^{-1} \cdot P^{11}$. We have $C_{\pi\pi} = \varsigma \cdot vw^T \otimes H$ as desired.

\medskip 

\noindent
It remains to prove the technical statements (1) and (2). To see (1), note that $\scalp{D^{\cngc 0;\kappa}, P^{\nu\kappa}\row j \bullet P^{\mu\kappa}_{\row i}}$ translates to
$\scalp{D^{\cngc 0;\kappa}, H\row j \bullet H_{\row i}} = H(D^{\cngc 0;\kappa})\cj H^T$.
Statement (1) now follows by equation \eqref{eq:31031902}.

We shall show (2), that is, the linear dependence of the $D^{\cngc 0;\mu}$.
Define the $(r-1) \times r$ matrix $H'$ by $H'_{ij} = H_{ij}\cj H_{r j}$ for all $j \in [r]$ and $i \in [r-1]$ and let $\diag(D^{\cngc 0;\mu})$ be the vector of diagonal entries of $D^{\cngc 0;\mu}$ then (1) implies that
$(H D^{\cngc 0;\mu} \cj H^T)_{r j} = 0 \text{ for all } j \neq r$ and all $\mu \in [m]$. 
This is equivalent to $H'\diag(D^{\cngc 0;\mu}) = 0$ for all $\mu \in [m]$. As $H$ is non-singular we have $\rank H' = r-1$ and therefore all solutions to this system of linear equations are linearly dependent. This proves (2).
\end{proof}

\subsection{The Remaining Conditions \cond{C2}, \cond{D2} and \cond{D3}}

We now turn to the part of the proof devoted to the remaining conditions \cond{C2}, \cond{D2} and \cond{D3}.

\begin{lemma}\label{lem:cond_C2_D2}
Let $C, \mfam D$ satisfy conditions \cond{C1} and \cond{D1} then either 
$\evalk(C, \mfam D)$ is $\#\PP$-hard or conditions \cond{C2} and \cond{D2} hold.
\end{lemma}

\begin{lemma}\label{lem:cond_D3}
Let $C, \mfam D$ satisfy conditions \cond{C1},\cond{C2} and \cond{D1} and 
\cond{D2} then either $\evalk(C, \mfam D)$ is $\#\PP$-hard 
or condition \cond{D3} holds.
\end{lemma}
The proofs of both lemmas are technically similar. Therefore we start by discussing the technical details they have in common. The most technical part is given by the Pre-Uniform Diagonal Lemma~\ref{lem:pre-unif_diag} below.
Further, as the $\#\PP$-hardness proofs in both, Lemmas~\ref{lem:cond_C2_D2} and \ref{lem:cond_D3} are based on the same reduction, we discuss this reduction in Lemma~\ref{lem:struct_template}.

\paragraph*{A Technical Tool.} We call a diagonal-matrix $d \in \C^{m\times m}$ \sdef{pre-uniform} if there is a $d \in \C$ such that for all $i \in [m]$ we have $D_{ii} \in \{0,d\}$.

\begin{lemma}[Pre-Uniform Diagonal Lemma]\label{lem:pre-unif_diag}
Let $H \in \U_{\denom}^{r \times r}$ be a non-singular matrix and $D \in \Ralg^{r \times r}$ a non-negative diagonal matrix. Let $K = \{ k \in [r] \mid D_{kk} \neq 0\}$ be the set of non-zero diagonal entries in $D$ then one of the following is true
\begin{itemize}
 \item There is a computable $p_0 \in \Nat$ such that for all $p \ge p_0$ the matrix $\abs{H D^p \cj H^T}$ contains a block of rank at least $2$.
 \item $D$ is pre-uniform and for all $I \subseteq [r]$ the following holds. If $H_{IK}$ is non-singular then it is a complex Hadamard matrix.
\end{itemize}
\end{lemma}
\begin{proof}\marc{proof OK}
The matrix $B : = H D^p \cj H^T$ satisfies, for all $i,j \in [r]$:
\begin{equation}\label{eq:pr-u_diag_lem_def_B}
B_{ij} = \sum_{k=1}^r H_{ik} \cj H_{jk} D^p_{kk} = \sum_{k\in K} H_{ik} \cj H_{jk} D^p_{kk} = (H\col K D^p_{KK} \cj H\col K^T)_{ij}. 
\end{equation}
That is, for every $I \subseteq [r]$ we have $B_{II} = H_{IK} D^p_{KK} \cj H_{IK}^T$. Let $\kappa = |K|$ and fix a set $I \subseteq [r]$ such that $|I| = \kappa$ and $H_{IK}$ has rank $\kappa$.
Note that every principal $2 \times 2$ submatrix of $\abs {B_{II}}$ has non-zero determinant. To see this, fix an arbitrary such submatrix induced by distinct indices $i,j \in [r]$. 
Define a matrix $X = H_{IK}D^{(1/2)}_{KK}$ which is non-singular since $H_{IK}$ is. We have $B = X \cj{X}^T$. The determinant of the submatrix just defined is thus
$$
\left| \begin{array}{c c} 
        |\scalp{X\row i, X \row i}| & |\scalp{X\row i, X \row j}| \\
        |\scalp{X\row j, X \row i}| & |\scalp{X\row i, X \row j}|
       \end{array}\right|
= |\scalp{X\row i, X \row i}||\scalp{X\row j, X \row j}| - |\scalp{X\row i, X \row j}|^2
$$
Using the Cauchy-Schwarz inequality, the non-singularity of $X$ implies that the above determinant is non-zero since
$$
|\scalp{X\row i, X \row j}|^2   < |\scalp{X\row i, X \row i}||\scalp{X\row j, X \row j}|.
$$
Therefore, every principal $2 \times 2$ submatrix of $\abs{ B_{II}}$ has non-zero determinant. The existence of a block of rank at least two in $\abs{B_{II}}$ is thus guaranteed, if we find a non-zero off-diagonal entry. Hence, to finish the proof it remains to show the following:
If there is no computable $p_0$ such that for all $p \ge p_0$ the matrix $B_{II}$ contains a non-zero off-diagonal entry, then $D$ is pre-uniform and $H_{IK}$ complex Hadamard.

For every pair of values $i,j \in I$ define a family $\m Z_{ij}$ of equivalence classes $Z \in \m Z_{ij}$ such that $k,k' \in Z$ if, and only if $H_{ik}\cj H_{jk} =  H_{ik'}\cj H_{jk'}$. Define for each such $Z \in \m Z_{ij}$ the value $\zeta_Z : = H_{ik}\cj H_{jk}$ for some $k \in Z$.
Furthermore, define a family $\m J$ of equivalence classes induced by the equality of diagonal entries of $D_{KK}$. That is, for each $J \in \m J$ we have $j,j' \in J$ iff $D_{jj} = D_{j'j'}$ and denote by $d_J$ the corresponding diagonal value of $D$.
We obtain, for all $i,j \in I$
\begin{eqnarray*}
B_{ij} = \sum_{k \in K} H_{ik}\cj H_{jk} D^p_{kk} = \sum_{J \in \m J} d^p_{J} \sum_{Z \in \m Z_{ij}} \zeta_Z|J \cap Z|
\end{eqnarray*}
By Lemma~\ref{lem:coeff_zero_inf} there is some $p_0$ such that the following holds. For all $p \ge p_0$ if $B_{ij}$ is zero then, for every $J \in \m J$,
$$
0 = \sum_{Z \in \m Z_{ij}} \zeta_Z|J \cap Z| = \sum_{k \in J} H_{ik}\cj H_{jk} 
$$
As this is true for all $i\neq j \in I$ the matrix $H_{IJ}$ is non-singular and complex Hadamard. This implies that $J = K$, that is, $D$ is pre-uniform.
\end{proof}

\paragraph*{The $\#\PP$-hardness construction.}

\begin{lemma}\label{lem:struct_template}
Let $C \in \Calg^{m \times m}$ and $\mfam D$ a family of diagonal matrices which satisfy \cond{C1} and \cond{D1}. 
For all $p,q \in \Nat$, there is a diagonal matrix
\begin{equation}\label{eq:symm_def_Theta}
\Theta = \Theta(p,q) =\sum_{\mu = 1}^m w^2_\mu  v_{\mu}^{2pq} D^{\cngc 0;\mu} \left|\sum_{\kappa = 1}^m  v_{\kappa}^p D^{\cngc{p};\kappa}  \right|^{2q}
\end{equation}
such that the following is true.

If $\abs{H\Theta\cj H^T}$ contains a block of rank at least $2$ then 
$\evalk(C, \mfam D)$ is $\#\PP$-hard.
\end{lemma}
\begin{proof}\marc{proof OK}
We define a \emph{reduction template} $T_{p,q}$ which will be used in the following reductions. This template is a digraph $T_{p,q} = (V_{p,q}, E_{p,q})$ with two terminal vertices $u$ and $v$ which are connected by a length $2$ path with middle vertex $w$. The most important subgraph of this template is a graph $P(p)$, which  consists of two vertices $b$ and $a$ connected by $p$ many paths of length $2$ directed from $b$ to $a$. Then, $q$ many disjoint copies of $P(p)$ are attached to $w$ by identifying their terminal vertex $a$ with $w$ and further $q$ copies of $P(p)$ are attached by identifying their terminal vertex $b$ with $w$. Figure~\ref{fig:red_template} illustrates the construction.
\begin{figure}
\begin{center} 
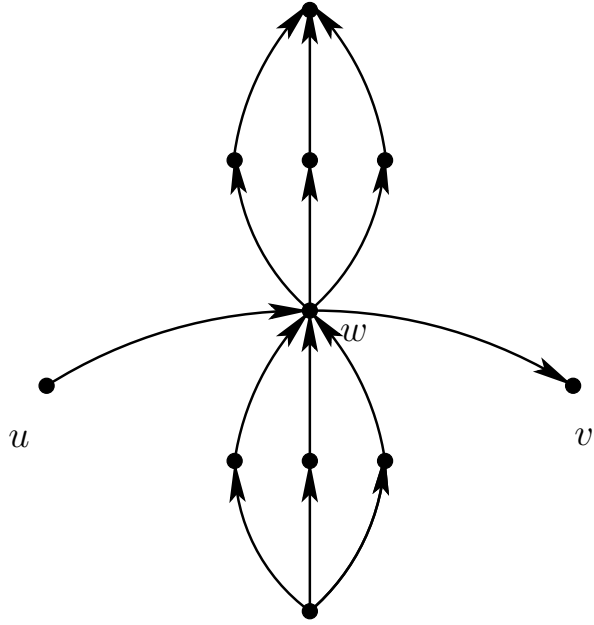
\end{center}
\caption{The reduction template $T_{p,q}$ for $p=3,q=1$.}
\label{fig:red_template}
\end{figure}
Its formal definition is given by
\begin{eqnarray*}
 V_{p,q} &=& \{u,v,w, x_i,x_{ij},y_{ij},y_{i} \mid i \in [q], j \in [p]\} \\
 E_{p,q} &=& \{uw, wv, x_ix_{ij}, x_{i j} w, w y_{i j}, y_{ij} y_i \mid i \in [q], j \in [p]\}
\end{eqnarray*}
For a given digraph $G$, let $G'$ be the graph obtained from $G$ by 
replacing every edge $uv$ by a distinct copy of $T_{p,q}$.
Let $\vpin$ be a pinning of $G$, then 
\begin{equation}\label{eq:1305091834}
Z_{C',\mfam D}(\vpin, G) = Z_{C,\mfam D}(\vpin,G')
\end{equation}
where $C'$ is a matrix whose structure we will analyze in the following. We start the analysis by observing a technical detail. We have, by Lemma~\ref{lem:cond_C1_D1}(1),
\begin{equation*}
CD^{\cngc{0}}C = (vv^T) \otimes \sum_{\nu = 1}^m w^{2}_\nu H D^{\cngc 0;\nu}\cj H^T
= (vv^T) \otimes \sum_{\nu = 1}^m w^{2}_\nu  \tr{D^{\cngc 0;\nu}} I_r
\end{equation*}
Condition \cond{D1} implies that $\tr{D^{\cngc 0;\nu}}$ is positive for all $\nu$ and thus there is some positive $\gamma$ such that
\begin{equation} \label{eq:symm_131208-1}
CD^{\cngc{0}}C = (vv^T) \otimes \gamma I_r.
\end{equation}
Here $I_r$ denotes the $r \times r$ identity matrix.
Denote by $Z_{C,\mfam D}(i,j; T_{p,q})$ the partition function of $T_{p,q}$ with vertex $u$ pinned to $i$ and $v$ pinned to $j$. 
We have $C'_{ij} = Z_{C,\mfam D}(i,j; T_{p,q})$ for all $i,j$. Further the definition of $T_{p,q}$ directly implies that $C' = C\Delta C$ for some diagonal matrix $\Delta= \Delta(p,q)$. In the following we will index $\Delta$ by pairs $(\mu,i) \in [m]\times[r]$ and further define $\Delta^{\mu}_{ii} = \Delta_{(\mu,i)(\mu,i)}$. We obtain
\begin{eqnarray*}
\Delta^{\mu}_{i,i} &=& D^{\cngc 0;\mu}_{i,i} \left(\sum_{\kappa = 1}^m \sum_{k=1}^r D^{\cngc{p};\kappa}_{k,k}\left(CD^{\cngc 0}C\right)^p_{(\mu,i),(\kappa,k)} \right)^q \left(\sum_{\kappa = 1}^m \sum_{k=1}^r D^{\cngc{-p};\kappa}_{k,k}\left(CD^{\cngc 0}C\right)^p_{(\kappa,k),(\mu,i)} \right)^q \\
\end{eqnarray*}
Applying equation \eqref{eq:symm_131208-1} we see that, for all  $(\kappa,k), (\mu,i) \in [m]\times [r]$,
$$
\left(CD^{\cngc 0}C\right)^p_{(\mu,i),(\kappa,k)} = \left(CD^{\cngc 0}C\right)^p_{(\kappa,k),(\mu,i)} = v_{\kappa}^p v_{\mu}^p\cdot \gamma^p \cdot (I_r)_{k,i} 
$$ 
which is zero unless $k=i$. By condition \cond{D1} we have $D^{\cngc{-p}} = \cj{ D^{\cngc{p}}}$ and thus we can rewrite
\begin{eqnarray*}
\Delta^{\mu}_{i,i} &=& D^{\cngc 0;\mu}_{i,i} \left(\sum_{\kappa = 1}^m  D^{\cngc{p};\kappa}_{i,i} v_{\kappa}^p v_{\mu}^p\cdot \gamma^p\right)^q \left(\sum_{\kappa = 1}^m \cj{D^{\cngc{p};\kappa}_{i,i}}v_{\kappa}^p v_{\mu}^p\cdot \gamma^p \right)^q \\
                  &=& v_{\mu}^{2pq}\gamma^{2pq} D^{\cngc 0;\mu}_{i,i} \left(\sum_{\kappa = 1}^m  v_{\kappa}^p D^{\cngc{p};\kappa}_{i,i}\right)^q \cj{\left(\sum_{\kappa = 1}^m v_{\kappa}^p D^{\cngc{p};\kappa}_{i,i}\right)}^q 
\end{eqnarray*}
And thus
\begin{equation}\label{eq.01041710}
\Delta^{\mu}_{i,i} = v_{\mu}^{2pq}\gamma^{2pq} D^{\cngc 0;\mu}_{i,i} \left|\sum_{\kappa = 1}^m  v_{\kappa}^p D^{\cngc{p};\kappa}_{i,i}  \right|^{2q} 
\end{equation}
A direct computation yields
\begin{equation}\label{eq:symm_131208-2}
C\Delta C = vv^T \otimes H \left(\sum_{\mu = 1}^m w^2_\mu  \Delta^{\mu}\right)\cj H^T.
\end{equation}
and by equation \eqref{eq.01041710} we obtain
\[\begin{aligned}
\sum_{\mu = 1}^m w^2_\mu  \Delta^{\mu} &=& \gamma^{2pq} \sum_{\mu = 1}^m w^2_\mu  v_{\mu}^{2pq} D^{\cngc 0;\mu} \left|\sum_{\kappa = 1}^m  v_{\kappa}^p D^{\cngc{p};\kappa}  \right|^{2q}
&=& \gamma^{2pq}\cdot \Theta
\end{aligned}\]
where the last equality follows from the definition of $\Theta = \Theta(p,q)$ in equation \eqref{eq:symm_def_Theta}. Equation \eqref{eq:symm_131208-2} then rephrases to
$$
C\Delta C = vv^T \otimes H \left(\gamma^{2pq} \Theta \right)\cj H^T = \gamma^{2pq} \cdot vv^T \otimes H\Theta\cj H^T.
$$
It thus follows that $\abs{C\Delta C}$ has a block of rank at least $2$ iff 
$\abs{H\Theta\cj H^T}$ does. Assume that $\abs{H\Theta\cj H^T}$ contains a 
block of rank at least $2$. Then $\evalk(C',\mfam D)$ is $\#\PP$-hard 
by Lemma~\ref{lem:cng_block2_hard}. Equation \eqref{eq:1305091834} implies 
that $\evalk(C',\mfam D) \Tle \evalk(C,\mfam D)$.
This finishes the proof.
\end{proof}

\paragraph*{The proofs of Lemma \ref{lem:cond_C2_D2} and \ref{lem:cond_D3}.}

\begin{proof}[of Lemma \ref{lem:cond_C2_D2}] \marc{proof OK}
Assume that $\evalk(C,\mfam D)$ is not $\#\PP$-hard, 
we shall show that \cond{C2} and \cond{D2} are satisfied. 
Fix a positive $p$ such that $\cngc{p} = 0$.
Let $\Theta = \Theta(p,q)$ the matrix defined in equation \eqref{eq:symm_def_Theta}. By Lemma~\ref{lem:struct_template}, for all $p,q \in \Nat$ the matrix $\abs{H \Theta \cj H^T}$ contains no blocks of rank at least $2$. Recall the definition of $\Theta = \Theta(p,q)$ and note that our choice of $p$ yields,
$$
\Theta = \sum_{\mu = 1}^m w^2_\mu  v_{\mu}^{2pq} D^{\cngc 0;\mu} \left|\sum_{\kappa = 1}^m  v_{\kappa}^p D^{\cngc{p};\kappa}  \right|^{2q} 
       = \sum_{\mu = 1}^m w^2_\mu  v_{\mu}^{2pq} D^{\cngc 0;\mu} \left(\sum_{\kappa = 1}^m  v_{\kappa}^p D^{\cngc 0;\kappa}  \right)^{2q}
$$
The second equality follows by condition \cond{D1} as the absolute values $|\cdot|$ are inessential for $D^{\cngc 0}$.
By Lemma~\ref{lem:cond_C1_D1}(2), we have
\begin{eqnarray*}
\Theta &=& \sum_{\mu = 1}^m w^2_\mu  v_{\mu}^{2pq} d_\mu D^{\cngc 0;1} \left(\sum_{\kappa = 1}^m  v_{\kappa}^p d_\kappa D^{\cngc 0;1}  \right)^{2q}.
\end{eqnarray*}
Define $f(p,q) : = \sum_{\mu = 1}^m w^2_\mu  v_{\mu}^{2pq} d_\mu\left(\sum_{\kappa = 1}^m  v_{\kappa}^p d_\kappa\right)^{2q}$ which is positive for all $p,q \in \Nat$. Then
\[\begin{aligned}
\Theta &=& \sum_{\mu = 1}^m w^2_\mu  v_{\mu}^{2pq} d_\mu \left(\sum_{\kappa = 1}^m  v_{\kappa}^p d_\kappa\right)^{2q} (D^{\cngc 0;1})^{2q+1} &=& f(p,q)(D^{\cngc 0;1})^{2q+1}.
\end{aligned}\]
For notational convenience let $D: = D^{\cngc 0;1}$. Then
$$
H \Theta H^T = H (f(p,q)\cdot D^{2q+1}) \cj H^T = f(p,q)H D^{2q+1} \cj H^T
$$
which implies that $\abs{H \Theta \cj H^T}$ contains a block of rank at least $2$ iff $\abs{H D^{2q+1} \cj H^T}$ does.

We therefore arrive at the assumption that for all $q \in \Nat$ the matrix $\abs{H D^{2q+1} \cj H^T}$ does not contain a block of row rank at least $2$. The Pre-Uniform Diagonal Lemma~\ref{lem:pre-unif_diag} thus implies that $H$ is complex Hadamard and $D = \alpha I_r$ for some $\alpha$.  This proves \cond{C2}.
Let $\Delta^{\cngc 0}_{\mu\mu} = d_{\mu}\alpha$ with the $d_\mu$ as given by Lemma~\ref{lem:cond_C1_D1}(2). Then $D^{\cngc 0} = \Delta^{\cngc 0} \otimes I_r$ which proves \cond{D2}.
\end{proof}

\begin{proof}[of Lemma \ref{lem:cond_D3}]\marc{proof OK}
Assume that $\evalk(C,\mfam D)$ is not $\#\PP$-hard, 
we shall show that \cond{D3} is satisfied for $D^{\cngc c}$ 
for all $c \in \Int_{\denom}$. Fix an arbitrary $c \in \Int_{\denom}$ and let 
$\Gamma_c =\{p \in \Nat \mid p \equiv c \mmod{\denom}\}$ be the set of natural 
numbers which are congruent to $c$ modulo $\denom$. That is, we have 
$\cngc{p} = c$ for all $p \in \Gamma_c$.
Let $\Theta = \Theta(p,q)$ be the matrix defined in equation \eqref{eq:symm_def_Theta}. By Lemma~\ref{lem:struct_template} we have
\begin{equation*}\label{eq:_cond_cond_C5}
\text{For all $p,q \in \Nat$ the matrix $\abs{H \Theta \cj H^T}$ contains no blocks of rank at least $2$.} 
\end{equation*}
We shall see how this implies condition \cond{D3} for $c$.
Assume from now on that $p \in \Gamma_c$. Then by the definition of $\Theta = \Theta(p,q)$ in equation \eqref{eq:symm_def_Theta} we have
\begin{eqnarray*}
\Theta &=& \sum_{\mu = 1}^m w^2_\mu  v_{\mu}^{2pq} D^{\cngc 0;\mu} \left|\sum_{\kappa = 1}^m  v_{\kappa}^p D^{\cngc{p};\kappa}  \right|^{2q}\\
&=& \sum_{\mu = 1}^m w^2_\mu  v_{\mu}^{2pq} \Delta^{\cngc 0}_{\mu\mu} \left|\sum_{\kappa = 1}^m  v_{\kappa}^p D^{\cngc c;\kappa}  \right|^{2q}
\end{eqnarray*}
where the last equality follows from $\cngc{p} = c = \cngc c$ and our assumption that condition \cond{D2} is satisfied.
Let $f(x) := \left(\sum_{\mu = 1}^m w^2_\mu  v_{\mu}^{x} \Delta^{\cngc 0}_{\mu\mu}\right)$ and define
\begin{equation}\label{eq:symm_def_delta_cond_C5}
\Delta = \left|\sum_{\kappa = 1}^m  v_\kappa^{p} D^{\cngc c;\kappa}\right| 
\end{equation}
Then $\Theta = f(2pq) \cdot \Delta^{2q}$ and we have
$H \Theta \cj H^T = H (f(2pq) \cdot \Delta^{2q}) \cj H^T = f(2pq) H \Delta^{2q} \cj H^T$.
That is, the existence of a block of rank $2$ in $\abs{H \Theta \cj H^T}$ is equivalent to the existence of such a block in $\abs{H\Delta^{2q} H^T}$ and thus it follows from our assumption that 
\begin{equation}\label{eq:_2cond_cond_C5}
\text{ For all $p,q \in \Nat$ the matrix $\abs{H\Delta^{2q} \cj H^T}$ does not contain a block of rank at least $2$. }
\end{equation}

\begin{claim}\label{cl:150509-1}
For all $p \in \Gamma_c$ the matrix $\Delta = \Delta(p)$ is pre-uniform.
\end{claim}
\begin{clproof}
Assume otherwise. That is, fix a $p \in \Gamma_c$ such that $\Delta(p)$ is not pre-uniform. Then the Pre-Uniform Diagonal Lemma~\ref{lem:pre-unif_diag} implies that there is a $q$ such that $\abs{H\Delta^{2q} \cj H^T}$ contains a block of rank at least $2$, in contradiction to equation \eqref{eq:_2cond_cond_C5}.
\end{clproof}
In the above claim, we used the parameter $q$ to show that $\Delta$ is pre-uniform irrespective of our choice of $p\in \Gamma_c$. In the following we will show that choosing $p$ large enough we can infer condition \cond{D3} for $c$. To achieve this, we will prove the following two statements
\begin{claim}\label{cl:131208_s_0}
There is a $p_0$ such that for all $p \in \Gamma_c$ with $p \ge p_0$, 
$$
 \text{If } \Delta_{ii} = \Delta(p)_{ii} = 0 \text{ then for all } \mu \in [m] \text{ we have } D^{\cngc c;\mu}_{ii} = 0.
$$
\end{claim}
\begin{clproof}
Fix an $i \in [r]$. By the definition of $\Delta$ in equation \eqref{eq:symm_def_delta_cond_C5} we see that $\Delta_{ii} = 0$ if, and only if,
$$
0 = \sum_{\kappa = 1}^m  v_\kappa^{p} D^{\cngc c;\kappa}_{ii}.
$$
If this equation is satisfied for all $p$, Lemma~\ref{lem:coeff_zero_inf} implies the existence of values $p_i$ such that for all $p \in \Gamma_c$ with $p \ge p_i$ we have $D^{\cngc c;\mu}_{ii} = 0$ for all $\mu \in [m]$. We derive such a $p_i$ for each $i \in [r]$ and the claim follows for $p_0 = \max\{p_1,\ldots, p_r\}$.
\end{clproof}

\begin{claim}\label{cl:131208_s_eq}
There is a $p_{=} \in \Nat$ such that for all $p \in \Gamma_c$ with $p \ge p_{=}$ and all $i,j \in [r]$ we have that
$$
\Delta_{ii} = \Delta_{jj} \text{ implies that there is a } \zeta_{ij}\in \U \text{ such that } D^{\cngc c;\mu}_{ii} = \zeta_{ij} D^{\cngc c;\mu}_{jj} \text{ for all } \mu \in [m].
$$
\end{claim}

\begin{clproof}
By the definition of $\Delta$ the assumption $\Delta_{ii} = \Delta_{jj}$ is equivalent to
$$
\left|\sum_{\kappa = 1}^m  v_\kappa^{p} D^{\cngc c;\kappa}_{ii}\right| = \left|\sum_{\kappa = 1}^m  v_\kappa^{p} D^{\cngc c;\kappa}_{jj}\right|
$$
Or likewise, there is a $\zeta_{ij} \in \U$ such that
$$
0 = \sum_{\kappa = 1}^m  v_\kappa^{p} \left(D^{\cngc c;\kappa}_{ii} - \zeta_{ij}D^{\cngc c;\kappa}_{jj}\right).
$$
By Lemma~\ref{lem:coeff_zero_inf} there is a $p_{ij}$ such that for all $p \ge p_{ij}$, the above implies that all coefficients 
$D^{\cngc c;\kappa}_{ii} - \zeta_{ij}D^{\cngc c;\kappa}_{jj}$ are zero.
The claim follows with $p_{=} = \max\{p_{ij} \mid i\neq j \in [r]\}$.
\end{clproof}
With these three claims we are now able to finish the proof of the lemma. Let $p \in \Gamma_c$ such that $p \ge \max\{p_0,p_=\}$ with $p_0$ and $p_=$ as in Claims~\ref{cl:131208_s_0} and \ref{cl:131208_s_eq}. We will argue by considering the diagonal entries of $\Delta = \Delta(p)$.
Define $M = \{\mu \in [m] \mid D^{\cngc c;\mu} \neq 0\}$ and let $\Lambda = \{ i \in [r] \mid \Delta_{ii} \neq 0\}$. 

If $\Lambda = \emptyset$ then  Claim~\ref{cl:131208_s_0} implies that $D^{\cngc c} = 0$ and the proof follows with $\Delta^{\cngc c} = 0$ and $U^{\cngc c} = 0$.

Assume therefore that $\Lambda \neq \emptyset$ and fix some $a \in \Lambda$. As $\Delta$ is pre-uniform by Claim~\ref{cl:150509-1} we have $\Delta_{ii}= \Delta_{aa}$ for all $i \in \Lambda$. 
By Claim~\ref{cl:131208_s_eq} this implies that 
$$
D^{\cngc c;\mu}_{ii} = \zeta_{ia}D^{\cngc c;\mu}_{aa} \text{ for all } \mu \in M,\, i \in \Lambda.
$$
Define $\Delta^{\cngc c}_{\mu\mu} = D^{\cngc c;\mu}_{aa}$ for every $\mu$ and note that $D^{\cngc c;\mu}_{aa} = 0$ if, and only if $\mu \notin M$. Define $U^{\cngc c}_{ii} = \zeta_{ia}$ for all $i \in [r]$.
Then $ D^{\cngc c} = \Delta^{\cngc c} \otimes U^{\cngc c}$.
\end{proof}

\subsection{Finishing the Proof of Lemma \ref{lem:gen_red_to_Had_non-bip}}
Let $A$ be a connected non-bipartite $\denom$-algebraic Hermitian matrix and $D$ 
a diagonal matrix of positive vertex weights.
By Lemma~\ref{lem:cond_C1_D1} either $\evalk(A,D)$ is $\#\PP$-hard 
or the following holds.
There is a matrix $C$ satisfying condition \cond{C1} and a family 
$\mfam D$ of diagonal matrices which satisfies condition \cond{D1} such that,
$$
\evalk(A,D) \Tequiv \evalk(C, \mfam D).
$$
Further, by Lemmas~\ref{lem:cond_C2_D2} and \ref{lem:cond_D3} the problem 
$\evalk(C,\mfam D)$ either is $\#\PP$-hard or the matrix $C$ satisfies 
conditions \cond{C1}--\cond{C2} and the family $\mfam D$ satisfies 
\cond{D1}--\cond{D3}. By Lemma~\ref{lem:cng_non-bip_end_decomp} we then have
$$
\evalk(C, \mfam D) \Tequiv \evalk(H, \mfam U)
$$
where $H$ and $\mfam U$ define an \cond{H--STD} problem. This finishes the proof.

\section{Hadamard Components}\label{sec:herm_hadmard}
\noindent 
In this section we will prove Theorem~\ref{thm:hadamard_reduction} 
by proving the following two lemmas.

\begin{lemma}\label{lem:hadamard_reduction_non-bip}
Let $H$ be an $\denom$-algebraic $n \times n$ matrix and $\mfam D$ a family of diagonal matrices defining an \cond{H--STD}-problem.
Then either $\evalk(H, \mfam D)$ is $\#\PP$-hard or $H$ and $\mfam D$ satisfy conditions \cond{GC},\cond{R1} through \cond{R5} and the Affinity Condition \cond{AF}.
\end{lemma}

\begin{lemma}\label{lem:hadamard_reduction_bip}
Let $A$ be $\denom$-algebraic with underlying $n \times n$ block $H$ and $\mfam D$ a family of diagonal matrices defining a \cond{B--H--STD}-problem.
Then either $\evalk(A, \mfam D)$ is $\#\PP$-hard or $H$ and $\mfam D$ satisfy conditions \cond{GC},\cond{B--R1} through \cond{B--R5} and the Affinity Condition \cond{B--AF}.
\end{lemma}

\begin{proof} [of Theorem \ref{thm:hadamard_reduction}]
Let $A$ be an $\denom$-algebraic matrix and $\mfam D$ a family of diagonal matrices defining either an \cond{H--STD} problem or a \cond{B--H--STD} problem. 

If $\evalk(A, \mfam D)$ is not $\#\PP$-hard then the following holds.
If $A$ and $\mfam D$ define an \cond{H--STD} problem then Lemma~\ref{lem:hadamard_reduction_non-bip} implies that $A$ and $\mfam D$ satisfy conditions \cond{GC}, \cond{R1} through \cond{R5} and the Affinity Condition \cond{AF}. If $A$ and $\mfam D$ define a \cond{B--H--STD} problem then Lemma~\ref{lem:hadamard_reduction_bip} implies that $A$ and $\mfam D$ satisfy conditions \cond{GC}, \cond{B--R1} through \cond{B--R5} and the Affinity Condition \cond{B--AF}.
This finishes the proof of the Theorem. 
\end{proof}
Throughout the section, $n$ and $\denom$ will have a fixed meaning. That is $H$ will be an $\denom$-algebraic $n \times n$ matrix. 
We will discuss the non-bipartite case in order to prove Lemma~\ref{lem:hadamard_reduction_non-bip}. In a sequence of several steps the proof will show that the problem $\evalk(H, \mfam D)$ is $\#\PP$-hard unless the conditions \cond{GC}, \cond{R1} through \cond{R5} and the Affinity Condition \cond{AF} are satisfied. This will be done in an inductive manner always relying on the conditions for which we have shown this so far.
The proof of Lemma~\ref{lem:hadamard_reduction_bip} follows by similar means, 
therefore we omit it. The interested reader may find the proof in \cite{thu09}.

\subsection{The Group Condition \cond{GC}}
The first step of the proof is to enable a group theoretic description 
of the matrix $H$ underlying the \cond{H--STD} problem at hand.
\begin{lemma}\label{lem:non-bip_non_gc_hardness}
Let $H$ be a matrix and $\mfam D$ a family of diagonal matrices defining an \cond{H--STD}-problem. 
If $H$ does not satisfy the group condition \cond{GC} then $\evalk(H, \mfam D)$ is $\#\PP$-hard.
\end{lemma}
\begin{proof}\marc{proof OK}
Let $G=(V,E)$ be a given digraph and $\vpin$ a pinning. Let $G_q = (V_q,E_q)$ be the digraph obtained from $G$ by
\begin{eqnarray*}
 V_q & := & V \, \cup \, \{v_e, v'_e, v_{e,1} \ldots, v_{e,2q} \, \vert \, e \in E\} \\
 E_q & := &\{\,uv_{e,2i},\,v_{e,2i-1}u,\,v_{e,2i}v,\,vv_{e,2i-1} \, \vert \, i \in [q],\; e=uv \in E\}\\
     & & \cup \,\{\,v_ev_{e,2i},\,v_{e,2i-1}v_e,\,v_{e,2i}v'_e,\,v'_ev_{e,2i-1} \, \vert \, i \in [q],\; e \in E\}
\end{eqnarray*}
We can think of $G_{q}$ being obtained
from $G$ by replacing every edge by a distinct copy of a graph $\Gamma_q$
which, for $q=1$ is illustrated in Figure \ref{fig:gc_gadget}.
\begin{figure}
\begin{center} 
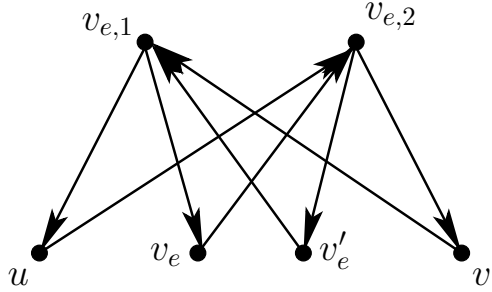
\end{center}
\caption{The gadget for $q=1$}
\label{fig:gc_gadget}
\end{figure}
Let $i,j,a,b$ be spins assigned to $u,v,v_e,v'_{e}$.  
This witnesses a reduction $\evalk(C^{[q]})\Tle \evalk(H,\mfam D)$ with
\begin{eqnarray*}
C^{[q]}_{ij} &=& \sum_{a = 1}^n \sum_{b = 1}^n\left(\sum_{c = 1}^n H_{ic} H_{cj} H_{ac} H_{cb} \right)^q\left(\sum_{c = 1}^n H_{ci} H_{jc} H_{ca} H_{bc} \right)^q \\
             &=& \sum_{a = 1}^n \sum_{b = 1}^n\left|\scalp{H\row a \bullet H\row b, H\row j \bullet H\row i}\right|^{2q}
\end{eqnarray*}
By an argument based on $C^{[q]}_{ij}$ for $i, j \in [n]$ we will see in the following that $\evalk(C^{[q]})$ is $\#\PP$-hard if $H$ does not satisfy the group condition. 
We will not give a direct proof, but we will rather show that,
\begin{center}
if there are $a',b' \in [n]$ such that $H\row {a'} \bullet H\row {b' }\notin R(H)$ then $\evalk(H,\mfam D)$ is $\#\PP$-hard.
\end{center}
To see that this finishes our proof, we shall prove that
\begin{claim}
If $H\row i \bullet H\row j \in R(H)$ for all $i,j\in [n]$ then $H$ satisfies \cond{GC}.
\end{claim}
\begin{clproof}
If $H\row i \bullet H\row j \in R(H)$ for all $i,j, \in [n]$, then there is a mapping $\tau: [n]^2 \rightarrow [n]$ such that, for all $i,j$ we have $H\row i \bullet H\row j = H\row{\tau(i,j)}$. Fix an arbitrary $j \in [n]$ and consider the map $\tau_j := \tau(\cdot,j)$. This map is bijective. To see this, assume that $\tau(a,j) = \tau (c,j)$, which implies that $H\row a \bullet H\row j = H\row c \bullet H\row j$ which yields $H_a = H_c$.
Therefore, we see that, for all $i,j \in [n]$ there is a $c \in [n]$ such that $\tau(c,j) = i$, that is, $H\row c \bullet H\row j = H\row i$ which yields $H\row i \circ H\row j = H\row c$.
Since $H$ is Hermitian this also proves the second part of the group condition.
\end{clproof}
Assume that there are $a',b' \in [n]$ such that $H\row {a'} \bullet H\row {b' }\notin R(H)$. Fix such $a',b'$ and let $i=1$ then we have 
\[
C^{[q]}_{1j} = \sum_{a = 1}^n \sum_{b = 1}^n\left|\scalp{H\row a \bullet H\row b, H\row j \bullet H\row 1}\right|^{2q}
             = \sum_{a = 1}^n \sum_{b = 1}^n\left|\scalp{H\row a \bullet H\row b, H\row j}\right|^{2q}\\ 
\]
Note that for all $a,b$ we have $|\scalp{H\row a \bullet H\row b, H\row j}| \le n$ and further
\begin{claim}\label{cl:non-bip_gc_two}
$|\scalp{H\row{ a'} \bullet H\row {b'}, H\row j}| < n$ for all $j \in [n]$.
\end{claim}
\begin{clproof}
Assume that $|\scalp{H\row{ a'} \bullet H\row {b'}, H\row j}| = n$. 
Then the Cauchy-Schwarz inequality implies that there is a $\zeta$ such that $H\row{ a'} \bullet H\row {b'} = \zeta H\row j$ but as $H$ is normalized we have $H_{j1} = (H\row{ a'} \bullet H\row{b'})_1 = 1 $ and therefore $\zeta = 1$ which implies $H\row{ a'} \bullet H\row {b'} \in R(H)$ in contradiction to our assumption.
\end{clproof}

\begin{claim}\label{cl:non-bip_gc_three}
There is a $j \in [n]$ such that $\scalp{H\row{ a'} \bullet H\row {b'}, H\row j} \neq 0$.
\end{claim}
\begin{clproof}
Otherwise, $H\row{ a'} \bullet H\row {b'}$ would be orthogonal to all vectors in $R(H)$ and by $H\row{ a'} \bullet H\row {b'} \notin R(H)$ this would imply an $n+1$ element basis of an $n$-dimensional vector space.
\end{clproof}
By construction we have $\abs{C^{[q]}} = C^{[q]}$ and $C^{[q]}$ is symmetric for all $q$. Therefore the proof follows from Lemma~\ref{lem:cng_block2_hard}  if we can show that there is a $j \in [n]$ such that the $2 \times 2$ submatrix of $C^{[q]}$ induced by $1,j$ is indecomposable and has rank $2$.
We have $C^{[q]}_{ii} = n^{2q+1}$ for all $i \in [n]$, as
$$
C^{[q]}_{ii} = \sum_{a = 1}^n \sum_{b = 1}^n\left|\scalp{H\row a \bullet H\row b, H\row i \bullet H\row i}\right|^{2q} = \sum_{a = 1}^n \sum_{b = 1}^n\left|\scalp{H\row a,H\row b}\right|^{2q}
$$
and the term $|\scalp{H\row a,H\row b}|$ is non-zero iff $a = b$.
Fix $j\in [n]$ as in Claim~\ref{cl:non-bip_gc_three}. We see that the $2\times 2$ submatrix of $C^{[q]}$ induced by $1,j$ satisfies
$$
\left(\begin{array}{c c}
       C^{[q]}_{11} & C^{[q]}_{1j} \\
       C^{[q]}_{j1} & C^{[q]}_{jj} \\
      \end{array}
 \right)
=
\left(\begin{array}{c c}
       n^{2q+1} & C^{[q]}_{1j} \\
       C^{[q]}_{j1} & n^{2q+1} \\
      \end{array}
 \right).
$$
Claim~\ref{cl:non-bip_gc_three} further implies that it is indecomposable.
To show that this submatrix is a witness of a block of rank at least $2$ it remains to show that its determinant $n^{4q+2} - (C^{[q]}_{1j})^2$ is non-zero for some $q$.
Define values $c_0, \ldots c_n$ by $c_\nu = \left|\{ (a,b)\in [n]^2 : |\scalp{H\row a \bullet H\row b, H\row j}| =\nu\}\right|$. Then we can rewrite
\begin{eqnarray*}
C^{[q]}_{1j} = \sum_{a = 1}^n \sum_{b = 1}^n\left|\scalp{H\row a \bullet H\row b, H\row j}\right|^{2q} = \sum_{\nu = 1}^n c_{\nu} \nu^{2q}
\end{eqnarray*}
Assume, for contradiction that the above mentioned determinant is zero for all $q$. This implies that
$$
C^{[q]}_{1j} = n^{2q + 1} = \sum_{\nu = 0}^n c'_\nu \nu^{2q}
$$
for $c'_0,\ldots, c'_n$ such that $c'_0 = c'_{n-1} = 0$ and $c'_n = n$.
By Lemma~\ref{lem:coeff_zero_inf} we see that for some large enough computable $q$  this implies that $c_\nu = c'_\nu$ for all $\nu \in [n]$. In particular $n = c'_n = c_n$ which contradicts the fact that Claim~\ref{cl:non-bip_gc_two} implies $c_n = 0$.
\end{proof}

\subsection{The Representation Conditions \cond{R1} through \cond{R5}}

Now we shall describe how to satisfy the Representation Conditions \cond{R1}--\cond{R5}.

\begin{lemma}\label{lem:non-bip_coset_supp}
Let $\evalk(H,\mfam D)$ be an \cond{H--STD} problem such that $H$ and $\mfam D$ satisfy \cond{R1}--\cond{R2}.
Then $\evalk(H,\mfam D)$ is $\#\PP$-hard unless conditions \cond{R3} and \cond{R4} are satisfied. 
\end{lemma}
\begin{proof}\marc{proof OK}
Fix some $c \in \Int_{\denom}$, we may assume that $c > 0$ as for $c = 0$ we have $\beta_0 = 0$ and $\gG_0 = \gG$ by the fact that $D^{\cngc 0} = I_n$. Further, for $-c > 0$ we know that $\cj D^{\cngc c} = D^{\cngc{-c}}$ by \cond{H--STD} and thus $\Lambda_c = \Lambda_{-c}$.

For any $c > 0$ we will construct a reduction as follows. Let $G = (V,E)$ be a given digraph and $\vpin$ a pinning. We construct a digraph $G^c$ from $G$ by replacing each edge $e = uv$ by a distinct length $2$ path and call the middle vertex of this path $v_e$. Then we add another vertex $v'_e$ to the graph and connect $v_e$ to $v'_e$ (in this direction) by $c$ distinct length $2$ paths.
Formally, we have $G^c = (V^c, E^c)$ with
\begin{eqnarray*}
 V^c &=& V \cup \{v_e,v'_e, v_{e,1}, \ldots, v_{e,c} \mid e \in E\} \\
 E^c &=& \{v_ev_{e,i},v_{e,i}v'_e \mid i \in [c],\; e \in E\}
\end{eqnarray*}
We have $c = \grade(v_e) = - \grade(v'_e)$ and therefore the vertex weights of $v_e$ and $v'_e$ have the same support $\Lambda_c$.
The above construction witnesses a reduction $\evalk(C, \mfam D) \Tle \evalk(H, \mfam D)$ for a matrix $C$ which we will describe now. We label the rows and columns of $C$ symmetrically by the elements of $\gG$.
Let $a,b,g,h \in \gG$ denote the spins of $u,v,v_e$ and $v'_e$ respectively. Then
\begin{eqnarray*}
C_{a,b} &=& \sum_{g\in \gG} H_{a,-g}H_{g,-b}D^{\cngc c}_{g,g}\sum_{h\in \gG}D^{\cngc {-c}}_{h,h} \left(\sum_{\mu \in \gG} H_{g,-\mu}H_{\mu,-h}D^{\cngc 0}_{\mu,\mu} \right)^c \\
&=& \sum_{g,h\in \Lambda_c} H_{a,-g}\cj H_{b,-g}D^{\cngc c}_{g,g}\cj{D^{\cngc {c}}_{h,h}} \scalp{H\row g, H\row h}^c 
\end{eqnarray*}
The second equality follows from the fact that $D^{\cngc 0} = I_n$ and $H_{\mu,-h} = \cj{H}_{h,-\mu}$. We know that $\scalp{H\row g, H\row h} = 0$ unless $g=h$. And if $g=h$, then $\scalp{H\row g, H\row h} = n$. Thus, we have
\[
C_{a,b} = \sum_{g\in \Lambda_c} H_{a,-g}\cj H_{b,-g}D^{\cngc c}_{g,g}\cj{D^{\cngc {c}}_{g,g}} n^c 
       = n^c\sum_{g\in \Lambda_c} H_{b,g}\cj H_{a,g}.
\]
Note that the right equality uses $H_{a,-g} = \cj{H}_{a,g}$. 
We know that $\Lambda_c \subseteq \gG$ but it might be the case that the neutral element $\gneut$ of $\gG$ is not contained in $\Lambda_c$. Fix some $\lambda \in \Lambda_c$ and define $S = \{g-\lambda \mid g \in \Lambda_c\}$ which clearly contains $\gneut$. 
We have
\begin{equation}\label{eq:160109-1}
C_{ab} = n^c\sum_{g\in \Lambda_c} H_{b-a,g}
       = n^c\sum_{x\in S} H_{b-a,x+\lambda} 
        = n^c H_{b-a,\lambda} \cdot \sum_{x\in S} H_{b-a,x}
\end{equation}
Let $\chi : \gG \rightarrow \{0,1\}$ denote the characteristic function of $S$, then row $\gneut$ of $C$ satisfies
\[
C_{\gneut b} = n^c H_{b,\lambda} \sum_{x\in \gG} H_{b,x}\chi(x) 
       = n^c H_{b,\lambda} \scalp{H\row {b},\chi}.
\]
Define $\mathcal C_0 = \{b \in \gG \mid \bil{b, g} = \gneut \, \text{ for all } g \in S\}$ and recall that $\caniso[0] = 1$. Thus
\begin{equation}\label{eq:1905091601}
\scalp{H\row {b},\chi} = \sum_{g\in S} \caniso\left[\bil{b,-g}\right] = \sum_{g\in S} \caniso\left[-\bil{b,g}\right] = \sum_{g\in S} \caniso\left[\gneut\right] = |S| \quad \text{ for all } b \in \mathcal C_0. 
\end{equation} 

\begin{claim}\label{cl:1905091601}
If $\eval(C,\mfam D)$ is not \#\PP-hard then for all $b \in \gG \setminus \mathcal C_0$ we have $\scalp{H\row b, \chi} = 0$.
\end{claim}
\begin{clproof}
Fix an element $b \in \gG \setminus \mathcal C_0$. 
Equation \eqref{eq:160109-1} straightforwardly implies that $C_{aa} = n^c |S|$ for all $a \in \gG$. Consider a $2 \times 2$ submatrix of $\abs{C}$ induced by $0$ and $b$,
$$
\left(\begin{array}{c c}
 |C_{00}| & |C_{0b}| \\
 |C_{b0}| & |C_{bb}| 
\end{array}\right) 
=
n^c \cdot \left(\begin{array}{c c}
 |S| & |\scalp{H\row b,\chi}| \\
 |\scalp{H\row b,\chi}| & |S|
\end{array}\right).
$$
By Lemma~\ref{lem:cng_block2_hard} the problem $\evalk(C, \mfam D)$ is \#\PP-hard, if $\abs{C}$ contains a block of row rank at least $2$. Therefore no such $2\times 2$ submatrix can be a witness for the existence of a block of rank at least $2$ in $\abs{C}$. This however is possible only if either $|\scalp{H\row b,\chi}| = 0$ or the determinant of this submatrix is zero. 
The latter would imply $|S| = |\scalp{H\row b,\chi}|$ which is impossible for the following reasons.
First of all, as $H$ is normalized, and $\gneut \in S$ the expression $\scalp{H\row b,\chi} = \sum_{x\in S} H_{b,x}$ contains at least one $1$ entry.
However, as $b \notin \mathcal C_0$ we see that $H\row b = \caniso\left[\bil{b,\absent}\right]$ is not constantly $1$ on $S$ which implies $|\scalp{H\row b,\chi}| < |S|$ and the claim follows.
\end{clproof}
Define $\gS$ as the smallest subgroup of $\gG$ containing $S$.
Recall that by the subgroup criterion of finite groups, each element of $\gS$ is of the form $\sum_{g\in S} \lambda_g g$ for appropriate $\lambda_g \in \Int$. Thus for every $b \in \mathcal C_0$ and every $g' \in \gS$ we have $\bil{b,g'} = \bil{b,\left(\sum_{g \in S} \lambda_g g\right)} = \sum_{g \in S} \lambda_g \bil{b,g} = 0$.

Let $\chi_{\gS} : \gG \rightarrow \{0,1\}$ be the characteristic function of $\gS$. Analogously to the above we see that for all $b \in \mathcal C_0$,
\begin{equation}\label{eq:1905091602}
\begin{aligned}
\scalp{H\row b,\chi_{\gS}} &=& \sum_{g' \in \gS}\caniso\left[-\bil{b, g'}\right]
    &=& \sum_{g' \in \gS}\caniso\left[-\sum_{g \in S} \lambda_g \bil{b, g}\right] &=& \sum_{g' \in \gS}\caniso\left[\gneut\right] &=& |\gS|.
  \end{aligned}
\end{equation}
We will furthermore see that
\begin{claim}\label{cl:160109-3}
For all $b \in \gG \setminus \mathcal C_0$ we have $\scalp{H\row b,\chi_{\gS}} = 0$.
\end{claim}
Before we prove this claim we will argue that this finishes the proof of the lemma.
Note first that, the basic properties of Fourier analysis of Abelian groups imply that
$$
\chi(x) = \dfrac{1}{|\gG|}\sum_{g \in \gG} \scalp{H\row g,\chi}\cdot \cj{H}_{g,x}.
$$
and 
$$
\chi_{\gS}(x) = \dfrac{1}{|\gS|}\sum_{g \in \gG} \scalp{H\row g, \chi_{\gS}}\cdot \cj H_{g,x}.
$$
Combining Claims~\ref{cl:1905091601} and \ref{cl:160109-3} and equations \eqref{eq:1905091601} and \eqref{eq:1905091602} we see that for all $g \in \gG$ we have $\scalp{H\row g,\chi_{\gS}}|S| = \scalp{H\row g,\chi}|\gS|$. Which implies that $\chi(x)|\gS| = \chi_{\gS}(x)|S|$ for all $x \in \gG$, that is $\gS = S$.
It follows straightforwardly that $\Lambda_c = \lambda + \gS$. That is, it is a coset of $\gS$. This proves condition \cond{R3}. Recall then from the discussion of conditions \cond{R1} through \cond{R5} in Section~\ref{pg:repres_criteria} that condition \cond{R4} is satisfied, as well. This finishes the proof.

\smallskip 
\subparagraph*{The Proof of Claim~\ref{cl:160109-3}.}
Let $b \in \gG \setminus \mathcal C_0$ and define $\gS^b_{\alpha} = \{g \in \gS \mid \bil{b,g} = \alpha\}$ for all $\alpha \in \Int_{\denom}$.
As $\bil{b, \absent}$ defines a homomorphism, each nonempty set $\gS^b_{\alpha}$ is just a fiber of this homomorphism and $\gS^b_0$ is its kernel. In particular, $\gS^b_0$ is a subgroup of $\gS$, it is non-empty as $\gneut \in \gS^c_\gneut$ and the other non-empty sets $\gS^c_\alpha$ are its cosets.

Let $\gA = \{\alpha \in \Int_{\denom} \mid \gS^b_\alpha \neq \emptyset\}$ be the subset of $\Int_{\denom}$ which selects all non-empty $\gS^b_\alpha$. The set $\gA$ is an Abelian group. To see this, recall that $0 \in \gA$ and further the existence of elements $g \in \gS^b_\alpha$ and $h \in \gS^b_{\alpha'}$ denotes 
$\bil{b,g} = \alpha$ and $\bil{b, h} = \alpha'$.
We have $g+h \in \gS$ and by the bilinearity of the $\bil{\absent,\absent}$ operator $\alpha + \alpha' = \bil{b, g} + \bil{b, h} = \bil{b,g+h}$.
Thus $\gG^b_{\alpha + \alpha'}$ is non-empty and therefore the subgroup criterion implies that $\gA$ is a group.
We have
\[\begin{aligned}
\scalp{H\row b,\chi_{\gS}} &=& \sum_{g\in \gS} \caniso\left[\bil{b, -g} \right]
&=& \sum_{\alpha\in \gA} \sum_{\substack{g\in \gS\\ \bil{b, g} = -\alpha}} \caniso\left[\alpha\right]
&=& \sum_{\alpha\in \gA} \caniso\left[\alpha\right] \cdot \left|\gS^b_{-\alpha}\right|
\end{aligned}
\]
As the $\gS^b_\alpha$ are cosets of $\gS^b_0$ they all have the same cardinality 
and we obtain
\[\begin{aligned}
\scalp{H\row b,\chi_{\gS}} 
&=& \left|\gS^b_{\gneut}\right| \cdot \sum_{\alpha\in \gA} \caniso\left[\alpha\right].
\end{aligned}
\]
It remains to show that the right hand side sum is zero. 
By the Fundamental Theorem of Finitely Generated Abelian Groups, there is a direct sum decomposition $\gA \cong \gC_1 \oplus \ldots \oplus \gC_z$ and each of these cyclic groups has all elements of the form $\lambda_i h_i$ for some $\lambda_i \in \Int_{\ord(h_i)}$ and $h_i$ for $i \in [z]$. 
Therefore,
\begin{equation}\label{eq:1805091634}
\begin{aligned}
\sum_{\alpha\in \gA} \caniso\left[\alpha\right] 
&=& \sum_{\lambda_1 \in \Int_{\ord(h_1)}} \cdots \sum_{\lambda_z \in \Int_{\ord(h_z)}}  \prod_{i=1}^{z} \caniso\left[h_i\right]^{\lambda_i}
&=& \prod_{i=1}^{z} \sum_{\lambda_i \in \Int_{\ord(h_i)}}\caniso\left[h_i\right]^{\lambda_i}
\end{aligned}
\end{equation}
And each of the sums on the right hand side satisfies
$$
\sum_{\lambda_i \in \Int_{\ord(h_i)}}\caniso\left[h_i\right]^{\lambda_i} = 0
$$
provided that $h_i \neq \gneut$, 
as $\caniso\left[h_i\right]$ is some $\ord(h_i)$-th root of unity. Further, since $b \in \gG \setminus \mathcal C_0$ there is at least one element in $\gA$ which is not the neutral element. Therefore in the decomposition of $\gA$ at least one non-trivial generator $h_i \neq \gneut$ exists. This proves that the term in equation \eqref{eq:1805091634} is zero and thus finishes the proof of the claim.
\end{proof}

\begin{lemma}\label{lem:non-bip_condR5}
Let $\evalk(H,\mfam D)$ be an \cond{H--STD} problem such that $H$ satisfies \cond{GC}.
Assume that $H$ and $\mfam D$ have a representation as given in \cond{R1}--\cond{R4}. 
There is an \cond{H--STD}-problem $\evalk(H,\mfam D')$ which satisfies conditions \cond{R1}--\cond{R5} such that 
$$
\evalk(H,\mfam D) \Tequiv \evalk(H,\mfam D').
$$
\end{lemma}
\begin{proof}
Define, for each $c \in \Int_\denom$ values $\lambda_c$ as follows.
If $D^{\cngc c} = 0$ let $\lambda_c = 0$. Otherwise, we know by condition \cond{R4} that $D^{\cngc c}_{\beta_c, \beta_c} = \caniso[\rcwf_c(\gneut)]$. We define $\lambda_c = \caniso[\rcwf_{c}(\gneut)]^{-1}$ and let $\mfam D' = (\lambda_c \cdot D^{\cngc c})_{c \in \Int_\denom}$. 
Let $G=(V,E)$ be a digraph and $\vpin$ a pinning. Let, for each $c \in \Int_\denom$ be $n_c$ the number of vertices $v \in V\setminus \df(\vpin)$ such that $\cngc{\grade(v)} = c$. Then
$$
Z_{H,\mfam D}(\vpin, G) = \left(\prod_{c = 0}^{\denom-1} \lambda_c^{n_c} \right)Z_{H,\mfam D'}(\vpin, G).
$$
This proves the reducibility in both directions.
\end{proof}

\subsection{The Affinity Condition \cond{AF}}

In the present situation we are faced with an \cond{H--STD} problem $\evalk(H,\mfam D)$ which already satisfies the group condition \cond{GC} and the representability conditions \cond{R1}--\cond{R5}. Our aim will be now to provide the last big step necessary for proving Lemma~\ref{lem:hadamard_reduction_non-bip}. This step is given by the following Lemma.

\begin{lemma}[The Non-Bipartite Affinity Lemma]\label{lem:non-bip_AF_proof}
Let $\evalk(H,\mfam D)$ be some \cond{H--STD}-problem such that $H$ satisfies \cond{GC}.
Assume that $H$ and $\mfam D$ have a representation as given in \cond{R1}--\cond{R5}.
If \cond{AF} is not satisfied then $\evalk(H, \mfam D)$ is \#\PP-hard.
\end{lemma}
The proof relies on a construction which we will analyze separately in Lemma~\ref{lem:non-bip_nymph_ana} so as to make it more digestible.
We shall describe the reduction first.

\paragraph*{The Nymphaea Reduction.} We consider the problem $\evalk(H,\mfam D)$ as above. Let $G =(V,E)$ be a given digraph and $\vpin$ a pinning.
\begin{figure}
\begin{center} 
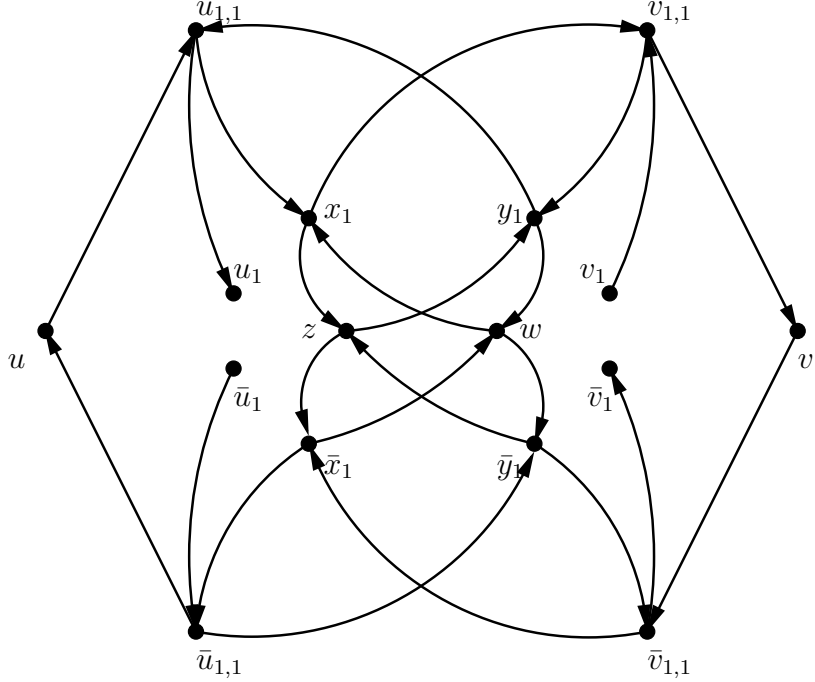
\end{center}
\caption{Nymphaea for $p=1,q=1$}
\label{fig:nymphaea}
\end{figure}
For parameters $p$ and $q$, we construct a digraph $G' = G'(p,q)$ by replacing each edge in $G$ by a graph $\Gamma_{p,q}$ which has two distinguished terminals $u$ and $v$. We refer to this graph as the "Nymphaea" (see Figure \ref{fig:nymphaea}).
The vertex set of $\Gamma_{p,q}$ is 
\begin{equation}\label{eq:verts_nymph}
\{u,v,z,w,u_{i,j}, \bar u_{i,j}, u_i, \bar u_i, v_{i,j}, \bar v_{i,j},v_i, \bar v_i ,x_i,\bar x_i, y_i, \bar y_i \mid i \in [p], j \in [q]\} 
\end{equation}
Its edge set is
\begin{equation}\label{eq:edges_nymph}
\begin{aligned}
& &\{ uu_{i,j} ,\, y_iu_{i,j},\, u_{i,j}u_i,\, u_{i,j}x_i \mid i \in [p],\, j \in [q] \}
&\cup& \{ \bar u_{i,j} u ,\, \bar u_{i,j}\bar y_i,\, \bar u_i\bar u_{i,j},\,\bar x_i \bar u_{i,j} \mid i \in [p],\, j \in [q] \} \\
&\cup& \{v_{i,j}v,\, v_{i,j}y_i,\, v_iv_{i,j},\, x_iv_{i,j} \mid i \in [p],\, j \in [q] \} &\cup & \{ v \bar v_{i,j},\, \bar y_i\bar v_{i,j},\, \bar v_{i,j} \bar v_i,\, \bar v_{i,j} \bar x_i \mid i \in [p],\, j \in [q] \} \\
&\cup &\{ x_iz,\, wx_i,\, zy_i,\, y_iw \mid i \in [p],\, j \in [q] \} &\cup & \{ z \bar x_i,\, \bar x_i w,\, \bar y_i z,\, w \bar y_i \mid i \in [p],\, j \in [q] \} 
\end{aligned}
\end{equation}

\begin{lemma}\label{lem:non-bip_nymph_ana}
Let $\evalk(H,\mfam D)$ be an \cond{H--STD} problem such that $H$ satisfies \cond{GC}.
Assume that $H$ and $\mfam D$ have a representation as given in \cond{R1}--\cond{R4}.

Then for all $p \in \Nat$ and $q \in \Int_{\denom}$ the Nymphaea reduction witnesses
$$
\evalk(C) \Tle \evalk(H,\mfam D)
$$
for a non-negative real valued $n \times n$ matrix $C = C(p,q)$ which satisfies the following.
For all $u,v \in \beta_q + \gG_q$ we have
$$
C_{u,v} = \sum_{g \in \gG} \left\vert\sum_{x \in \gG_{q}}
          \caniso\left[\rcwf_{q}(v- \beta_{q}+x) - \rcwf_{q}(u- \beta_{q}+x) + \bil{g, x}\right]\right\vert^{2p} 
$$
\end{lemma}
\begin{proof}\marc{proof OK}
Let $G = (V,E)$ be a digraph and $\vpin$ a pinning. For parameters $p$ and $q$ let $G' = G'(p,q)$ by the digraph of the Nymphaea reduction as described above.

Observe first that there is a certain axis of symmetry in the Nymphaea which has also been made explicit in Figure \ref{fig:nymphaea}. The vertices $u,v,z,w$ lie on this axis and all other vertices occur in two forms, without the bar ( e.g. $u_i$) and with the bar ($\bar u_i$). Further, the connections of the unbared vertices are the same as for the bared ones except for the fact that the directions are reversed. By $H$ being Hermitian, this amounts to a conjugation of the corresponding contributions and will become important in the following.
First of all, due to this symmetry, it will suffice to analyze the unbared part of the gadget. The results for the bared part then follow clearly by conjugation.

\paragraph*{Analyzing the Nymphaea.}
To simplify the analysis further we will focus on one of its subgraphs first.
For some $i \in [p]$, $j \in [q]$ consider the subgraph induced by vertices $u, u_{i,j}, u_i,x_i,y_i$.
We will fix the spins of $u, u_i, x_i$, and $y_i$ for the moment and --- slightly abusing notation --- denote these by just the vertices they correspond to. We denote the contribution of this subgraph excluding the vertex weights of $u,u_i,x_i,y_i$ by $Z_1(u,u_i,x_i,y_i)$.
The vertex $u_{i,j}$ will not be fixed and its spin will be denoted by $\mu$. Note that $\grade(u_{i,j}) = 0$ which implies by \cond{H--STD} that the vertex weights on $u_{i,j}$ are inessential. Then --- recall that $H$ is indexed by elements of the group $\gG$ --- it follows that
\begin{eqnarray*}
Z_1(u,u_i,x_i,y_i) &=& \sum_{\mu \in \gG} H_{u, -\mu}H_{\mu, -x_i} H_{y_i, -\mu} H_{\mu, -u_i} \\
                   &=& \sum_{\mu \in \gG} H_{u,- \mu}\cj H_{x_i,- \mu } H_{y_i,- \mu} \cj H_{u_i,- \mu} \\
                   &=& \sum_{\mu \in \gG} H_{u - x_i + y_i - u_i, - \mu}
\end{eqnarray*}
This can be rephrased as $Z_1(u,u_i,x_i,y_i) = \scalp{H\row {u - x_i + y_i - u_i}, H\row \gneut}$
which by the Hadamard property of $H$ implies that $Z_1(u,u_i,x_i,y_i) = 0$ unless $u - x_i + y_i - u_i = \gneut$ and if so, we have $Z_1(u,u_i,x_i,y_i) = n$.

Analogous reasoning on the subgraph induced by vertices $v, v_{i,j}, v_i,x_i,y_i$ shows that its contribution is
\begin{eqnarray*}
Z_2(v,v_i,x_i,y_i) &=& \scalp{H\row {v_i - v + x_i - y_i}, H\row \gneut}.
\end{eqnarray*}
Similarly, this is zero unless $v_i - v + x_i - y_i = 0$, and if so we have $Z_2(v,v_i,x_i,y_i) = n$.

Let now $Z^q_1(u,u_i,x_i,y_i)$ denote the contribution of the subgraph defined by $u, u_{i,j}, u_i,x_i,y_i$ for all $j \in [q]$. Then straightforwardly $Z^q_1(u,u_i,x_i,y_i) = Z_1(u,u_i,x_i,y_i)^q$. Defining $Z^q_2(v,v_i,x_i,y_i)$ analogously on $v, v_{i,j}, v_i,x_i,y_i$ for all $j \in [q]$ we obtain $Z^q_2(v,v_i,x_i,y_i) = Z_2(v,v_i,x_i,y_i)^q$.
Furthermore, for the subgraph on vertices $z,w,x_i,y_i$ we have a contribution of
$$
H_{x_i,-z}H_{z,-y_i}H_{w,-x_i}H_{y_i,-w} = H_{x_i,-z}\cj H_{y_i,-z}\cj H_{x_i,-w}H_{y_i,-w}  = \caniso\left[ \bil{x_i - y_i,z - w}\right]
$$
Let us now turn to the complete gadget. Note that $0 = \grade(z) =\grade(w) = 0$ and $\grade(x_i) = \grade(y_i) = 0$ for all $i \in [q]$. Therefore the corresponding vertex weights are inessential. The remaining vertices $u_i,v_i$ satisfy $-\grade(u_i) = \grade(v_i) = q$. By condition \cond{H--STD} we know that $D^{\cngc {-q}} = \cj{D^{\cngc q}}$ which particularly implies that the support of these two is the same, namely $\Lambda_q = \Lambda_{ - q}$. Let $\mu,\nu$ be the spins of $u_i$ and $v_i$ respectively. For every fixed $i \in [p]$ and every configuration of $u,v,x_i,y_i$ 
the subgraph defined by
$u,v,x_i,y_i,u_i,v_i,u_{i,j},v_{i,j}$ for all $j\in [q]$ then yields a contribution
$$
Z(u,v,x_i,y_i) = \sum_{\mu,\nu \in \Lambda_{q}} \cj{D^{\cngc{q}}_{\mu,\mu}}	D^{\cngc{q}}_{\nu,\nu} 
Z_1(u,\mu,x_i,y_i)^qZ_2(v,\nu,x_i,y_i)^q
$$
and by the above reasoning on $Z_1$ and $Z_2$ we have
\begin{eqnarray}
Z(u,v,x_i,y_i) &=& \sum_{\mu,\nu \in \Lambda_{q}} \cj{D^{\cngc{q}}_{\mu,\mu}}	D^{\cngc{q}}_{\nu,\nu} 
\scalp{H\row {u - x_i + y_i - \mu}, H\row \gneut}^q\scalp{H\row {\nu - v + x_i - y_i}, H\row \gneut}^q\\
&=& n^{2q} \cj{D^{\cngc{q}}_{u - x_i + y_i,u - x_i + y_i}} D^{\cngc{q}}_{v - x_i + y_i,v - x_i + y_i}
\label{eq:1905091739}
\end{eqnarray}
where we now included all vertex weights. 
By the fact that the bared part of the Nymphaea has all edges in reverse direction we obtain by the same arguments as above that
$$
H_{z,-\bar x_i}H_{\bar y_i,-z}H_{\bar x_i,-w}H_{w,-\bar y_i} = \cj{\caniso\left[ \bil{\bar x_i - \bar y_i,z - w}\right]}
$$
Further the the subgraph defined by
$u,v,\bar x_i,\bar y_i,\bar u_i,\bar v_i,\bar u_{i,j},\bar v_{i,j}$ for all $j\in [q]$ then yields a contribution
\begin{eqnarray*}
\bar Z(u,v,\bar x_i,\bar y_i) &=& \sum_{\mu,\nu \in \Lambda_{q}} D^{\cngc{q}}_{\mu,\mu}\cj{D^{\cngc{q}}_{\nu,\nu}} 
\cj{\scalp{H\row {u - \bar x_i + \bar y_i - \mu}, H\row \gneut}}^q\cj{\scalp{H\row {\nu - v + \bar x_i - \bar y_i}, H\row \gneut}}^q
\end{eqnarray*}
That is
\begin{equation}\label{eq:non-bip_nymph_z_barz}
 \bar Z(u,v,\bar x_i,\bar y_i) = \cj{Z(u,v,\bar x_i,\bar y_i)}
\end{equation}
Note further, that by construction all original vertices have grade $0$.
Altogether, the Nymphaea therefore witnesses a reduction
$\evalk(C') \Tle \evalk(H,\mfam D)$ for a symmetric non-negative matrix $C'$ defined by
\begin{eqnarray*}
C'_{u,v} &=&\sum_{z,w \in \gG} \phantom{\times} \prod_{i=1}^p\left(\sum_{x_i,y_i \in \gG}  Z(u,v,x_i,y_i) \caniso\left[\bil{x_i - y_i,z - w}\right]\right) \\
        & &\phantom{\sum_{z,w \in \gG}} \times \prod_{i=1}^p\left(\sum_{\bar x_i, \bar y_i \in \gG}  \bar Z(u,v,\bar x_i,\bar y_i) \cj {\caniso\left[\bil{\bar x_i - \bar y_i, z - w}\right]}\right)
\end{eqnarray*}
Furthermore as the different parts of the $p$-adic product terms in the definition of $C'$ are all independent, we can replace all $x_i, \bar x_i$ and $y_i, \bar y_i$ by only two variables $x,y$. By application of equation \eqref{eq:non-bip_nymph_z_barz} we obtain
\begin{eqnarray*}
C'_{u,v} &=&\sum_{z,w \in \gG} \phantom{\times} \prod_{i=1}^p\left(\sum_{x,y \in \gG}  Z(u,v,x,y) \caniso\left[\bil{x - y,z - w}\right]\right) \\
        & &\phantom{\sum_{z,w \in \gG}} \times \prod_{i=1}^p\left( \cj {\sum_{x, y \in \gG} Z(u,v,x,y) \caniso\left[\bil{x - y, z - w}\right]}\right)\\
&=&\sum_{z,w \in \gG} \left\vert\sum_{x,y \in \gG}  Z(u,v,x,y) \caniso\left[\bil{x - y,z - w}\right]\right\vert^{2p}
\end{eqnarray*}
Note that for all $z,g \in \gG$ the equation $z - w = g$ has a unique solution $w \in \gG$. Therefore we can simplify
\begin{eqnarray*}
C'_{u,v} &=& n\sum_{g \in \gG} \left\vert\sum_{x,y \in \gG}  Z(u,v,x,y) \caniso\left[\bil{x - y, g}\right]\right\vert^{2p} \\
\end{eqnarray*}
And with the definition of $Z(u,v,x,y)$ as given in equation \eqref{eq:1905091739} we get
\begin{eqnarray*}
C'_{u,v} &=& n\sum_{g \in \gG} \left\vert\sum_{x,y \in \gG} n^{2q} \cj{D^{\cngc{q}}_{u - x + y,u - x + y}} D^{\cngc{q}}_{v-x+y,v-x+y} \caniso\left[\bil{x - y, g}\right]\right\vert^{2p} \\
\end{eqnarray*}
Now, for all $x, a \in \gG$ there is an unique solution $y$ to the equation $x - y = a$. Thus
\begin{eqnarray*}
C'_{u,v} &=&n^{4pq + 1}\sum_{g \in \gG} \left\vert n\sum_{x \in \gG} 
          \cj{D^{\cngc{q}}_{u - x,u - x}}D^{\cngc{q}}_{v - x,v - x} \cdot \caniso\left[\bil{x,g}\right]\right\vert^{2p} \\
&=&n^{4pq + 2p + 1}\sum_{g \in \gG} \left\vert\sum_{x \in \gG}
          \cj{D^{\cngc{q}}_{u-x,u-x}}D^{\cngc{q}}_{v-x,v-x} \cdot \caniso\left[\bil{x,g}\right]\right\vert^{2p} 
\end{eqnarray*}
Defining $C = n^{-(4pq + 2p + 1)} \cdot C'$ we have $\evalk(C) \Tequiv \evalk(C')$. We will show that $C$ satisfies the statement of the lemma.
By condition \cond{R3} we have $\Lambda_{q} = \beta_{q} + \gG_{q}$ for some $\beta_{q} \in \gG$ and $\gG_q$ a subgroup of $\gG$.
In the following we will make the additional assumption that $u - \beta_{q},v - \beta_{q} \in \gG_{q}$. Recall that the vertex weights $\cj{D^{\cngc{q}}_{u-x,u-x}}D^{\cngc{q}}_{v-x,v-x}$ are non-zero iff $v-x,u-x \in \Lambda_{q}$. This thus is satisfied iff $x \in \gG_{q}$ and we may rewrite
\begin{eqnarray*}
C_{u,v} &=&\sum_{g \in \gG} \left\vert\sum_{x \in \gG_{q}}
          \cj{D^{\cngc{q}}_{u-x,u-x}}D^{\cngc{q}}_{v-x,v-x} \cdot \caniso\left[\bil{x,g}\right]\right\vert^{2p} \\
       &=& \sum_{g \in \gG} \left\vert\sum_{x \in \gG_{q}}
          \caniso\left[\rcwf_{q}(v- \beta_{q} -x) - \rcwf_{q}(u- \beta_{q}-x) + \bil{g,-x}\right]\right\vert^{2p} \\
&=& \sum_{g \in \gG} \left\vert\sum_{x \in \gG_{q}}
          \caniso\left[\rcwf_{q}(v- \beta_{q}+x) - \rcwf_{q}(u- \beta_{q}+x) + \bil{g,x} \right]\right\vert^{2p} 
\end{eqnarray*}
where the second equality follows from the group theoretic definition of $D^{\cngc q}$ as given in \cond{R4} and from the skew-bilinearity: $\bil{x,g} = -\bil{g,x} = \bil{g,-x}$. The last equality follows from the fact that we are summing over all $x\in \gG_{q}$ which renders the inversion $-x$ of $x$ irrelevant.
 \end{proof}
We are now in a position to prove the Non-Bipartite Affinity Lemma~\ref{lem:non-bip_AF_proof}.

\begin{proof}[of Lemma \ref{lem:non-bip_AF_proof}]\marc{proof OK}
Fix some $q \in \Int_{\denom}$ such that $q \ge 0$. For every $p \in \Nat$ Lemma~\ref{lem:non-bip_nymph_ana} just proved shows that the Nymphaea reduction witnesses
$$
\evalk(C) \Tle \evalk(H,\mfam D)
$$
for a non-negative real valued $n \times n$ matrix $C = C(p,q)$ which satisfies that  for all $u,v \in \beta_q + \gG_q$ we have
$$
C_{u,v} = \sum_{g \in \gG} \left\vert\sum_{x \in \gG_{q}}
          \caniso\left[\rcwf_{q}(v- \beta_{q}+x) - \rcwf_{q}(u- \beta_{q}+x) + \bil{g,x}\right]\right\vert^{2p} 
$$
We need some small simplification and some further preparation before we can give the proof. Assume in the following that $u,v \in \beta_q + \gG_q$. Consider the term $\bil{g,x}$ in the above equation, where $g \in \gG$ and $x \in \gG_q$. Define a set $K = \{g \in \gG \mid \bil{g, x} = \gneut \,\text{ for all } x \in \gG_q\}$ which clearly is a subgroup of $\gG$ and let $\gG' = \gG/K$ be the factor group of $\gG$ modulo $K$. Let $f: \gG \rightarrow \gG'$ be the canonical homomorphism defined by $g \mapsto g + K$, which thus has kernel $\ker f = K$. Then the following holds
\begin{claim}\label{cl:1905091148}
For all $g' \in \gG'$ and all $a,b \in f^{-1}(g')$ we have
$$ \bil{a, x} = \bil{b, x} \text{ for all } x \in \gG_{q}.$$
And $\vert \gG' \vert = \vert \gG_q \vert$.
\end{claim}
\begin{clproof}
For the first part, we have $a + K = b + K$. Thus $b = a + k$ for some $k \in K$ which yields $\bil{b,x} = \bil{a,x} + \bil{k,x} = \bil{a,x}$.
Let us now show that $\vert \gG' \vert = \vert \gG_q \vert$. Note first that the columns $H\col {-g} = \caniso[\bil{\absent,g}]$ for all $g \in \gG_q$ form an independent set of columns. Thus there is a set $I \subseteq \gG$ such that the submatrix $H_{I,\gG_q}$ is non-singular. Note that by the first part of the claim, all distinct $a,b \in I$ are contained in distinct fibers of $f$. That is, there are distinct $g',g'' \in \gG'$ such that $a \in f^{-1}(g')$ and $b \in f^{-1}(g'')$. This proves $|\gG' | = |I| = |\gG_q|$.
\end{clproof}
By this claim, we may, for each $g \in \gG'$, fix an element $\hat g \in f^{-1}(g)$ (which is always possible as $f$ is surjective) and assume particularly that $\hat \gneut = \gneut$. We obtain
\begin{eqnarray*}
C_{u,v} &=& |K|\sum_{g \in \gG'}\left\vert\sum_{x \in \gG_q} \caniso\left[\rcwf_{q}(v- \beta_{q}+x) - \rcwf_{q}(u- \beta_{q} + x) + \bil{\hat g,x}\right]\right\vert^{2p}
\end{eqnarray*}
Define now $C' = C'(p,q) = |K|^{-1} \cdot C$. We have $\evalk(C') \Tequiv \evalk(C)$ and for $u,v \in \beta_{q} + \gG_{q}$,
\begin{equation}\label{eq:190109-1731}
C'_{uv} = \sum_{g \in \gG'}\left\vert\sum_{x \in \gG_q} \caniso\left[\rcwf_{q}(v- \beta_{q}+x) - \rcwf_{q}(u- \beta_{q} + x) + \bil{\hat g,x}\right]\right\vert^{2p}
\end{equation}
so we may found our further considerations on $C'$ which makes our reasoning a bit more convenient.

\begin{claim}\label{cl:190109-1735}
For all $g \in \gG'$ we have the following:
\[
\begin{aligned}
 \text{(1) If } g = \gneut &\text{ then  }& \sum_{x \in \gG_q}& \caniso \left[\bil{\hat g,x}\right] = |\gG_q|.\\
 \text{(2) If } g \neq \gneut &\text{ then  }& \sum_{x \in \gG_q}& \caniso\left[\bil{\hat g,x}\right] = 0.
\end{aligned}
\]
In particular, the mappings $\{\chi_g: \gG_{ q} \rightarrow \U_\denom \mid \chi_g : =   \caniso\left[\bil{\hat g,\absent}\right],\; g \in \gG'\}$ form a basis of the $\vert \gG_{ q}\vert$ dimensional vector space $L^2(\gG_{ q})$ of functions $ f: \gG_{ q} \rightarrow \C$.
\end{claim}
\begin{clproof}
The statement (1) is clear from the definition of $K$. For statement (2) recall that $\gG_q$ has a decomposition into cyclic groups $\gC_1 \oplus \ldots \oplus \gC_z$ and each $x\in \gG_q$ has a representation $x = \sum_{i=1}^z\lambda_i h_i$ with each $h_i$ being a generator of $\gC_i$ and each $\lambda_i \in \Int_{\ord(h_i)}$.
Thus
\begin{eqnarray*}
\sum_{x \in \gG_q} \caniso\left[\bil{\hat g,x}\right] &=& \sum_{\lambda_1 \in \Int_{\ord(h_1)}} \cdots \sum_{\lambda_z \in \Int_{\ord(h_z)}} \caniso\left[\bil{\hat g, \sum_{i=1}^z\lambda_i h_i}\right]\\
&=& \sum_{\lambda_1 \in \Int_{\ord(h_1)}} \cdots \sum_{\lambda_z \in \Int_{\ord(h_z)}} \prod_{i=1}^z \caniso\left[\bil{\hat g,h_i}\right]^{\lambda_i}\\
&=& \prod_{i=1}^z \sum_{\lambda_i \in \Int_{\ord(h_i)}} \caniso\left[\bil{\hat g, h_i}\right]^{\lambda_i}
\end{eqnarray*}
By $g \neq \gneut$ we know that there is an $x \in \gG_{ q}$ such that $\bil{\hat g,x} \neq \gneut$. Hence by $x = \sum_{i=1}^z\lambda_i h_i$ there is at least one $h_i$ such that $\bil{\hat g,h_i} \neq \gneut$. 
Let $N$ be a multiple of the order of $h_i$. By the homomorphism property of the $\bil{\absent,\absent}$ operator we have
$$
\caniso\left[ \bil{\hat g,h_i}\right]^N = \caniso\left[ N \bil{\hat g,h_i}\right]
= \caniso\left[\bil{\hat g,Nh_i}\right] = \caniso\left[\bil{\hat g,\gneut}\right] = 1
$$
That is $\caniso\left[ \bil{\hat g,h_i}\right]$ is an $\ord(h_i)$-th root of unity an therefore we have
$$
\sum_{\lambda_i \in \Int_{\ord(h_i)}} \caniso\left[\bil{\hat g,h_i}\right]^{\lambda_i} = 0.
$$
This finishes the first part of the claim. 
For the second part note that we have, for $g\neq h \in \gG'$,
$$
\scalp{\chi_g, \chi_h} = \sum_{x \in \gG_{ q}}\chi_g(x)\cdot \cj{\chi_h(x)}
= \sum_{x \in \gG_{ q}}\caniso\left[\bil{\hat g, x} - \bil{\hat h, x}\right] =
\sum_{x \in \gG_{ q}}\caniso\left[\bil{\hat g - \hat h ,x}\right].
$$
By the first part of the claim, this value is zero iff $\hat g - \hat h \in K$ which implies $g = h$ by the definition of the elements $\hat g$. Therefore all distinct $\chi_g,\chi_h$ are pairwise orthogonal. The fact that they form a basis of $L^2(\gG_q)$ now follows from $|\gG'| = |\gG_q|$ as given by Claim~\ref{cl:1905091148}.
\end{clproof}

\begin{claim}\label{cl:190109-1926}
If $\evalk(H,\mfam D)$ is not \#\PP-hard then the following is true. 
For all $a \in  \gG_{ q}$ there is exactly one $g \in \gG'$ such that
$$
\vert \gG_{ q}\vert = \left\vert\sum_{x \in \gG_q} \caniso\left[\rcwf_{q}(a+x) - \rcwf_{q}(x) + \bil{\hat g, x}\right]\right\vert
$$
and for all other $g' \in \gG'$ we have
$$
0 = \left\vert\sum_{x \in \gG_q} \caniso\left[\rcwf_{q}(a+x) - \rcwf_{q}(x) + \bil{\hat g',x}\right]\right\vert
$$ 
\end{claim}
\begin{clproof}
As $\evalk(H,\mfam D)$ not being \#\PP-hard implies the same for $\evalk(C')$ we can, for all choices of $p\in \Nat$, derive the following by Lemma~\ref{lem:cng_block2_hard}:
\begin{equation*}
\text{ For all } p \in \Nat, \text{ the matrix } C' = C'(p,q) \text{ does not contain a block of rank at least} \ge 2.
\end{equation*}
This is particularly true for all submatrices of $C'$ induced by indices from the coset $\beta_{ q} + \gG_{ q}$. Note that equation \eqref{eq:190109-1731}, Claim~\ref{cl:190109-1735} and $\hat \gneut = \gneut $, entail that
for all $u \in \beta_{ q} + \gG_{ q}$
$$
C'_{u,u} = \sum_{g \in \gG'}\left\vert\sum_{x \in \gG_q} \caniso\left[\bil{\hat g,x}\right]\right\vert^{2p} = \left\vert\sum_{x \in \gG_q} \caniso\left[\bil{\hat \gneut,x}\right]\right\vert^{2p} = \vert \gG_{ q}\vert^{2p}.
$$
Consider a $2 \times 2$ submatrix of $C'$ induced by $u = \beta_{ q} + \gneut$ and some $v = \beta_{ q} + a \in \gG_{ q}$,
$$
\left(\begin{array}{c c}
       C'_{u,u} & C'_{u,v} \\
       C'_{v,u} & C'_{v,v}
      \end{array}\right)
=
\left(\begin{array}{c c}
       \vert \gG_{ q}\vert^{2p} & C'_{u,v} \\
       C'_{u,v} & \vert \gG_{ q}\vert^{2p}
      \end{array}\right).
$$
As this submatrix is either not part of a single block, or it is of rank at most $1$ we see that

\begin{equation} 
\text{ For all $p\in \Nat$ either } C'_{u,v} = 0 \text{ or } (C'_{u,v})^2 =\vert \gG_{ q}\vert^{4p}.
\end{equation}
By the choice of $u,v$ we have
$$
C'_{u,v} = \sum_{g \in \gG'}\left\vert\sum_{x \in \gG_q} \caniso\left[\rcwf_{q}(a+x) - \rcwf_{q}(x) + \bil{\hat g,x}\right]\right\vert^{2p}
$$
and note that the inner term $\left\vert\sum_{x \in \gG_q} \caniso\left[\rcwf_{q}(a+x) - \rcwf_{q}(x) + \bil{\hat g,x}\right]\right\vert$ is always at most $\vert \gG_{ q}\vert$.
For all $0 \le i \le \vert \gG_{ q}\vert$ define 
$$
c_{a,i} = \left|\left\{g\in \gG' : \left\vert\sum_{x \in \gG_q} \caniso\left[\rcwf_{q}(a+x) - \rcwf_{q}(x) + \bil{\hat g,x}\right]\right\vert = i  \right\}\right|.
$$ 
This enables us to rewrite the expression for $C'_{u,v}$ by
$$
C'_{u,v} = \sum_{\nu = 0}^{\vert \gG_{ q}\vert} c_{a,\nu} (\nu^2)^p.
$$
And $C'_{u,u} = C'_{v,v} = |\gG_{q}|^{2p}$.
Assume first that $C'_{u,v} = 0$ holds for infinitely many $p$. Then Lemma~\ref{lem:coeff_zero_inf} implies that $c_{a,0} = \vert \gG' \vert$ and $c_{a,1} = \ldots = c_{a,\vert \gG_{ q}\vert} = 0$. However this would mean that
the mapping $f : \gG_{ q} \rightarrow \U$ defined by
$$
f(x) := \caniso\left[\rcwf_{q}(a+x) - \rcwf_{q}(x) \right]
$$
is orthogonal to all $\chi_g$ for $g\in \gG'$ in contradiction to these being a basis of $L^2(\gG_{ q})$ --- as we have seen in Claim~\ref{cl:190109-1735}.
Therefore, there are infinitely many $p$ such that $(C'_{u,v})^2 =\vert \gG_{ q}\vert^{4p}$. That is, there are infinitely many $p$ such that
$$
|\gG_{q}|^{2p} = \sum_{\nu = 0}^{\vert \gG_{ q}\vert} c_{a,\nu} (\nu^2)^p
$$
which by Lemma~\ref{lem:coeff_zero_inf} implies that $c_{a,0} = \vert \gG' \vert-1$ and $c_{a,1} = \ldots = c_{a,\vert \gG_{ q}\vert - 1} = 0$ and $c_{a,\vert \gG_{ q}\vert} = 1$. This finishes the proof.
\end{clproof}
Assume from now on that $\evalk(H,\mfam D)$ is not $\#\PP$-hard. Then the uniqueness statement of Claim~\ref{cl:190109-1926}, implies that there is a mapping $\gamma_{ q} : \gG_{ q} \rightarrow \gG$ such that for all $a \in \gG_{ q}$ we have
$$
\vert \gG_{ q}\vert = \left\vert\sum_{x \in \gG_q} \caniso\left[\rcwf_{q}(a+x) - \rcwf_{q}(x) - \bil{\gamma_{ q}(a),x}\right]\right\vert
$$
Further, by
\begin{eqnarray*}
\left\vert\sum_{x \in \gG_q} \caniso\left[\rcwf_{q}(a+x) - \rcwf_{q}(x) - \bil{\gamma_{ q}(a),x}\right]\right\vert
&\le &\\
\sum_{x \in \gG_q} \left\vert\caniso\left[\rcwf_{q}(a+x) - \rcwf_{q}(x) - \bil{\gamma_{ q}(a), x}\right]\right\vert
&=& \vert \gG_{ q}\vert
\end{eqnarray*}
we see that equality can be achieved only, if there is a constant $\alpha_a \in \Omega$ such that for all $x \in \gG_{ q}$ we have
$$
\rcwf_{q}(a+x) - \rcwf_{q}(x) - \bil{\gamma_{ q}(a), x} = \alpha_{a}.
$$
With $x = \gneut$ and $\rcwf_q(\gneut) = \gneut$ we see that $\alpha_a = \rcwf_{c}(a)$. This finishes the proof.
\end{proof}

\subsection{The Proof of the Non-Bipartite Case Lemma \ref{lem:hadamard_reduction_non-bip}}

Let $H$ be an $\denom$-algebraic $n \times n$ matrix and $\mfam D$ a family of diagonal matrices defining an \cond{H--STD}-problem. 
Lemma~\ref{lem:non-bip_non_gc_hardness} implies that $\evalk(H,\mfam D)$ is $\#\PP$-hard unless the group condition \cond{GC} is satisfied.
 
Assume therefore that \cond{GC} is satisfied and recall the definitions of the Representability Conditions \cond{R1}--\cond{R5} on page \pageref{pg:repres_criteria}. 
We know that $H$ and $\mfam D$ satisfy \cond{R1} and \cond{R2}. By Lemma~\ref{lem:non-bip_coset_supp} the problem $\evalk(H,\mfam D)$ is $\#\PP$-hard unless the conditions \cond{R3} and \cond{R4} are satisfied.
By Lemma~\ref{lem:non-bip_condR5} we may further assume w.l.o.g. that \cond{R5} is satisfied. Therefore the proof follows as Lemma~\ref{lem:non-bip_AF_proof} proves $\#\PP$-hardness of $\evalk(H,\mfam D)$ unless the Affinity Condition \cond{AF} holds.

\section{Polynomial Time Computable Partition Functions}\label{sec:ptime}
\noindent In this section we will prove Theorem~\ref{thm:herm_ptime_partition_functions}. The following two lemmas yield the proof.

\begin{lemma}\label{lem:herm_ptime_case_non-bip}
Let $H$ be an $\denom$-algebraic $n \times n$ matrix and $\mfam D$ a family of diagonal matrices defining an \cond{H--STD}-problem which has a representation as given in conditions \cond{R1} through \cond{R5}. Assume further that the Affinity Condition \cond{AF} is satisfied.
Then the problem $\evalk(H,\mfam D)$ is polynomial time computable.
\end{lemma}

\begin{lemma}\label{lem:herm_ptime_case_bip}
Let $A$ be an $\denom$-algebraic and $\mfam D$ a family of diagonal matrices defining a \cond{B--H--STD}-problem which has a representation as given in conditions \cond{B--R1} through \cond{B--R5}. Assume further that the Affinity Condition \cond{B--AF} is satisfied.
Then the problem $\evalk(A,\mfam D)$ is polynomial time computable.
\end{lemma}
As in the previous sections, we will give the proof only 
of Lemma~\ref{lem:herm_ptime_case_non-bip}. The proof of 
Lemma~\ref{lem:herm_ptime_case_bip} is omitted due to its similarity to 
the former. For completeness, we refer to \cite{thu09} for a proof. 

The proof of Lemma~\ref{lem:herm_ptime_case_non-bip} relies on a reduction of 
the given partition function to functions $Z_q(f)$ closely related to degree 
$2$ polynomials $f$ over the ring $\Int_q$  ($q$ being a prime power). 
In the first part of this section we will introduce these functions 
and show how they can be computed in polynomial time.
Afterwards, we will show how to reduce the computational problems satisfying 
the preconditions of the Lemma to these problems. This will prove both results.

\subsection{A Polynomial Time Computable Problem}\label{sec:ptime_algo}
Let $q$ be a prime power and $f \in \Int_q[X_1, \ldots, X_n]$ a polynomial. 
Define $\zeta_q = \exp(\itpi\cdot q^{-1})$ -- we fix this definition for the 
rest of this section. Define the function
$$
Z_q(f) = \sum_{X_1,\ldots, X_n \in \Int_q} \zeta_q^{f(X_1,\ldots, X_n)}.
$$
We say that $f$ is of \sdefi{degree $2$}{degree $2$ polynomial}, 
if there are constants $c_{ij}, c_i \in \Int_q$ such that
\begin{equation}\label{eq:2205091417}
f(X_1, \ldots, X_n) = \sum_{i\le j} c_{ij}X_iX_j + \sum_{i=1}^n c_i X_i + c_0. 
\end{equation}
Let $\eval(q)$ denote the problem of
computing $Z_{q}(f)$ given a degree $2$ polynomial $f \in \Int_q[X_1,\ldots, X_n]$.
Since it has been shown in \cite{caichelu09} that this problem is polynomial
time computable, it will be central in the proof of polynomial time 
computability of partition functions.
\begin{thm}[Theorem 12.1 in \cite{caichelu09}]\label{thm:evalq_algo_ptime}
Let $q$ be a prime power. The problem $\eval(q)$ is polynomial time computable.
\end{thm}

\subsection{Computing Partition Functions --- The Non-Bipartite Case}

We will now prove Lemma~\ref{lem:herm_ptime_case_non-bip}. Recall that we are given an $\denom$-algebraic $n \times n$ matrix $H$ and a family $\mfam D$ of diagonal matrices defining an \cond{H--STD}-problem which has a representation as given in conditions \cond{R1} through \cond{R5} and the Affinity Condition \cond{AF} is satisfied. We shall show that the problem $\evalk(H,\mfam D)$ is polynomial time computable. 

The proof relies on the reduction of $\evalk(H,\mfam D)$ to the problems $\eval(q)$ which we have introduced in the preceding section. To make our partition functions at hand more amenable for this reduction, we will first analyze the structure of the vertex weights represented by the family $\mfam D$ a bit more.

\subsubsection{The Structure of the Mappings $\rcwf_c$}

Recall the mappings $\rcwf_{c}$ as given in condition \cond{R4} and assume further that they satisfy condition \cond{R5} and \cond{AF}.
In this section we will derive an important property of these functions. Before we state it, a word of preparation is in order. Recall that the image of all mappings $\rcwf_c$ lies in the Abelian group $\Omega$ defined in condition \cond{R1}. We will work with binomial coefficients which we define by $\binom{\lambda}{2} := \frac{\lambda(\lambda-1)}{2}$ for all $\lambda \in \Int$. Observe in particular that $\binom{\lambda}{2}$ is defined also for negative values and for each $g \in \Omega$ the expression $\binom{\lambda}{2} \cdot g$ is well-defined.

\begin{lemma}\label{lem:200109-2054}
For every $c \in \Int_\denom$ the mapping $\rcwf_c$ satisfies the following. Let $\lambda_1,\ldots, \lambda_z \in \Int$ and $h_1,\ldots,h_z \in \gG_c$ then
$$
\rcwf_{c}\left(\sum_{i=1}^z \lambda_ih_i\right) 
= \sum_{i=1}^z \lambda_i\rcwf_{c}(h_i) + \binom{\lambda_i}{2}\bil{\gamma_c(h_i), h_i} + \sum_{j < i} \lambda_i\lambda_j\bil{\gamma_c(h_i), h_j}
$$
Furthermore, for all $g \in \gG_c$ the following holds.
If $\ord(g)$ is odd, then the order of $\rcwf_{c}(g)$ divides 
$\ord(g)$.
If $\ord(g)$ is even, then the order of $\rcwf_{c}(g)$ divides 
$2\cdot\ord(g)$.
\end{lemma}
Recall the Affinity Condition \cond{AF}: For all $c \in \Int_{\denom}$ there is 
a $\gamma_c : \gG_c \rightarrow \gG$ such that 
\begin{equation}\label{eq:200109-1655}
\rcwf_{c}(y + x) - \rcwf_{c}(x) - \rcwf_{c}(y) = \bil{\gamma_c(y),x}  \text{ for all } x,y \in \gG_c.
\end{equation}
\begin{lemma}\label{lem:non-bip_hom_prop_gamma_c}
The following holds for all $c \in \Int_\denom$ and $y,y',x \in \gG_c$.
$$
\bil{\gamma_c(y+y'), x} = \bil{\gamma_c(y),x} + \bil{\gamma_c(y'),x} \quad \text{ and } \quad \bil{\gamma_c(x),y} = \bil{\gamma_c(y),x}. $$
\end{lemma}
\begin{proof}\marc{proof OK}
By commutativity of the addition $x+y$ equation \eqref{eq:200109-1655} yields
\begin{equation}\label{eq:200109-1741}
\bil{\gamma_c(y),x} = 
\rcwf_{c}(y + x) - \rcwf_{c}(x) - \rcwf_{c}(y)
=
\bil{\gamma_c(x),y}
\end{equation}
By the bilinearity of $\scalp{\absent, \absent}$ and equation \eqref{eq:200109-1741}
we have, for all $y,y' \in \gG_{c}$,
$\bil{\gamma_c(y+y'),x} =  \bil{\gamma_c(x),y+y'} =  \bil{\gamma_c(x),y} +  \bil{\gamma_c(x),y'}.$
A second application of \eqref{eq:200109-1741} on the right hand side of this 
yields $\bil{\gamma_c(y+y'),x} =  \bil{\gamma_c(y),x}  +\bil{\gamma_c(y'),x}$.
\end{proof}

\begin{proof}[of Lemma \ref{lem:200109-2054}]\marc{proof OK}
We will first prove that, for all $h_1, \ldots, h_z \in \gG_c$,
\begin{equation}\label{eq:040109-1}
\rcwf_{c}\left(\sum_{i=1}^z h_i\right) = \sum_{i=1}^z \rcwf_{c}(h_i) + \sum_{j < i} \bil{\gamma_c(h_i), h_j}.
\end{equation}
The proof is a straightforward induction on $z$. The case $z = 1$ is trivially true. For $z > 1$, recall that by equation \eqref{eq:200109-1655}
$$
\rcwf_{c}(y+x) = \rcwf_{c}(y) + \rcwf_{c}(x) + \bil{\gamma_c(x),y} \text{ for all } x,y \in \gG_c.
$$
This yields
\begin{eqnarray*}
\rcwf_{c}\left(\sum_{i=1}^z h_i\right) &=& \rcwf_{c}\left(\sum_{i=1}^{z-1} h_i\right) + \rcwf_{c}(h_z) + \bil{\gamma_c (h_z),\left(\sum_{i=1}^{z-1} h_i\right)} \\
&=& \sum_{i=1}^{z-1} \rcwf_{c}(h_i)  + \sum_{ j < i } \bil{\gamma_c(h_i),h_j } + \left(\rcwf_{c}(h_z) + \sum_{j=1}^{z-1}\bil{\gamma_c(h_z),h_j}\right).
\end{eqnarray*}
Here, we used the induction hypothesis for $\rcwf_{c}\left(\sum_{i=1}^{z-1} h_i\right)$ to obtains the second equality. This proves \eqref{eq:040109-1}. We claim that the following is true.
\begin{equation}\label{cl:040109-1}
 \text{ For all $\lambda \in \Int$ and $h \in \gG_c$, we have }
\rcwf_{c}(\lambda h) = \lambda\rcwf_{c}(h)  + \binom{\lambda}{2} \bil{\gamma_c(h),h}.
\end{equation}
Before we prove this, let us see how it helps proving the Lemma. By equation \eqref{eq:040109-1} we have
\begin{equation*}
\rcwf_{c}\left(\sum_{i=1}^z \lambda_ih_i\right) 
= \sum_{i=1}^z \rcwf_{c}(\lambda_ih_i) + \sum_{j < i} \lambda_i\lambda_j\bil{\gamma_c(h_i),h_j}.
\end{equation*}
And substituting $\rcwf_{c}(\lambda_ih_i)$, for all $i \in [z]$, with the expression given by equation \eqref{cl:040109-1} yields
\begin{eqnarray*}
\rcwf_{c}\left(\sum_{i=1}^z \lambda_ih_i\right)
&=& \sum_{i=1}^z \lambda_i\rcwf_{c}(h_i) + \binom{\lambda_i}{2}\bil{\gamma_c(h_i), h_i} + \sum_{j < i} \lambda_i\lambda_j\bil{\gamma_c(h_i), h_j}
\end{eqnarray*}
just as in the statement of the lemma.

\paragraph*{The proof of equation~\eqref{cl:040109-1}.} For $\lambda = 0$ this is trivial and for $\lambda > 0$ recall equation \eqref{eq:040109-1}. We have
\begin{equation}\label{eq:200109-1821}
\rcwf_{c}\left(\lambda h\right) = \rcwf_{c}\left(\sum_{i=1}^\lambda h\right) = \sum_{i=1}^\lambda \rcwf_{c}(h) + \binom{\lambda}{2} \bil{\gamma_c(h), h}.
\end{equation}
This proves equation~\eqref{cl:040109-1} for $\lambda\ge 0$. It remains to prove the case $\lambda < 0$. Before we do this we need to prove the second statement of the Lemma, namely
\begin{claim}\label{cl:040109-2}
Let $g \in \gG_c$. 
If $\ord(g)$ is odd, then the order of $\rcwf_{c}(g)$ divides the order of $g$.
If $\ord(g)$ is even, then the order of $\rcwf_{c}(g)$ divides $2\cdot \ord(g)$.
\end{claim}
\begin{clproof}
Let $N \ge 0$ be the order of $g$. That is, $N\cdot g = \gneut$ and equation \eqref{eq:200109-1821} implies
$$
\gneut = \rcwf_{c}(\gneut) = \rcwf_{c}(N\cdot g) 
= N \rcwf_{c}(g) + \binom{N}{2}\bil{\gamma_c(g), g}.
$$
If $N$ is odd then it divides $\binom{N}{2}$ and therefore $\binom{N}{2} g = \gneut$. Hence $\binom{N}{2}\bil{\gamma_c(g), g} = \bil{\gamma_c(g), \binom{N}{2}g} = \bil{\gamma_c(g), \gneut} = \gneut$ and the above equation thus implies $\gneut = N \rcwf_{c}(g)$. If $N$ is even then the same argument using $2N$ yields the proof.
\end{clproof}
Let us now prove equation~\eqref{cl:040109-1} for $\lambda < 0$. Define $N$ as the smallest multiple of $2\cdot \ord(h)$ such that $N+\lambda > 0$, then $\lambda h = (N+\lambda)h$ and
$ \rcwf_{c}(\lambda h) = \rcwf_{c}((N+\lambda)h)$. We may thus apply the part of equation~\eqref{cl:040109-1} which has been proved already and obtain
\begin{eqnarray*}
\rcwf_{c}((N+\lambda)h) &=& (N+\lambda)\rcwf_{c}(h)  + \binom{N+\lambda}{2} \bil{\gamma_c(h), h} \\
&=& \lambda\rcwf_{c}(h) + \binom{\lambda}{2} \bil{\gamma_c(h),h}.
\end{eqnarray*}
The second equality follows from Claim~\ref{cl:040109-2}.
\end{proof}
From Lemma~\ref{lem:200109-2054} we can derive a rather technical result which will become necessary in the proof of Lemma~\ref{lem:herm_ptime_case_non-bip}.
\begin{lemma}\label{lem:order_of_rho_plus_bil}
Let $g \in \gG_c$ be an element of order $q = 2^k$. The order of the element
$2\cdot \rcwf_{c}(g) + \bil{\gamma_c(g), g}$ in $\Int_{\denom}$ is at most $q/2$.
\end{lemma}
\begin{proof}\marc{proof OK}
By Lemma~\ref{lem:200109-2054} we have
\begin{equation}\label{eq:2205091713}
\gneut = \rcwf_{c}(\gneut) = \rcwf_{c}(q\cdot g) 
= q \rcwf_{c}(g) + \binom{q}{2}\bil{\gamma_c(g), g}. 
\end{equation}
and the order of $\rcwf_c(g)$ divides $2q$. Note that the order of $\bil{\gamma_c(g), g}$ divides $q$ by $q\bil{\gamma_c(g), g} = \bil{\gamma_c(g), q\cdot g}= \bil{\gamma_c(g), \gneut} = \gneut$.

Assume first that the order of $\bil{\gamma_c(g), g}$ equals $q$. Then the element
$2\cdot \rcwf_{c}(g) + \bil{\gamma_c(g), g}$ has order at most $q/2$ only if the order of $2\rcwf_{c}(g)$ is $q$, as well. Assume therefore, for contradiction, that the order of $2\rcwf_{c}(g)$ divides $q/2$.  Equation \eqref{eq:2205091713} implies
$$
\gneut = \binom{q}{2}\bil{\gamma_c(g),g}.
$$
As $q-1$ is odd, $q$ does not divide $\binom{q}{2}$ --- a contradiction.

Assume now that the order of $\bil{\gamma_c(g),g}$ is strictly smaller than $q$, that is $q/2 \cdot \bil{\gamma_c(g),g} = 0$. As $q/2$ divides $\binom{q}{2}$ equation \eqref{eq:2205091713} implies that 
$\gneut = q \rcwf_{c}(g)$. This finishes the proof.
\end{proof}

\subsubsection{The Final Reduction}
We will now prove Lemma~\ref{lem:herm_ptime_case_non-bip}. We start with 
some preliminaries. Consider $\Int_q$-polynomial $f$ of degree $2$ as in equation
\eqref{eq:2205091417}.
With each variable $X_i$ we associate an order $\ord(X_i)$ which divides $q$. A degree $2$ monomial $c_{ij}X_iX_j$ for $i \neq j$ is called \sdefi{consistent}{consistent monomial} if the values $c_{ij}\cdot \ord(X_i)$ and $c_{ij}\cdot \ord(X_j)$ are divisible by $q$. Similarly, a monomial $c_iX_i$ is consistent if $q$ divides $c_i\cdot \ord(X_i)$. Finally, a \sdef{square monomial} $c_{ii}X^2_i$ is consistent if $2 \cdot c_{ii}\cdot \ord(X_i)$ is divisible by $q$.

\begin{lemma}\label{lem:inhom_reduce_to_hom_eval}
Let $q$ be a prime power. The following problem can be reduced to $\eval(q)$ in polynomial time.
The input consists of a degree $2$ polynomial $f \in \Int_q[X_1,\ldots, X_n]$ of the form given by equation \eqref{eq:2205091417}. Each variable $X_i$ has an order $\ord(X_i)$ which divides $q$ and all non-constant monomials are consistent. The task is to compute the value
$$
\sum_{X_1 \in \Int_{\ord(X_1)}} \cdots \sum_{X_n \in \Int_{\ord(X_n)}} \zeta_q^{f(X_1,\ldots,X_n)}.
$$
\end{lemma}
\begin{proof}\marc{proof OK}
To perform the reduction of the problem at hand to $\eval(q)$ we will need to show how to ``lift'' the order of each variable $X_i$ to $q$. Assume w.l.o.g. that the order $x_i := \ord(X_i)$ of each variable is greater than $1$. Fix some variable $X_\nu$. We will have a look at the situation after the order of $X_\nu$ has been lifted. For all $i \in [n]$ define $q_i = q \cdot x_{i}^{-1}$. We have
\begin{equation}\label{eq:2205091350}
\sum_{\forall i \neq \nu :\; X_i \in \Int_{x_i}}\sum_{X_\nu \in \Int_{q}}\zeta_q^{f(X_1,\ldots, X_\nu, \ldots, X_n)}  = 
\sum_{\forall i \neq \nu :\; X_i \in \Int_{x_i}}\sum_{X_\nu \in \Int_{x_\nu}}
\sum_{j = 0}^{q_\nu - 1} \zeta^{f(X_1,\ldots, X_\nu + j\cdot x_\nu, \ldots , X_n)}.
\end{equation}
Let $\ell(X_1,\ldots,\hat X_\nu,\ldots, X_n) = \sum_{i < \nu} c_{i \nu}X_i + \sum_{\nu < i} c_{\nu i}X_{i}$ and fix the degree $2$ polynomial $h$ which captures the part of $f$ not involving $X_\nu$. We have
$$
f(X_1, \ldots, X_n) = c_{\nu\nu}X_\nu^2 + X_\nu\cdot \ell(X_1,\ldots,\hat X_\nu,\ldots, X_n) + h(X_1,\ldots,\hat X_\nu,\ldots, X_n).
$$
Further
\begin{eqnarray*}
f(X_1,\ldots, X_\nu + j \cdot x_\nu, \ldots , X_n) 
&=& c_{\nu\nu}(X_\nu + j\cdot x_\nu)^2 + (X_\nu + j\cdot x_\nu)\cdot \ell + h \\
&=& c_{\nu\nu} X^2_\nu + 2c_{\nu\nu}x_{\nu}\cdot j X_\nu + c_{\nu\nu}x_{\nu}^2\cdot j^2 + X_{\nu}\ell + jx_{\nu}\ell + h
\end{eqnarray*}
By the consistency of the monomials $c_iX_i$ and that of $c_{ij}X_iX_j$ for all $i \neq j$ we see that all coefficients of $x_{\nu}\ell$ are divisible by $q$. This is also true for $2c_{\nu\nu}x_\nu$. To see this for $c_{\nu\nu}x^2_{\nu}$ note first that if $q$ is a power of $2$ then $c_{\nu\nu}x^2_{\nu}$ is divisible by $2c_{\nu\nu}x_{\nu}$ as $2$ divides $x_\nu$. Otherwise the condition that $2c_{\nu\nu}x_{\nu}$ be divisible by $q$ is equivalent to $c_{\nu\nu}x_{\nu}$ being divisible by $q$.
Therefore, for all $j \in \{0,\ldots,q_{\nu} - 1\}$ we have 
$f(X_1,\ldots, X_\nu + j \cdot x_\nu, \ldots , X_n) \equiv c_{\nu\nu} X^2_\nu + X_{\nu}\ell + h \mmod{q}$. Using equation \eqref{eq:2205091350}, we see that
$$
\sum_{\forall i \neq \nu :\; X_i \in \Int_{x_i}}\sum_{X_{\nu} \in \Int_{q}}\zeta_q^{f(X_1,\ldots, X_n)}
= q_\nu \cdot \sum_{X_1 \in \Int_{x_1}} \cdots \sum_{X_n \in \Int_{x_n}} \zeta_q^{f(X_1,\ldots, X_n)}.
$$
As this holds independently for all $\nu \in [n]$, it follows that
$$
Z_{q}(f(X_1,\ldots,X_n)) = \left(\prod_{i=1}^n q_i\right) \sum_{X_1 \in \Int_{x_1}} \cdots \sum_{X_n \in \Int_{x_n}} \zeta_q^{f(X_1,\ldots,X_n)}.
$$
This finishes the proof.
\end{proof}

\subsubsection{The proof of Lemma \ref{lem:herm_ptime_case_non-bip}}
\begin{proof}[of Lemma \ref{lem:herm_ptime_case_non-bip}]\marc{proof OK}
Let now $\evalk(H,\mfam D)$ be an \cond{H--STD} problem which meets all conditions \cond{R1}--\cond{R4} and \cond{AF}. Let $G = (V,E)$ be a digraph an $\vpin$ a pinning. With $V' = V \setminus \df(\vpin)$ we have
\begin{equation}\label{eq:2305091414}
Z_{H,\mfam{D}}(\vpin,G) = \sum_{\vpin \subseteq \vcfg:V \rightarrow \gG} \prod_{uv \in E} H_{\vcfg(u),- \vcfg(v)} \prod_{v \in V'}D^{\cngc{\grade(v)}}_{\vcfg(v),\vcfg(v)}
\end{equation}
We shall prove that $Z_{H,\mfam{D}}(\vpin,G)$ can be computed in polynomial time. To do this, we will present a polynomial time reduction to the problems $\eval(q)$. Most of this proof will be devoted to rewriting the expression of this partition function into a form which resembles the expression $Z_q(f)$ used in $\eval(q)$.

Recall the representation of $H$ and $\mfam D$ as given in conditions \cond{R1}--\cond{R5} on page \pageref{pg:repres_criteria}. Note that a configuration $\vcfg$ which maps some vertex $v \in V'$ to $\vcfg(v) \notin \Lambda_{\cngc{\grade(v)}}$ does not contribute to this partition function. Assume therefore that $\vcfg(v) \in \Lambda_{\cngc{\grade(v)}}$ is satisfied for all $v\in V$. Slightly abusing notation we will denote $\Lambda_{\cngc{\grade(v)}}$ by $\Lambda_v = \beta_v + \gG_v$. We introduce, for each $v \in V$ a variable $\vcfg_v \in \gG_{v}$. For all $v \in \df(\vpin)$ we assume that the value of $\vcfg_v$ is fixed to $\vcfg_v = \vpin(v)$.
Application of the representation \cond{R1}--\cond{R5} thus yields
\begin{eqnarray*}
 Z_{H,\mfam{D}}(\vpin,G) &=& \sum_{v \in V'}\sum_{\vcfg_v \in \gG_{v}} \prod_{uv \in E} H_{\beta_u + \vcfg_u,-(\beta_v + \vcfg_v)} \prod_{v \in V'}D^{\cngc{\grade(v)}}_{\beta_{v} + \vcfg_v,\beta_{v} + \vcfg_v} \\
&=& \sum_{v \in V'}\sum_{\vcfg_v \in \gG_{v}} \caniso\left[\sum_{uv \in E} \bil{\beta_u+\vcfg_u,\beta_v+\vcfg_v} + \sum_{v \in V'} \rcwf_{\cngc{\grade(v)}}(\vcfg_v)\right]
\end{eqnarray*}
Let us rewrite the expression $\sum_{uv \in E} \bil{\beta_u+\vcfg_u,\beta_v+\vcfg_v}$ a little bit. Denote by $\mu_{uv}$ for all $u,v \in V$ the multiplicity of the edge $uv$ in $E$. Then, using the skew-bilinearity of the $\bil{\absent, \absent}$ operator,
\begin{eqnarray*}
\sum_{uv \in E} \bil{\beta_u+\vcfg_u,\beta_v+\vcfg_v}
& = & \sum_{u,v \in V} \mu_{uv} \bil{\beta_u+\vcfg_u,\beta_v+\vcfg_v} \\
& = & \sum_{u,v \in V} \mu_{uv} \left(\bil{\beta_u,\beta_v} 
      + \bil{\beta_u,\vcfg_v}
      - \bil{\beta_v,\vcfg_u}
      + \bil{\vcfg_u,\vcfg_v}\right) 
\end{eqnarray*}
For each $v \in V$ define the value $\eta_v = \sum_{u \in V} (\mu_{uv} - \mu_{vu}) \beta_u$. Then
\begin{eqnarray*}
\sum_{uv \in E} \bil{\beta_u+\vcfg_u,\beta_v+\vcfg_v}
& = & \sum_{u,v \in V} \mu_{uv}\bil{\beta_u,\beta_v} 
      + \mu_{uv}\bil{\vcfg_u,\vcfg_v}
      + \sum_{v \in V} \bil{\eta_v,\vcfg_v}
\end{eqnarray*}
Let $\rcwf_{v}$ denote the mapping $\rcwf_{\cngc{\grade(v)}}$ and define a value
$$
c = \caniso\left[\sum_{u,v \in V} \mu_{uv} \bil{\beta_u,\beta_v} + \sum_{v \in \df(\vpin)} \bil{\eta_v,\vcfg_v} \right]
$$
The value $c$ is constant for all different values the $\vcfg_v$ corresponding to $v \in V'$ may assume. We have
$$
 Z_{H,\mfam{D}}(\vpin,G) 
=\sum_{v \in V'}\sum_{\vcfg_v \in \gG_{v}} 
\caniso\left[\sum_{u,v \in V} \mu_{uv}\bil{\beta_u,\beta_v} 
      + \mu_{uv}\bil{\vcfg_u,\vcfg_v}
      + \sum_{v \in V} \bil{\eta_v,\vcfg_v} + \sum_{v \in V'} \rcwf_{v}(\vcfg_v)\right] \\
$$
which, by a straightforward computation yields
\begin{equation}\label{eq:2305091528}
Z_{H,\mfam{D}}(\vpin,G) 
= c \cdot \sum_{v \in V'}\sum_{\vcfg_v \in \gG_{v}} 
\caniso\left[\sum_{u,v \in V} \mu_{uv}\bil{\vcfg_u,\vcfg_v}
      +  \sum_{v \in V'} \bil{\eta_v,\vcfg_v} + \rcwf_{v}(\vcfg_v)\right]
\end{equation}
We will proceed by further rewriting the terms of the expression
\begin{equation}\label{eq:1204091227}
\sum_{u,v \in V} \mu_{uv} \bil{\vcfg_u,\vcfg_v} + \sum_{v \in V'} \bil{\eta_v,\vcfg_v} + \rcwf_{v}(\vcfg_v).
\end{equation}
For some technical reason which will become clear later, we now decompose $2\denom$ into prime powers. Let $\denom_1 \cdots \denom_t = 2\denom$ be a decomposition of $2\denom$ into powers of pairwise distinct primes. Denote by $p_\nu$ the prime corresponding to $\denom_\nu$ and assume w.l.o.g. that $\denom_1$ is a power of $2$ --- if $\denom$ is not divisible by $2$ then we let $\denom_1 = 2$. 

Recall once more the representation given by \cond{R1}--\cond{R5} on page \pageref{pg:repres_criteria}. We have $\gG = \Int_{q_1} \oplus \ldots \oplus \Int_{q_z}$ for some prime powers $q_i$. It will be convenient to write this decomposition of $\gG$ in terms of the above prime decomposition of $2\denom$. Define sets $P_\nu = \{ i \in [z] \mid p_{\nu} \text{ divides } q_i\}$ for each $\nu \in [t]$. By condition \cond{R2}, the order of each $g \in \gG$ divides $\denom$ implying that the $P_1, \ldots, P_t$ form a partition (with possibly empty parts) of $[z]$.

Therefore, we rewrite the decomposition of $\gG$ by $\gG = \bigoplus_{\nu = 1}^t \bigoplus_{i \in P_\nu} \Int_{q_i}$. Further, with each $\gG_c$ being a subgroup of $\gG$ we see that they also have decompositions $\gG_c = \bigoplus_{\nu = 1}^t \bigoplus_{i \in P_\nu} \Int_{q_{c,i}}$ such that w.l.o.g. $q_{c,i}$ divides $q_i$ for all $c \in \Int_{\denom}$ and all $i \in [z]$. 
Let $g_{c,i}$ be a generator of $\Int_{q_{c,i}}$ for all $c \in \Int_\denom$, $\nu \in [t]$ and $i \in P_\nu$. For notational convenience, for every $v \in V$ we denote by $g_{v,i}$ the element $g_{\cngc {\grade(v)},i}$ for all $i \in [z]$.

Define a vector $\vec X = (X_{(v,i)})_{v\in V, i \in [z]}$. Every $\vcfg_v \in \gG_{v}$ has a representation in terms of a tuple $(X_{(v,1)},\ldots,X_{(v,z)})$ of variables each satisfying $X_{(v,i)} \in \Int_{q_{v,i}}$ such that
\begin{equation}\label{eq:2305091521}
\vcfg_v = \sum_{i=1}^{z} X_{(v,i)} g_{v,i}.
\end{equation}
Define matrices $\Gamma$, $\Delta$ and a vector $\Lambda$, with indices in $V \times [z]$ such that, for all $(u,i),(v,j) \in V\times [z]$,
\begin{eqnarray}
\label{eq:nonbip_def_form_delta}
\Delta_{(u,i)(v,j)} &=& \left\lbrace \begin{array}{l l}
                                     \bil{\gamma_v(g_{v,i}),g_{v,j}} \quad\quad&, \text{ if } u=v \in V' \text{ and } j < i \\
                                     0 &, \text{ otherwise.}
                                   \end{array}\right.\\
\label{eq:nonbip_def_form_Gamma}
\Gamma_{(u,i)(v,j)} &=& \mu_{uv}\cdot \bil{g_{u,i}, g_{v,j}} \\
\label{eq:nonbip_def_form_lambda}
\Lambda_{(v,i)} &=&\left\lbrace \begin{array}{l l}
                                     \rcwf_v(g_{v,i}) + \bil{\eta_v, g_{v,i}} &, \text{ if } v \in V', i \in [z]\\
                                     0 &, \text{ otherwise.}
                                   \end{array}\right. 
\end{eqnarray}
Further, define vectors $\vec B(\vec X)$ and $\Xi$ such that for all $(v,i) \in V\times [z]$,
\begin{eqnarray}
\vec B(\vec X)_{(v,i)} &=& \left\lbrace \begin{array}{l l}
                                     \binom{X_{(v,i)}}{2} \quad \quad\;\,\quad&, \text{ if } v \in V', i \in [z]\\
                                     0 &, \text{ otherwise.}
                                   \end{array}\right. \\
\Xi_{(v,i)} &=& \left\lbrace \begin{array}{l l}
                                     \bil{\gamma_v(g_{v,i}), g_{v,i}} &, \text{ if } v \in V', i \in [z]\\
                                     0 &, \text{ otherwise.}
                                   \end{array}\right.
\label{eq:2305091538}
\end{eqnarray}
We can now rewrite the terms of equation \eqref{eq:1204091227}. Applying equation \eqref{eq:2305091521} and the definition of $\Gamma$ in \eqref{eq:nonbip_def_form_Gamma}, we find that
$$
\sum_{u,v \in V} \mu_{uv} \bil{\vcfg_u,\vcfg_v} = \sum_{u,v \in V}\sum_{i,j=1}^{z} X_{(u,i)}X_{(v,j)} \cdot \mu_{uv}\bil{g_{u,i}, g_{v,j}}
= \vec X \Gamma \vec{X}^T.
$$
Further we have, for each $v \in V'$
\begin{eqnarray*}
\bil{\eta_v,\vcfg_v} + \rcwf_{v}(\vcfg_v) &=& \sum_{i=1}^{z} X_{(v,i)} \bil{\eta_v, g_{v,i}} + \rcwf_{v}\left(\sum_{i=1}^{z} X_{(v,i)}g_{v,i}\right) \\
&=& \sum_{i=1}^{z} X_{(v,i)}\left(\rcwf_{v}(g_{v,i}) + \bil{\eta_v, g_{v,i}}\right) + \binom{X_{(v,i)}}{2}\bil{\gamma_v(g_{v,i}), g_{v,i}} \\
& & + \sum_{j < i} X_{(v,i)}X_{(v,j)}\bil{\gamma_v(g_{v,i}), g_{v,j}}.
\end{eqnarray*}
where the second equality follows from Lemma~\ref{lem:200109-2054}.
By the definition of $\Delta$,$\Lambda$, $\vec B( \vec X)$ and $\Xi$ this can be rewritten into
\begin{eqnarray*}
\sum_{v \in V'} \bil{\eta_v,\vcfg_v} + \rcwf_{v}(\vcfg_v) 
&=& \vec X \cdot \Lambda^T  + \vec X \Delta \vec X^T + \vec B( \vec X) \cdot \Xi^T
\end{eqnarray*}
Define $f(\vec X) = \vec X \Gamma \vec{X}^T + \vec X \cdot \Lambda^T  + \vec X \Delta \vec X^T + \vec B( \vec X) \cdot \Xi^T$.
By equation \eqref{eq:2305091528} we therefore have
\begin{equation}
Z_{H,\mfam{D}}(\vpin,G) = c \cdot \sum_{\vec X} \caniso\left[f\left( \vec X\right) \right]
\end{equation}
where the sum is over all vectors $\vec X = (X_{(v,i)})_{(v,i) \in V \times [z]}$ such that $X_{(v,i)} \in \Int_{q_{v,i}}$ for all $(v,i) \in V' \times [z]$. For all $(v,i) \in \df(\vpin) \times [z]$ the values $X_{(v,i)}$ are fixed so as to satisfy $\vcfg_v = \vpin(v)$. Note that this is always possible by the definition of $\vcfg_v$ in equation \eqref{eq:2305091521}. We call any vector $\vec X$ satisfying these conditions a \emph{valid assignment}.

We will simplify the description of $f$ further. Define $\m I = \{ I_{\nu} \mid \nu \in [t]\}$ with $I_{\nu} = \{ (v,i) \mid v\in V,\; i \in P_{\nu}\}$ --- That is, the sets $I_\nu$ form a partition of $V \times [z]$. Recall that $X_I$ and $\Delta_{IJ}$ denote subvectors and submatrices of $X$ and $\Delta$ induced by the indices in $I$ and $J$. Then
$$
f(\vec X) = \sum_{I,J \in \m I} \vec X_I \Gamma_{IJ} \vec{X}_J^T + \vec X_I \Delta_{IJ} \vec X_J^T + \sum_{I \in \m I} \vec X_I \cdot \Lambda_I^T  + \vec B( \vec X)_I \cdot \Xi_I^T
$$
We will claim that the "cross-terms" in this expression are zero, i.e.
$$f(\vec X) = \sum_{I \in \m I} \vec X_I \Gamma_{II} \vec{X}_I^T + \vec X_I \Delta_{II} \vec X_I^T + \vec X_I \cdot \Lambda_I^T  + \vec B( \vec X)_I \cdot \Xi_I^T.
$$
To see this, we shall prove that, for all $I,J \in \m I$ with $I \neq J$ we have $\Gamma_{IJ} = 0$ and $\Delta_{IJ} = 0$. We show this for $\Delta_{IJ}$, the result follows analogously for $\Gamma$. By definition (cf. equation \eqref{eq:nonbip_def_form_delta}), for $v \in V$, $i \in I$ and $j \in J$ we have $\Delta_{(v,i)(v,j)} = \bil{\gamma_v(g_{v,i}),g_{v,j}}$. By the properties of $\gamma_v$ (cf. Lemma~\ref{lem:non-bip_hom_prop_gamma_c}) and the bilinearity of $\bil{\absent,\absent}$ we have $N\bil{\gamma_v(g_{v,i}),g_{v,j}} = \bil{\gamma_v(Ng_{v,i}),g_{v,j}} = \bil{\gamma_v(g_{v,i}),Ng_{v,j}}$.
Therefore the order of $\Delta_{(u,i)(v,j)}$ divides both, $\ord_{\gG}(g_{v,i})$ and $\ord_{\gG}(g_{v,j})$. By our choice of $I$ and $J$ these orders are co-prime and the claim follows. Define
\begin{equation}\label{eq:2305091714}
f_\nu(\vec X_{I}) = \vec X_I \Gamma_{II} \vec{X}_{I}^T + \vec X_I \Delta_{II} \vec X_I^T + \vec X_I \cdot \Lambda_I^T  + \vec B( \vec X)_I \cdot \Xi_I^T 
\text{ for every $\nu$ and $I = I_\nu$}.
\end{equation}
Thus 
\begin{equation*}
f(\vec X) = \sum_{\nu = 1}^{t} f_\nu(\vec X_{I_\nu}) 
\end{equation*}

\begin{claim}\label{cl:1604091735}
The order of all entries of $\Gamma, \Delta$ and $\Xi$ divides $\denom$ and the order of all entries of $\Lambda$ divides $2\denom$. 
\end{claim}
\begin{clproof}
For $\Delta, \Gamma$ and $\Xi$ this follows directly from their definitions in equations \eqref{eq:nonbip_def_form_delta}, \eqref{eq:nonbip_def_form_Gamma} and \eqref{eq:2305091538}. Every non-zero entry of $\Lambda$ satisfies $\Lambda_{(v,i)} =  \rcwf_v(g_{v,i}) + \bil{\eta_v, g_{v,i}}$ and the claim follows from Lemma~\ref{lem:200109-2054}.
\end{clproof}
\begin{claim}\label{cl:2305091701}
Let $\nu \in [t]$ and $I = I_\nu$, then all non-zero entries of the matrices $\Gamma_{I I}$, $\Delta_{II}$ and the vectors $\Lambda_{I}$ and $\Xi_I$ are divisible by $\denom_\mu$ for all $\mu \neq \nu$.
\end{claim}
\begin{clproof}
For $(u,i), (v,j) \in I_\nu$ we have $\Gamma_{(u,i), (v,j)} = \mu_{uv}\cdot \bil{g_{u,i}, g_{v,j}}$. The order of this element (in $\Int_{2\denom}$) in is some divisor of $\denom_\nu$ and therefore the element itself is divisible by $2\denom \cdot (\denom_{\nu})^{-1}$.
The proofs of the other statements follow analogously.
\end{clproof}
Claim~\ref{cl:1604091735} implies that $f(\vec X) \in \Int_{2\denom}$ for all valid assignments of $\vec X$. Define $\caniso_{2\denom}(x) = \exp(\frac{\itpi \cdot x }{2\denom})$. We see that
\begin{eqnarray*}
Z_{H,\mfam{D}}(\vpin,G) &=& c \cdot \sum_{\alpha \in \Int_{2\denom}} \caniso_{2\denom}\left[\alpha\right]\cdot  \left|\left\{\vec X \,:\, f(\vec X) \equiv \alpha \mmod{2\denom}\right\}\right|
\end{eqnarray*}
By Claim~\ref{cl:2305091701} we see further that
$f(\vec X) \equiv \alpha \mmod{2\denom}$
is equivalent to a system of simultaneous congruences given by
$f_{\nu}(\vec X_{I_{\nu}}) \equiv \alpha \mmod{\denom_{\nu}} \text{ for all } \nu \in [t]$.
We have
\begin{eqnarray*}
Z_{H,\mfam{D}}(\vpin,G) &=& c \cdot \sum_{\alpha \in \Int_{2\denom}} \caniso_{2\denom}\left[\alpha\right] \prod_{\nu = 1}^t \left|\left\{ \vec X_{I_\nu} \,:\, f_{\nu}(\vec X_{I_\nu}) \equiv \alpha \mmod{\denom_{\nu}}\right\}\right| 
\end{eqnarray*}
By the Chinese Remainder Theorem each $\alpha$ can be written uniquely as $\alpha = \sum_{\nu=1}^t \frac{2\denom}{\denom_{\nu}} \cdot \alpha_\nu$ for values $\alpha_\nu \in [0,\denom_\nu-1]$.  Recall that by $\zeta_{q}$ we denote the value $\zeta_q = \exp(\itpi\cdot q^{-1})$. Then
\begin{eqnarray*}
Z_{H,\mfam{D}}(\vpin,G) &=& c \cdot \sum_{\alpha_1 \in \Int_{\denom_1}} \cdots \sum_{\alpha_t \in \Int_{\denom_t}} \prod_{\nu = 1}^t \caniso_{2\denom}\left[\frac{2\denom}{\denom_{\nu}}\cdot \alpha_\nu\right] \cdot  \left|\left\{ \vec X_{I_\nu} \,:\, f_{\nu}(\vec X_{I_\nu}) \equiv \alpha_\nu \mmod{\denom_{\nu}}\right\}\right| \\
&=& c \cdot\prod_{\nu = 1}^t \sum_{\alpha_\nu \in \Int_{\denom_\nu}} \zeta_{\denom_{\nu}}^{\alpha_\nu} \cdot \left|\left\{ \vec X_{I_\nu} \,:\, f_{\nu}(\vec X_{I_\nu}) \equiv \alpha_\nu \mmod{\denom_{\nu}}\right\}\right|  \\
\end{eqnarray*}
The partition function $Z_{H,\mfam{D}}(\vpin,G)$ thus simplifies to
\begin{eqnarray*}
Z_{H,\mfam{D}}(\vpin,G) &=& c \cdot\prod_{\nu = 1}^t \sum_{\vec X_{I_\nu}} \zeta_{\denom_{\nu}}^{f_{\nu}(\vec X_{I_\nu})} 
\end{eqnarray*}
and its computation reduces to the computation of $t$ many independent values of the form
\begin{equation}\label{eq:2305091733}
\sum_{\vec X_{I_\nu}} \zeta_{\denom_{\nu}}^{f_{\nu}(\vec X_{I_\nu})}.
\end{equation}
We shall show now, that each of these problems satisfies the preconditions of Lemma~\ref{lem:inhom_reduce_to_hom_eval}. This then implies that each problem is reducible to $\eval(\denom_\nu)$ and therefore polynomial time computable by Theorem~\ref{thm:evalq_algo_ptime}.

Fix some $\nu \in [t]$. To show that the problem \eqref{eq:2305091733} satisfies the preconditions of Lemma~\ref{lem:inhom_reduce_to_hom_eval}, we have to show that the monomials in $f_\nu$ are consistent.
In particular, the square monomials have to be consistent. Recall that this means that they are of the form $b \cdot X^2_{(v,i)}$ such that $2 \cdot b \cdot \ord(X_{(v,i)}) = 2 \cdot b \cdot q_{v,i}$ is divisible by $q$. By equation \eqref{eq:2305091714} and $I = I_\nu$, we have 
$$
f_\nu(\vec X_{I}) = \vec X_I \Gamma_{II} \vec{X}_{I}^T + \vec X_I \Delta_{II} \vec X_I^T + \vec X_I \cdot \Lambda_I^T  + \vec B( \vec X)_I \cdot \Xi_I^T.
$$
Observe that the condition $\vcfg_v = \vpin(v)$ for all $v \in \df(\vpin)$ turns some variables $X_{(v,i)}$ into constants. Therefore, some degree $2$ monomials turn into linear ones and others of these and some linear ones turn into constants. Therefore, to show that the consistency preconditions of Lemma~\ref{lem:inhom_reduce_to_hom_eval} are satisfied, it suffices to do this for the original polynomial $f_{\nu}$.
For the terms $\vec X_I \Gamma_{II} \vec{X}_{I}^T + \vec X_I \Delta_{II} \vec X_I^T$ the consistency of the corresponding monomial follows from their definition.
It remains to consider the monomials arising from $\vec X_I \cdot \Lambda_I^T  + \vec B( \vec X)_I \cdot \Xi_I^T$.
Recall that
\begin{equation}\label{eq:2305091751}
\vec X_I \cdot \Lambda_I^T + \vec B( \vec X)_I \cdot \Xi_I^T 
= \sum_{(v,i) \in I} X_{(v,i)} \cdot (\rcwf_v(g_{v,i}) + \bil{\eta_v, g_{v,i}}) + \binom{X_{(v,i)}}{2}\bil{\gamma_v(g_{v,i}), g_{v,i}}
\end{equation}
\paragraph*{Case A. $\nu \neq 1$.} We have $\denom_{\nu}$ a prime power of some odd prime $p_{\nu}$. Note that $2$ has a multiplicative inverse in $\Int_{\denom_{\nu}}$ and therefore the term
$$
\binom{X_{(v,i)}}{2}\bil{\gamma_v(g_{v,i}), g_{v,i}} = \dfrac{X_{(v,i)}\left(X_{(v,i)} - 1\right)}{2}\bil{\gamma_v(g_{v,i}), g_{v,i}}
$$
in equation \eqref{eq:2305091751} produces consistent monomials.
Further, the term $X_{(v,i)} \cdot (\rcwf_v(g_{v,i}) + \bil{\eta_v, g_{v,i}})$ is consistent as well. To see this, note that $X_{(v,i)} \cdot \bil{\eta_v, g_{v,i}}$ is consistent by definition of $X_{(v,i)}$ since $\ord(X_{(v,i)}) = \ord(g_{v,i})$. By Lemma~\ref{lem:200109-2054} the term
$X_{(v,i)} \cdot \rcwf_v(g_{v,i})$ is consistent as the order of $g_{v,i}$ is odd.

\paragraph*{Case B. $\nu =1$.}
By equation \eqref{eq:2305091751} we have
\begin{eqnarray*}
\vec X_I \cdot \Lambda_I^T + \vec B( \vec X)_I \cdot \Xi_I^T 
&=&  \sum_{(v,i) \in I} X_{(v,i)} \cdot (\rcwf_v(g_{v,i}) + \bil{\eta_v, g_{v,i}} - 2^{-1}\cdot \bil{\gamma_v(g_{v,i}),g_{v,i}}) \\
& & +\sum_{(v,i) \in I} X^2_{(v,i)} \cdot 2^{-1} \cdot \bil{\gamma_v(g_{v,i}), g_{v,i}} \\ 
\end{eqnarray*}
Note that the expression $2^{-1} \cdot \bil{\gamma_v(g_{v,i}), g_{v,i}}$ is well-defined, as $\bil{\gamma_v(g_{v,i}), g_{v,i}}$ has, by definition of $\denom_1$, order at most $\denom_1 / 2$. This proves that for the above square terms we have
$$
2 \cdot \ord(X_{(v,i)}) \cdot 2^{-1} \cdot \bil{\gamma_v(g_{v,i}), g_{v,i}} =
\ord(X_{(v,i)}) \cdot \bil{\gamma_v(g_{v,i}), g_{v,i}}
$$
is divisible by $\denom_1$.
Further, $X_{(v,i)} \cdot \bil{\eta_v, g_{v,i}}$ is consistent. And it remains to show this for $X_{(v,i)} \cdot (\rcwf_v(g_{v,i}) - 2^{-1}\cdot \bil{\gamma_v(g_{v,i}),g_{v,i}})$ --- which follows from Lemma~\ref{lem:order_of_rho_plus_bil}.
\end{proof}

\bibliographystyle{amsalpha}
\bibliography{bib}

\end{document}